\documentclass[aps,prd,nofootinbib]{revtex4}

\usepackage{graphicx}
\usepackage{slashed}

\begin{document}

\title{Heavy Quarkonium Production through the Semi-Exclusive $e^+ e^-$ Annihilation Channels around the $Z^0$ Peak}

\author{Zhan Sun}
\author{Xing-Gang Wu}
\email{wuxg@cqu.edu.cn}
\author{Gu Chen}
\author{Jun Jiang}
\author{Zhi Yang}

\address{Department of Physics, Chongqing University, Chongqing 401331, P.R. China}

\date{\today}

\begin{abstract}

Within the framework of the non-relativistic QCD, we present a detailed discussion on the heavy quarkonium production at the leading order in $\alpha_s$ at a $e^+ e^-$ collider with the collision energy around the $Z^0$ peak. Quarkonia are produced through the semi-exclusive channels $e^{+}e^{-} \rightarrow |H_{Q\bar{Q}}\rangle +X$ with $X=Q\bar{Q}$ or $gg$, where $Q$ indicates a heavy quark (respectively $b$ or $c$). It is noted that in addition to the color-singlet $1S$-level quarkonium states, the $2S$ and $1P$ color-singlet states and the color-octet $|(Q\bar{Q})[1^3S_1^{({\bf 8})}]g\rangle$ state also provide sizable contributions. The heavy quarkonium transverse momentum and rapidity distributions for the $e^+ e^-$ collision energy $E_{cm}=m_Z$ are presented. For both charmonium and bottomonium production via the $Z^0$ propagator, there is approximate ``spin degeneracy" between the spin-triplet and spin-singlet quarkonium states. Uncertainties for the total cross sections are estimated by taking $m_c=1.50\pm0.15$ GeV and $m_b=4.90\pm0.15$ GeV. Around $E_{cm}=m_{Z}$, due to the $Z^0$-boson resonance effect, total cross sections for the channels via the $Z^0$-propagator become much larger than the channels via the virtual photon propagator. We conclude that, in addition to the $B$ factories as BaBar and Belle and the hadronic colliders as Tevatron and LHC, such a super $Z$-factory will present an excellent platform for studying the heavy quarkonium properties. \\


\end{abstract}

\maketitle

\section{Introduction}

The investigation of the heavy quarkonium, e.g. charmonium and bottomonium can help us to achieve a deeper understanding of QCD in both the perturbative and nonperturbative sectors~\cite{Jpsi,Jpsi1,Jpsi2,Jpsi study1,Jpsi study2,Jpsi study3,Jpsi study4,Jpsi study5,Jpsi study6,Jpsi study7,Jpsi study8}. A comprehensive review of heavy quarkonium physics can be found in Ref.\cite{yellow}. In comparison to the hadronic colliders as Tevatron and LHC, an $e^+e^-$-collider has many advantages, as it provides a cleaner environment and the collision energy of the incoming electron-positron beams is well under control. Recently, a super $Z$ factory running at an energy around the $Z^0$-boson mass with a high luminosity ${\cal L}\simeq 10^{34-36}{\rm cm}^{-2}{\rm s}^{-1}$, similar to the GigaZ program of the Internal Linear Collider~\cite{ILC1,ILC2}, has been proposed~\cite{zfactory}. In the present paper, we will concentrate our attention on the heavy quarkonium production at such a super $Z$ factory. This can be a useful reference for experimental studies, complementing the present BaBar and Belle results on heavy quarkonia.

The non-relativistic QCD (NRQCD)~\cite{NRQCD1} provides a systematic approach for treating the inclusive decay and production of the heavy  quarkonia \cite{NRQCD1}. In this approach, effects of higher-Fock components of a quarkonium state can be considered systematically. Although the probability to find such higher-Fock components is suppressed, the effects of these higher-Fock states can be very significant. This has been shown in the explanation of the $\psi^{\prime}$-anomaly at Tevatron~\cite{crossexp}, where the inclusive $\psi^{\prime}$-production rate with large transverse momentum is in order of magnitude larger than the predicted if one only considers the color-singlet Fock state. By taking the color-octet $(c\bar{c})$ components into account, the Tevatron data can be explained~\cite{BrFl}. This is regarded as a great triumph of NRQCD.

Despite many successes of NRQCD, some problems still remain unsolved. Among them a crucial one is that the approach fails to predict the polarization of $J/\psi$ at the large transverse momentum ($p_T$) measured at Tevatron. The hadronic production of $J/\psi$ is dominated by the gluon fragmentation in which a gluon fragments into a color-octet state $(c\bar{c})[^{3}S^{\bf (8)}_1]$. If the spin-symmetry hold for charm quarks, as is usually adopted in the literature, then the prompt $J/\psi$ shall always show large transverse polarization~\cite{ChoWise}. But this prediction contradicts with the measured at Tevatron~\cite{oldcdf,expcdfII} and the newly LHC data~\cite{lhcn}. This is puzzling because the same mechanism explains the production of unpolarized $J/\psi$ but fails to explain the polarization of the produced $J/\psi$. On the one hand, this shows that the NRQCD itself could be immature. For example, in Ref.\cite{wu1} it has pointed out that a spin-flipping effect in the transition from the color-octet $(c\bar{c})$-pair to the produced $J/\psi$ could cure the polarization puzzle to a certain degree. On the other hand, it is helpful to find another platform in addition to the hadronic colliders, such as the cleaner $e^+e^-$ collider, to test NRQCD.

Within the framework of NRQCD, the production process can be factorized into a sum of products of short-distance coefficients and long-distance matrix elements~\cite{NRQCD1}. The short-distance coefficients are perturbatively calculable in a power series of $\alpha_s$ at the energy scale around the heavy quark mass. Generally, the non-perturbative long-distance matrix elements can be determined from lattice QCD calculations, or by fitting the prediction with the experimental data, or be roughly estimated by means of the NRQCD scaling rule. It has been shown that the matching of the color-octet matrix elements from the hadronic experiments strongly depends on the parton distribution function (PDF)~\cite{BrFl,COdata1,COdata3,COdata4,COdata5,COdata6}, and the PDF uncertainty usually provides one of the key uncertainties for the theoretical estimations. Regarding this point, the $e^+e^-$ collider provides a better platform for precise studies than the hadronic colliders and for testing the NRQCD formulas.

Within the framework of NRQCD, the physical state of a heavy quarkonium is described as a superposition of Fock states, and the relative importance among those infinite ingredients is evaluated by the velocity scaling rule~\cite{NRQCD1}. When the $(Q\bar{Q})$-pair preceding the formation of the hadron $|H_{Q\bar{Q}}\rangle$ is in color-singlet state, it usually gives the dominant contribution to the heavy quarkonium productions and decays; while the production via the color-octet $(Q\bar{Q})$-pair is suppressed by powers of $v$. Here $v$ stands for the typical velocity of the heavy quark and anti-quark in the quarkonium rest frame, $v^2\simeq 0.23$ for $(c\bar{c})$-quarkonium and $v^2\simeq 0.08$ for $(b\bar{b})$-quarkonium. For example, in the velocity expansion, we have
\begin{eqnarray}
|\eta_{Q}\rangle &=& {\cal O}(1) |Q\bar{Q}[^1S^{({\bf 1})}_{0}]\rangle +{\cal O}(v) |Q\bar{Q}[^1P^{({\bf 8})}_{1}]g \rangle + \cdots , \\
|\psi_{Q}\rangle &=& {\cal O}(1) |Q\bar{Q}[^3S^{({\bf 1})}_{1}]\rangle +{\cal O}(v) |Q\bar{Q}[^3P^{({\bf 8})}_{J}]g \rangle + \cdots , \\
|h_{Q}\rangle &=& {\cal O}(1) |Q\bar{Q}[^1P^{({\bf 1})}_{1}]\rangle +{\cal O}(v) |Q\bar{Q}[^1S^{({\bf 8})}_{0}]g \rangle + \cdots , \\
|\chi_{QJ}\rangle &=& {\cal O}(1) |Q\bar{Q}[^3P^{({\bf 1})}_{J}]\rangle +{\cal O}(v) |Q\bar{Q}[^3S^{({\bf 8})}_{1}]g \rangle + \cdots ,
\end{eqnarray}
where $Q=b$ or $c$. Throughout the paper we denote the pre-quarkonium color-octet and color-singlet $(Q\bar{Q})$ states with the extra superscripts ${\bf (8)}$ and ${\bf (1)}$, respectively. Later on we omit the superscript ${\bf (1)}$ for the color-singlet case. The angular momentum properties of the Fock states are defined in square brackets. The color-octet $(Q\bar{Q})$-pair can give sizable and observable contributions in certain cases or in certain kinematic regions when the color-singlet terms are highly suppressed by the hard scattering part. For example, for the $B_c$ meson decaying into leptons and inclusive light hadrons, the energy spectrum of the charged lepton for the color-octet components of the $B_c$ meson is dominant over its color-singlet component when the lepton has high energy~\cite{wuBc}.

In the present paper, we will make a detailed discussion on the heavy quarkonium production at the super $Z$ factory via the following two semi-exclusive channels: $e^{+}e^{-}\rightarrow \gamma^{*}/Z^0 \rightarrow |H_{Q\bar{Q}}\rangle +Q\bar{Q}$ and $e^{+}e^{-}\rightarrow \gamma^{*}/Z^0 \rightarrow |H_{Q\bar{Q}}\rangle +gg$. At present the BaBar and Belle measurements of these two processes are used to determine the color-octet components. However, the present estimations (especially for the charmonium case) are inconsistent with each other~\cite{co1,co2,co3}. Thus, it would be helpful to find a new platform, such as the super $Z$ factory, to learn more about these processes. More explicitly, we will deal with the following production processes:
\begin{widetext}
\begin{eqnarray}
&& e^+e^- \to Z^0,\gamma^* \to |(Q\bar{Q})[\left(1^{3}S_{1}\right), \left(2^{3}S_{1}\right), \left(1^{1}S_{0}\right), \left(2^{1}S_{0}\right)]\rangle +Q\bar{Q}, \nonumber \\
&& e^+e^- \to Z^0,\gamma^* \to |(Q\bar{Q})[\left(1^{1}P_{1}\right), \left(1^{3}P_{0}\right), \left(1^{3}P_{1}\right), \left(1^{3}P_{2}\right)]\rangle +Q\bar{Q}, \nonumber \\
&& e^+e^- \to Z^0 \to |(Q\bar{Q})[\left(1^1S_0^{\bf (8)}\right), \left(1^3S_1^{\bf (8)}\right)]g\rangle + Q\bar{Q} \to \psi_Q+Q\bar{Q} , \nonumber \\
&& e^+e^- \to Z^0 \to |(Q\bar{Q})[\left(1^1S_0^{\bf (8)}\right)]g\rangle + Q\bar{Q} \to h_Q+Q\bar{Q}, \nonumber \\
&& e^+e^- \to Z^0 \to |(Q\bar{Q})[\left(1^3S_1^{\bf (8)}\right)]g\rangle + Q\bar{Q} \to \chi_{QJ}+Q\bar{Q}\nonumber
\end{eqnarray}
and
\begin{eqnarray}
&& e^+e^- \to Z^0,\gamma^* \to |(Q\bar{Q})[\left(1^{3}S_{1}\right), \left(2^{3}S_{1}\right)]\rangle +gg, \nonumber \\
&& e^+e^- \to Z^0 \to |(Q\bar{Q})[\left(1^{1}S_{0}\right), \left(2^{1}S_{0}\right)]\rangle +gg, \nonumber\\
&& e^+e^- \to Z^0 \to |(Q\bar{Q})[\left(1^1S_0^{\bf (8)}\right), \left(1^3S_1^{\bf (8)}\right)]g\rangle + g \to \psi_Q+g ,\nonumber \\
&& e^+e^- \to Z^0 \to |(Q\bar{Q})[\left(1^1S_0^{\bf (8)}\right)]g\rangle + g \to h_Q+g , \nonumber \\
&& e^+e^- \to Z^0 \to |(Q\bar{Q})[\left(1^3S_1^{\bf (8)}\right)]g\rangle + g \to \chi_{QJ}+g ,\nonumber
\end{eqnarray}
\end{widetext}
where $Q$ indicates the heavy quark $c$ or $b$, respectively. It is noted that the unlisted color-singlet channels via the virtual photon are forbidden by considerations on angular momentum conservation and Bose statistics, as formalized in the Landau-Pomeranchuk-Yang theorem~\cite{lpy}. In the present study, we will focus on the dominant color-octet channels listed above, and the less important ones such as those via the virtual photon and those via the component $|(Q\bar{Q})[\left(1^3P_J^{\bf (8)}\right)]g\rangle$ (both color and $v^2$-suppressed with respect to the corresponding case of the color-singlet $S$-wave state) will not be discussed \footnote{There are other less important channels, either color suppressed or $v$ suppressed or phase-space suppressed. We will not discuss them in the present paper either. For example, we have numerically obtained small total cross-sections for the channel $e^{+}e^{-}\rightarrow |H_{Q\bar{Q}}\rangle +gg$ with $H_{Q\bar{Q}}$ in the color-singlet $P$-wave states, in agreement with the $v^2$-suppression with respect to the same channel with $H_{Q\bar{Q}}$ in the color-singlet $S$-wave states.}.

Principally, there are two approaches to deal with the heavy meson hadroproduction. One is the fragmentation approach, which automatically sums up the dominant contributions, including some important higher order effects, into the total/differential cross sections by using the Dokshitzer-Gribov-Lipatov-Altarelli-Parisi evolution equation. The fragmentation approach is comparatively simple, one can easily accomplish a leading logarithm order or even higher order calculation. However the fragmentation approach is satisfied only in the cases when one is only interested in the produced meson itself, i.e. it treats the co-produced objects inclusively, thus losing any information about the co-produced objects. The other one is the so-called complete calculation approach, in which we directly deal with the full hard scattering amplitude without any approximations. In this approach it is sometimes hard to derive the analytical expression even at the leading order, but the information on the accompanying quark or gluon jets is retained and can be compared to data. In the present paper, we will mainly adopt the complete pQCD calculation approach to deal with these channels.

In comparison to the $B$ factories as BaBar and Belle, we will show that a large number of heavy quarkonium events can be generated due to the $Z^0$-boson resonance effect. Some features of heavy quarkonium production at such super Z factory has already been discussed in Refs.\cite{zbc12,zbc2,zbc3,zbc4,zbc5,cc1,baryon}. In this paper we focus on some novel observations. For example, when the quarkonium is produced directly in color-singlet state in the channel $e^{+}e^{-}\rightarrow Z^0 \rightarrow |H_{Q\bar{Q}}\rangle +Q\bar{Q}$, the spin-singlet and the spin-triplet $S$-wave states are almost equally probable (approximate ``spin degeneracy").

The remaining part of the paper is organized as follows. In Sec.II, we present the calculation technique for dealing with the heavy quarkonium production processes $e^{+}e^{-}\rightarrow |H_{Q\bar{Q}}\rangle +Q\bar{Q}$ and $e^{+}e^{-}\rightarrow |H_{Q\bar{Q}}\rangle +gg$, where the intermediate $(Q\bar{Q})$-state is in either color-singlet or color-octet state respectively. In Sec.III, we present our numerical results. Total and differential cross sections are discussed, and an alternative proof of the above mentioned ``spin degeneracy" is provided in the framework of the fragmentation approach. Sec.IV is reserved for a summary.

\section {Formulation and Technique}

According to the NRQCD framework, the differential cross section for the process, $e^+e^-\rightarrow |H_{Q\bar{Q}}\rangle+X$, can be factorized as~\cite{NRQCD1,NRQCD2} :
\begin{eqnarray}
&{\rm d}\sigma =\sum\limits_{n} {\rm d}\hat\sigma \left(e^{+} e^{-}\rightarrow (Q\bar{Q})[n]+X\right) \frac{\langle 0|{\cal O}^H(n)|0 \rangle}{N_{col} N_{pol}} . \label{total}
\end{eqnarray}
The production matrix element $\langle 0|{\cal O}^H(n) |0\rangle$ is proportional to the inclusive transition probability of the intermediate perturbative $(Q\bar{Q})$-pair in $[n]$-state into the final bound-state $|H_{Q\bar{Q}}\rangle$. The symbol $[n]=[m^{2S+1}L^{({\bf 1}),({\bf 8})}_J]$ denotes the energy level $m$, the spin $S$, the orbital angular momentum $L$ and the total angular momentum $J$ of the intermediate $(Q\bar{Q})$-pair, i.e.,
\begin{displaymath}
n = 1^1S_0, 2^1S_0, 1^1P_1, 1^1S^{\bf (8)}_0; 1^3S_1, 2^3S_1, 1^3P_J, 1^3S^{ \bf (8)}_1 ,
\end{displaymath}
with $J=0,1,2$. These states provide the dominant contributions to the processes $e^{+}e^{-}\rightarrow \gamma^{*}/Z^0 \rightarrow |H_{Q\bar{Q}}\rangle+Q\bar{Q}$ and $e^{+}e^{-}\rightarrow \gamma^{*}/Z^0 \rightarrow |H_{Q\bar{Q}}\rangle+gg$ up to ${\cal O}(v^4)$. The parameters $N_{col}$ and $N_{pol}$ refer to the number of colors and polarization states of the intermediate $(Q\bar{Q})$-pair. $N_{col}=1$ for the color-singlet state or $N_{col}=8$ for the color-octet state. The color-singlet matrix elements can be directly related either to the wave function at the origin or (depending on the Fock state) to the first derivative of the wave function at the origin, which can be computed via potential models and/or potential NRQCD and/or lattice QCD. The color-octet matrix elements can be estimated by using the velocity scaling rule or be determined experimentally.

The short-distance cross section $d\hat\sigma(e^{+}(p_2) e^{-}(p_1) \rightarrow (Q\bar{Q})[n]+X)$ can be written in the following form:
\begin{eqnarray}
{\rm d}\hat\sigma\left(e^{+} e^{-}\rightarrow(Q\bar{Q})[n]+X \right) = \frac{\overline{\sum}|{\cal M}|^{2} {\rm d}\Phi_k}{4\sqrt{(p_1\cdot p_2)^2-m_e^4}} , \label{cs}
\end{eqnarray}
where $k$ stands for the number of final state particles, $\overline{\sum}$ means that we need to average over the spin states of the initial particles and to sum over the spin and color of all final particles \footnote{Because of the presence of (hereafter defined) projectors, the dimension of the short-distance cross section $\hat\sigma$ is $[{\rm pb}][{\rm GeV}^{-3}]$ for $S$-wave states and $[{\rm pb}][{\rm GeV}^{-5}]$ for $P$-wave states, which ensures the unit of the total cross section $\sigma$ be the wanted $[{\rm pb}]$.}. The phase space with $k$ final-state particles is
\begin{displaymath}
{\rm d}{\Phi_k}=(2\pi)^4 \delta^{4}\left(p_1+p_2 - \sum_{f=3}^{k+2} p_{f}\right)\prod_{f=3}^{k+2} \frac{{\rm d}^3 {p_f}}{(2\pi)^3 2 p_f^0}.
\end{displaymath}
The phase-space integration can be done with the help of a combination of the subroutines RAMBOS~\cite{rambos} and VEGAS~\cite{vegas}, which can be found in the generators GENXICC~\cite{genxicc1} and BCVEGPY~\cite{bcvegpy}. After generating proper phase-space points, one can calculate the numerical value for the squared amplitude $|{\cal M}|^2$. For the alternative calculations in the fragmentation approach, the phase space is calculated in a factorized form, as described in the appendix.

The hard scattering amplitude for those processes can be written as
\begin{equation}
i {\cal M}={\cal C} \; L_{rr^{\prime}}^{\mu}D_{\mu\nu}\sum^{j_{\rm max}}_{j=1}{\cal A}_j^{\nu} , \label{overall}
\end{equation}
where the leptonic current
\begin{equation}
L_{rr^{\prime}}^{\mu} = \bar{v}_r(p_2) \Gamma^{\mu} u_{r^{\prime}}(p_1) ,
\end{equation}
with the indices $r$ and $r'$ standing for the spin projections of the initial electron and positron. The value of $j_{\rm max}$ is process dependent, e.g.
\begin{eqnarray}
&& j_{\rm max}=4 \;\;{\rm for}\;\; e^{+} e^{-}\rightarrow (Q\bar{Q})[n]+Q\bar{Q} ,\nonumber\\
&& j_{\rm max}=6 \;\;{\rm for}\;\; e^{+} e^{-}\rightarrow (Q\bar{Q})[n]+gg , \nonumber
\end{eqnarray}
for $(Q\bar{Q})[n]$ in color-singlet state, while, for the case of color-octet production,
\begin{eqnarray}
&& j_{\rm max}=2 \;\;{\rm for}\;\; e^+e^- \to Z^0 \to |(Q\bar{Q}) [\left(1^1S_0^{\bf (8)}\right)]g\rangle + g , \nonumber\\
&& j_{\rm max}=2 \;\;{\rm for}\;\; e^+e^- \to Z^0 \to |(Q\bar{Q}) [\left(1^3S_1^{\bf (8)}\right)]g\rangle + g , \nonumber\\
&& j_{\rm max}=6 \;\;{\rm for}\;\; e^+e^- \to Z^0 \to |(Q\bar{Q})[\left(1^1S_0^{\bf (8)}\right)]g\rangle + Q\bar{Q} , \nonumber\\
&& j_{\rm max}=8 \;\;{\rm for}\;\; e^+e^- \to Z^0 \to |(Q\bar{Q})[\left(1^3S_1^{\bf (8)}\right)]g\rangle + Q\bar{Q}. \nonumber
\end{eqnarray}
For quarkonium production through the $Z^0$-boson propagator, the vertex is $\Gamma^{\mu}= \gamma^\mu(1-4\sin^2\theta_w-\gamma^5)$ and the propagator is $D_{\mu\nu}=\frac{i}{k^2-m^2_Z +im_Z\Gamma_z}\left(-g_{\mu\nu}+{k_\mu k_\nu}/{k^2}\right)$, where $\Gamma_z$ stands for the total decay width of the $Z^0$ boson. For quarkonium production through the virtual photon propagator, the vertex is $\Gamma^{\mu}=\gamma^\mu$ and the propagator is $D_{\mu\nu}=-\frac{i}{k^2}g_{\mu\nu}$.

The overall constant ${\cal C}$ is different for the production via color-singlet (${\cal C}_s$) and via color-octet $Q\bar{Q}$ state (${\cal C}_o$). Expressions for the reduced hard scattering amplitudes ${\cal A}^{\nu}_j$, which are process dependent, will be given in the following subsections.

\subsection{Color-Singlet Case}

\subsubsection{$e^{+}(p_2) e^{-}(p_1) \rightarrow |H_{Q\bar{Q}}\rangle(p_3) + Q(p_4) \bar{Q}(p_5)$}

\begin{figure}[htb]
\includegraphics[width=0.60\textwidth]{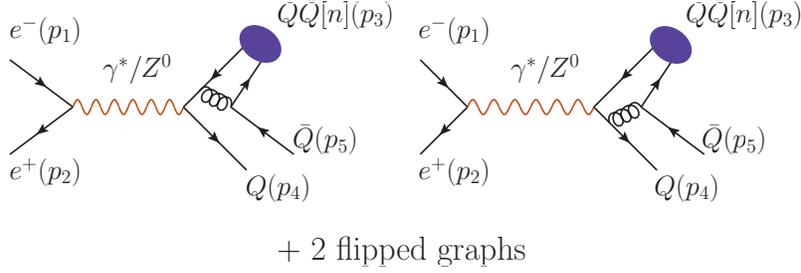}
\caption{Feynman diagrams for $e^{+} e^{-}\rightarrow |H_{Q\bar{Q}}\rangle+Q\bar{Q}$ via the perturbative state $(Q\bar{Q})[n]$. Here $Q$ stands for $c$ or $b$ quark, and $[n]$ indicates the quantum numbers
of the heavy quarkonium state.} \label{QQQQ}
\end{figure}

Typical Feynman diagrams for the process $e^{+} e^{-}\rightarrow |H_{Q\bar{Q}}\rangle + Q\bar{Q}$ through color-singlet $(Q\bar{Q})$-quarkonium states are presented in Fig. \ref{QQQQ}. According to the Feynman diagrams, we can write down the reduced hard scattering amplitudes ${\cal A}^{\nu}_j$ of the short-distance cross-section $d\hat\sigma(e^{+} e^{-}\rightarrow (Q\bar{Q})[n]+Q\bar{Q})$.

For the color-singlet production of $S$-wave states, we have
\begin{widetext}
\begin{eqnarray}
&&{\cal A}^{\nu}_1 = {\bar u_s}({p_4}){\Gamma^{\nu}_{Q\bar{Q}}}\frac{-\slashed{p}_3 -\slashed{p}_5 + m_{Q}}{(p_3 +p_5)^2 - m^2_{Q}}{\gamma_\rho} \frac{\Pi^{0(1)}_{(Q\bar{Q})}(p_3)} {(p_{31} + {p_5})^2}{\gamma^\rho}v_{s^{\prime}}(p_5), \label{A1}\\
&&{\cal A}^{\nu}_2= {\bar u_s}({p_4}){\gamma_\rho}\frac{\slashed{p}_4+\slashed{p}_5+\slashed{p}_{31} + m_{Q}}{(p_4+p_5+p_{31})^2 - m^2_{Q}}{\Gamma^{\nu}_{Q\bar{Q}}} \frac{\Pi^{0(1)}_{(Q\bar{Q})}(p_3)} {(p_{31} + {p_5})^2}{\gamma^\rho}v_{s^{\prime}}(p_5), \label{A2}  \\
&&{\cal A}^{\nu}_3= {\bar u_s}({p_4}){\gamma_\rho}\frac{\Pi^{0(1)}_{(Q\bar{Q})}(p_3)} {(p_{32} + {p_4})^2}{\gamma^\rho} \frac{\slashed{p}_3 +\slashed{p}_4 + m_{Q}}{(p_3 +p_4)^2 - m^2_{Q}}{\Gamma^{\nu}_{Q\bar{Q}}}v_{s^{\prime}}(p_5), \label{A3} \\
&&{\cal A}^{\nu}_4= {\bar u_s}({p_4}){\gamma_\rho}\frac{\Pi^{0(1)}_{(Q\bar{Q})}(p_3)} {(p_{32} + {p_4})^2}{\Gamma^{\nu}_{Q\bar{Q}}} \frac{-\slashed{p}_4-\slashed{p}_5-\slashed{p}_{32}+m_{Q}}{(p_4 +p_5+p_{32})^2 - m^2_{Q}}{\gamma^\rho}v_{s^{\prime}}(p_5) . \label{A4}
\end{eqnarray}
\end{widetext}

For the color-singlet production of $P$-wave states, we have
\begin{widetext}
\begin{eqnarray}
&&{\cal A}^{\nu,S=0,L=1}_1 = {\bar u_s}({p_4})\epsilon_{\alpha}^l(p_3)\frac{d}{dq_{\alpha}} \left[{\Gamma^{\nu}_{Q\bar{Q}}}\frac{-\slashed{p}_3 -\slashed{p}_5 + m_{Q}}{(p_3 +p_5)^2 - m^2_{Q}}{\gamma_\rho}\frac{\Pi^{0}_{(Q\bar{Q})}(p_3)} {(p_{31} + {p_5})^2}{\gamma^\rho}\right]_{q=0}v_{s^{\prime}}(p_5),  \label{A1p11} \\
&&{\cal A}^{\nu,S=0,L=1}_2= {\bar u_s}({p_4})\epsilon_{\alpha}^l(p_3) \frac{d}{dq_{\alpha}}\left[{\gamma_\rho}\frac{\slashed{p}_4+\slashed{p}_5+\slashed{p}_{31} + m_{Q}}{(p_4+p_5+p_{31})^2 - m^2_{Q}}{\Gamma^{\nu}_{Q\bar{Q}}} \frac{\Pi^{0}_{(Q\bar{Q})}(p_3)} {(p_{31} + {p_5})^2}{\gamma^\rho} \right]_{q=0}v_{s^{\prime}}(p_5), \label{A1p12}  \\
&&{\cal A}^{\nu,S=0,L=1}_3= {\bar u_s}({p_4})\epsilon_{\alpha}^l(p_3) \frac{d}{dq_{\alpha}}\left[{\gamma_\rho}\frac{\Pi^{0}_{(Q\bar{Q})}(p_3)} {(p_{32} + {p_4})^2}{\gamma^\rho} \frac{\slashed{p}_3 +\slashed{p}_4 + m_{Q}}{(p_3 +p_4)^2 - m^2_{Q}}{\Gamma^{\nu}_{Q\bar{Q}}}\right]_{q=0}v_{s^{\prime}}(p_5), \label{A1p13} \\
&&{\cal A}^{\nu,S=0,L=1}_4= {\bar u_s}({p_4})\epsilon_{\alpha}^l(p_3) \frac{d}{dq_{\alpha}}\left[{\gamma_\rho}\frac{\Pi^{0}_{(Q\bar{Q})}(p_3)} {(p_{32} + {p_4})^2}{\Gamma^{\nu}_{Q\bar{Q}}}\frac{-\slashed{p}_4-\slashed{p}_5-\slashed{p}_{32}+m_{Q}}{(p_4 +p_5+p_{32})^2 - m^2_{Q}}{\gamma^\rho}\right]_{q=0}v_{s^{\prime}}(p_5)  \label{A1p14}
\end{eqnarray}
and
\begin{eqnarray}
&&{\cal A}^{\nu,S=1,L=1}_1 = {\bar u_s}({p_4})\varepsilon_{\alpha\beta}^{J_z}(p_3)\frac{d}{dq_{\alpha}}\left[{\Gamma^{\nu}_{Q\bar{Q}}}\frac{-\slashed{p}_3 -\slashed{p}_5 + m_{Q}}{(p_3 +p_5)^2 - m^2_{Q}}{\gamma_\rho}\frac{\Pi^{\beta}_{(Q\bar{Q})}(p_3)} {(p_{31} + {p_5})^2}{\gamma^\rho}\right]_{q=0}v_{s^{\prime}}(p_5),  \label{A3pj1} \\
&&{\cal A}^{\nu,S=1,L=1}_2= {\bar u_s}({p_4})\varepsilon_{\alpha\beta}^{J_z}(p_3)\frac{d}{dq_{\alpha}}\left[{\gamma_\rho}\frac{\slashed{p}_4+\slashed{p}_5+\slashed{p}_{31} + m_{Q}}{(p_4+p_5+p_{31})^2 - m^2_{Q}}{\Gamma^{\nu}_{Q\bar{Q}}}\frac{\Pi^{\beta}_{(Q\bar{Q})}(p_3)} {(p_{31} + {p_5})^2}{\gamma^\rho}\right]_{q=0}v_{s^{\prime}}(p_5), \label{A3pj2}  \\
&&{\cal A}^{\nu,S=1,L=1}_3= {\bar u_s}({p_4})\varepsilon_{\alpha\beta}^{J_z}(p_3)\frac{d}{dq_{\alpha}}\left[{\gamma_\rho}\frac{\Pi^{\beta}_{(Q\bar{Q})}(p_3)} {(p_{32} + {p_4})^2}{\gamma^\rho} \frac{\slashed{p}_3 +\slashed{p}_4 + m_{Q}}{(p_3 +p_4)^2 - m^2_{Q}}{\Gamma^{\nu}_{Q\bar{Q}}}\right]_{q=0}v_{s^{\prime}}(p_5), \label{A3pj3} \\
&&{\cal A}^{\nu,S=1,L=1}_4= {\bar u_s}({p_4})\varepsilon_{\alpha\beta}^{J_z}(p_3) \frac{d}{dq_{\alpha}}\left[{\gamma_\rho}\frac{\Pi^{\beta}_{(Q\bar{Q})}(p_3)} {(p_{32} + {p_4})^2}{\Gamma^{\nu}_{Q\bar{Q}}}\frac{-\slashed{p}_4-\slashed{p}_5-\slashed{p}_{32}+m_{Q}}{(p_4 +p_5+p_{32})^2 - m^2_{Q}}{\gamma^\rho}\right]_{q=0}v_{s^{\prime}}(p_5) . \label{A3pj4}
\end{eqnarray}
\end{widetext}
Throughout the paper, we adopt the convention that the dummy index indicates summation. The parameters $s$ and $s^{\prime}$ stand for the spin projections of the outgoing quark and antiquark respectively. The symbol $s$ stands for the spin angular momentum quantum number, $l$ stands for the radial angular momentum quantum number, $J_z=s_z + l_z$ stands for the $z$-component of the total angular momentum quantum number of the bound state, respectively.

For convenience, we have introduced a general interaction vertex
\begin{equation}
\Gamma^{\nu}_{Q\bar{Q}}=\gamma^{\nu}(\xi_1 P_L+\xi_2 P_R) ,
\end{equation}
where $P_L={(1-\gamma^5)}/{2}$ and $P_R={(1+\gamma^5)}/{2}$. Here $\xi_1=2-\frac{8}{3}\sin^2\theta_w$ and $\xi_2=-\frac{8}{3}\sin^2\theta_w$ for $(Z{c}{\bar{c}})$-vertex, $\xi_1=2-\frac{4}{3}\sin^2\theta_w$ and $\xi_2=-\frac{4}{3}\sin^2\theta_w$ for $(Z{b}{\bar{b}})$-vertex, $\xi_1=1$ and $\xi_2=1$ for $(\gamma^{*}{Q}{\bar{Q}})$-vertex, respectively.

The momenta of the constituent quarks are
\begin{equation}
p_{31} = \frac{m_Q}{M_{Q\bar{Q}}}{p_3} + q \;\;{\rm and}\;\;
p_{32} = \frac{m_{\bar Q}}{M_{Q\bar{Q}}}{p_3} - q,
\end{equation}
where $M_{Q\bar{Q}}= m_Q + m_{\bar{Q}}$ is implicitly adopted to ensure the gauge invariance of the hard scattering amplitude, $q$ is the relative momentum between the two constituent quarks inside the quarkonium.

The covariant forms of the projectors are
\begin{equation}
\Pi^0_{(Q\bar{Q})}(p_3) = \frac{-\sqrt{M_{Q\bar{Q}}}}{4m_Q m_{\bar{Q}}} (\slashed{p}_{32}-m_{\bar{Q}}) \gamma_5(\slashed{p}_{31}+m_{Q})
\end{equation}
and
\begin{equation}
\Pi^1_{(Q\bar{Q})}(p_3) = \epsilon_{\kappa}^s(p_3) \Pi^{\kappa}_{(Q\bar{Q})}(p_3) ,
\end{equation}
where $\epsilon^{s}(p_3)$ stands for the polarization vector of the spin-triplet $S$-wave state and
\begin{equation}
\Pi^{\kappa}_{(Q\bar{Q})}(p_3) = \frac{-\sqrt{M_{Q\bar{Q}}}}{4m_Q m_{\bar{Q}}} (\slashed{p}_{32} -m_{\bar{Q}}) \gamma^{\kappa} (\slashed{p}_{31}+m_{Q}).
\end{equation}
Inserting these projectors into the amplitude, the amplitude can be squared, summed over the spin in the final state and averaged over the ones in the initial state. The selection of the proper angular momentum is done by performing a suitable polarization sum. For examples, the sum over polarization for a spin-triplet $S$-wave state ($^3S_1$) or a spin-singlet $P$-wave state ($^1P_1$) is given by:
\begin{eqnarray}
\sum_{J_z}\epsilon_\alpha(p_3) \epsilon^*_{\alpha'}(p_3)=\Pi_{\alpha\alpha'},
\end{eqnarray}
where $\epsilon(p_3)$ stands for the polarization vector of the meson, $\epsilon(p_3)=\epsilon^{s}(p_3)$ and $J_z=s_z $ for the $^3S_1$ state, $\epsilon(p_3)=\epsilon^{l}(p_3)$ and $J_z=l_z$ for the spin-singlet $^1P_1$ state. And the sum over polarization for the spin-triplet $P$-wave states ($^3P_J$ with $J=0,1,2$) is given by \cite{NRQCD2}
\begin{widetext}
\begin{eqnarray}
\varepsilon_{\alpha\beta}^{(0)}(p_3) \varepsilon_{\alpha'\beta'}^{(0)*}(p_3) &=&\frac{1}{3}\Pi_{\alpha\beta}\Pi_{\alpha'\beta'} \nonumber\\
\sum_{J_Z}\varepsilon_{\alpha\beta}^{(1)}(p_3) \varepsilon_{\alpha'\beta'}^{(1)*}(p_3) &=&\frac{1}{2}\left[\Pi_{\alpha\alpha'}\Pi_{\beta\beta'}-\Pi_{\alpha\beta'} \Pi_{\alpha'\beta}\right] \nonumber \\
\sum_{J_Z}\varepsilon_{\alpha\beta}^{(2)}(p_3) \varepsilon_{\alpha'\beta'}^{(2)*}(p_3) &=&\frac{1}{2}\left[\Pi_{\alpha\alpha'}\Pi_{\beta\beta'}+ \Pi_{\alpha\beta'}\Pi_{\alpha'\beta}\right]-\frac{1}{3}\Pi_{\alpha\beta}\Pi_{\alpha'\beta'} , \nonumber
\end{eqnarray}
\end{widetext}
where $\varepsilon^{(J)}_{\alpha\beta}(p_3)$ stands for the polarization tensor of the spin-triplet $P$-wave states. In the above formulas, we have defined a short notation for the polarization sum, i.e.
\begin{eqnarray}
\Pi_{\rho_1 \rho_2}\equiv-g_{\rho_1 \rho_2}+\frac{p_{3\rho_1} p_{3\rho_2}}{M^{2}_{Q\bar{Q}}} ,
\end{eqnarray}
where $\rho_1$ and $\rho_2$, which equal to $0,\cdots,4$ respectively, are Lorentz indices of the meson momentum $p_3$.

For the overall color-singlet parameter (${\cal C}$ in Eq.(\ref{overall})), we have ${\cal C}_s=\frac{4}{3\sqrt{3}} \frac{e^2 g^2_s} {\sin^2\theta_w \cdot (4\cos\theta_w)^2}\delta_{ij}$ for the quarkonium production through $Z^0$-boson propagator and ${\cal C}_s=\frac{4}{3\sqrt{3}}{e_Q e^2 g^2_s}\delta_{ij}$ for the quarkonium production through virtual photon propagator, where $e_Q$ stands for the electric charge of $Q$ in unit $e$. The symbols $i,j=(1,2,3)$ in $\delta_{ij}$ are color-indices for the outgoing antiquark and quark, respectively. Here $g_s$ stands for the gauge coupling of QCD, which is connected with the strong coupling $\alpha_s$ via the relation $\alpha_s=g_s^2/4\pi$, and $\theta_w$ is the weak mixing angle. \\

\subsubsection{$e^{+}(p_2) e^{-}(p_1) \rightarrow |H_{Q\bar{Q}}\rangle(p_3) +g(p_4) g(p_5)$}

\begin{figure}[t]
\includegraphics[width=0.60\textwidth]{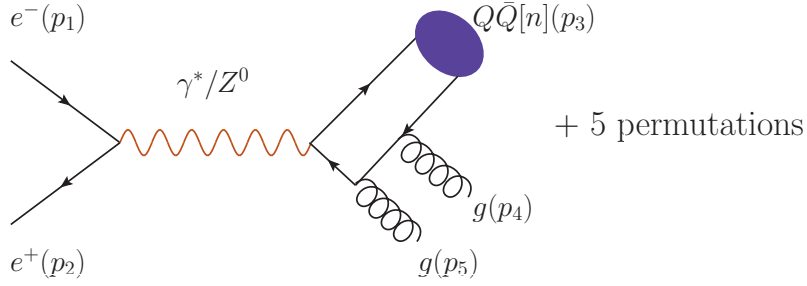}
\caption{Typical Feynman diagrams for $e^{+} e^{-}\rightarrow |H_{Q\bar{Q}}\rangle+gg$ through the perturbative state $(Q\bar{Q})[n]$. Here $Q$ stands for the $c$ or $b$ quark, and $[n]$ indicates the quantum numbers of the heavy quarkonium state.} \label{QQgg}
\end{figure}

Typical Feynman diagrams for the process $e^{+} e^{-}\rightarrow |H_{Q\bar{Q}}\rangle+gg$ are presented in Fig. \ref{QQgg}. According to the Feynman diagrams, one can write down the reduced hard scattering amplitudes ${\cal A}^{\nu}_j$ for the short-distance cross-section $d\hat\sigma(e^{+} e^{-}\rightarrow (Q\bar{Q})[n]+gg)$ as
\begin{widetext}
\begin{eqnarray}
{\cal A}_1^{\nu} &=& \textrm{Tr}\left[\left( \Pi^{0(1)}_{(Q\bar{Q})}(p_3) \right){\Gamma^{\nu}_{Q\bar{Q}}}\frac{-\slashed{p}_{32}-\slashed{p}_4-\slashed{p}_5 + m_{Q}}{(p_{32} +p_4+p_5)^2 - m^2_{Q}}\slashed{\epsilon}(p_{5}) \frac{-\slashed{p}_{32}-\slashed{p}_4+ m_{Q}}{(p_{32} +p_4)^2 - m^2_{Q}} \slashed{\epsilon}(p_{4}) \right],
\end{eqnarray}
\end{widetext}
and the amplitudes ${\cal A}_2^{\nu}$, $\cdots$, ${\cal A}_6^{\nu}$ can be derived by permutations.

For the overall color-singlet parameter, we have ${\cal C}_s=\frac{1}{2\sqrt{3}}\frac{e^2 g^2_s} {\sin^2\theta_w\cdot (4\cos\theta_w)^2}\delta_{ab}$ for the quarkonium production through $Z^0$-boson propagator and ${\cal C}_s=\frac{1}{2\sqrt{3}}{e_Q e^2 g^2_s}\delta_{ab}$ for the quarkonium production through virtual photon propagator, where $a,b=(1,\cdots,8)$ are color indices of the two outgoing gluons.

The reduced amplitudes for the case of color-singlet $P$-wave states can be written down in a similar way as for the process $e^{+} e^{-}\rightarrow |H_{Q\bar{Q}}\rangle + Q\bar{Q}$. Numerically, it is found that the total cross-sections for the color-singlet $P$-wave states are negligible, so we do not present their reduced amplitudes here.

\subsection{Color-Octet Case}

\subsubsection{$e^+(p_2)e^-(p_1) \to Z^0 \to |(Q\bar{Q})[\left(1^1S_0^{\bf (8)}\right), \left(1^3S_1^{\bf (8)}\right)]g\rangle(p_3) + Q(p_4)\bar{Q}(p_5)$}

\begin{figure}[htb]
\includegraphics[width=0.80\textwidth]{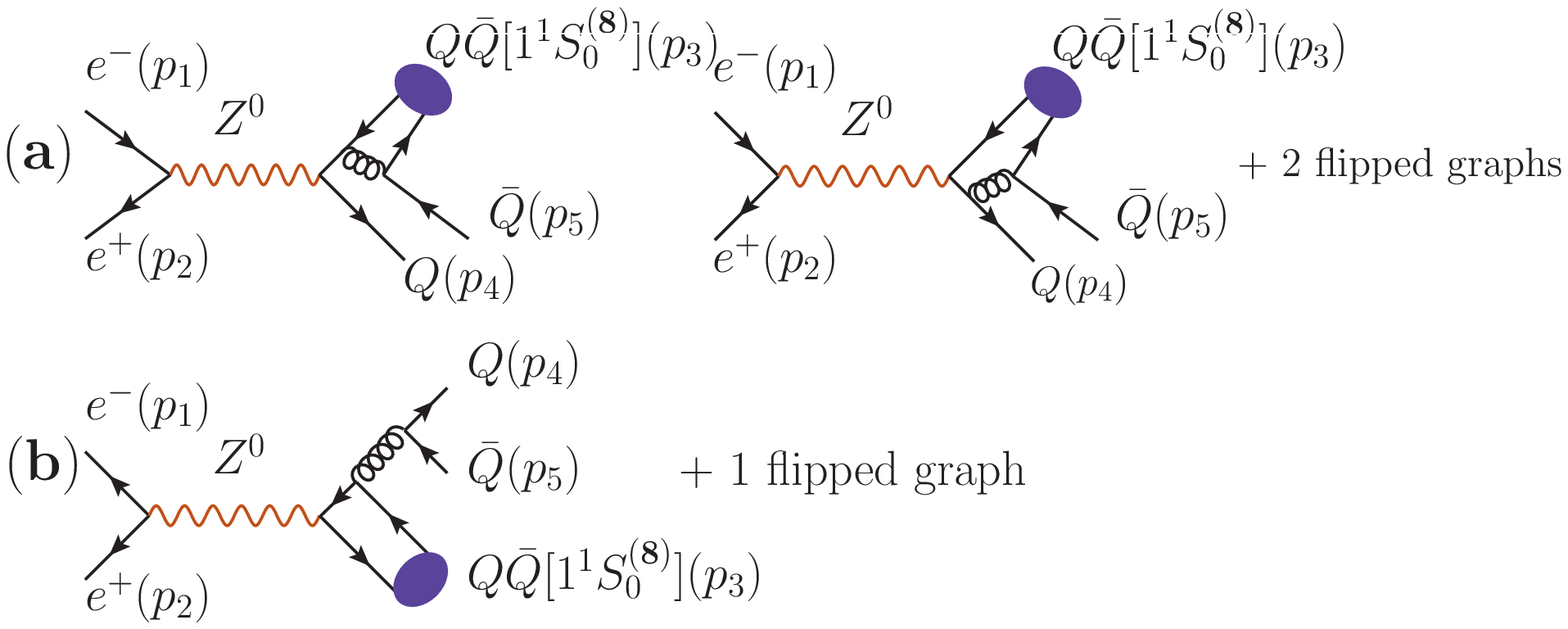}                                                                                                                           \caption{Typical Feynman diagrams for $e^+e^- \to Z^0 \to |(Q\bar{Q})[\left(1^1S_0^{\bf (8)}\right)]g\rangle + Q\bar{Q}$, where $Q$ stands for $c$ or $b$ quark.} \label{QQQQoctet1}
\end{figure}

\begin{figure}[htb]
\includegraphics[width=0.80\textwidth]{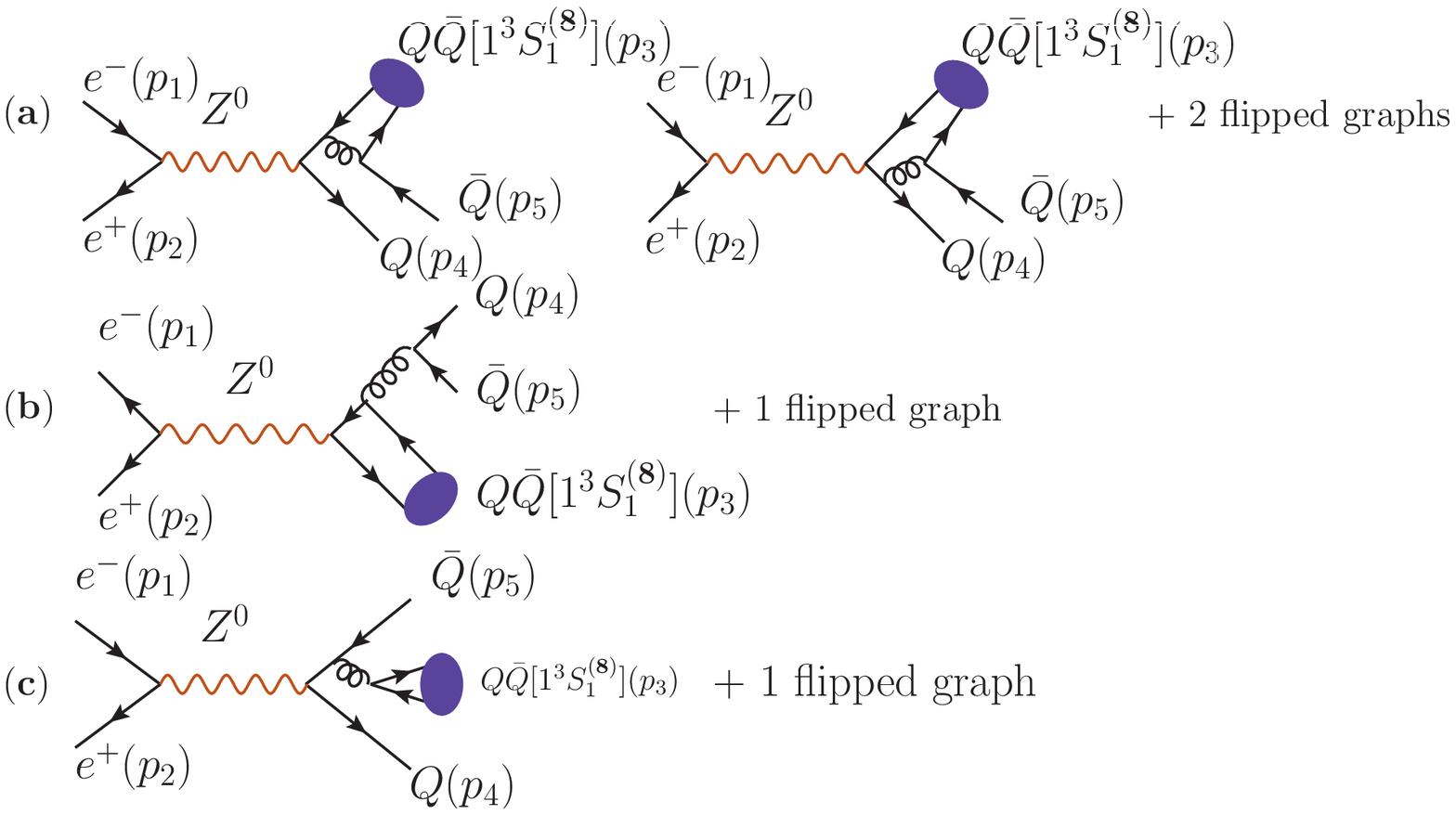}                                                                                                                           \caption{Typical Feynman diagrams for $e^+e^- \to Z^0 \to |(Q\bar{Q})[\left(1^3S_1^{\bf (8)}\right)]g\rangle + Q\bar{Q}$, where $Q$ stands for $c$ or $b$ quark.} \label{QQQQoctet2}
\end{figure}

Typical Feynman diagrams for the processes $e^+e^- \to Z^0 \to |(Q\bar{Q})[\left(1^1S_0^{\bf (8)}\right)]g\rangle + Q\bar{Q}$ and $e^+e^- \to Z^0 \to |(Q\bar{Q})[\left(1^3S_1^{\bf (8)}\right)]g\rangle + Q\bar{Q}$ are presented in Figs. \ref{QQQQoctet1} and \ref{QQQQoctet2}, respectively. It is found that
\begin{itemize}
\item The four Feynman diagrams shown in Fig. \ref{QQQQoctet1}a and Fig. \ref{QQQQoctet2}a have the same topologies as those of Fig. \ref{QQQQ}. Their reduced amplitudes ${\cal A}^{\nu}_1,\cdots,{\cal A}^{\nu}_4$ are the same as those of Eqs.(\ref{A1},\ref{A2},\ref{A3},\ref{A4}), with the overall parameter ${\cal C}_s$ replaced by ${\cal C}_o$, and the color-singlet matrix element replaced by the color-octet one.

\item The two Feynman diagrams shown in Fig. \ref{QQQQoctet1}b and Fig. \ref{QQQQoctet2}b have the same topologies, and their reduced amplitudes ${\cal A}^{\nu}_5$ and ${\cal A}^{\nu}_6$ are the same:
\begin{widetext}
\begin{eqnarray}
{\cal A}^{\nu}_5 &=&  {\bar u_s}({p_4}){\gamma_\rho}\textrm{Tr} \left[{\gamma^\rho} \frac{\Pi^{0(1)}_{(Q\bar{Q})}(p_3)}{(p_{4}+{p_5})^2}{\Gamma^{\nu}_{Q\bar{Q}}} \frac{-\slashed{p}_{32} -\slashed{p}_4 -\slashed{p}_5 + m_{Q}}{(p_{32} +p_4 +p_5)^2 - m^2_{Q}}\right]v_{s^{\prime}}(p_5) , \\
{\cal A}^{\nu}_6 &=&  {\bar u_s}({p_4}){\gamma_\rho} \textrm{Tr}\left[{\gamma^\rho} \frac{\slashed{p}_{31} +\slashed{p}_4 +\slashed{p}_5 + m_{Q}} {(p_{31} +p_4 +p_5)^2 - m^2_{Q}}{\Gamma^{\nu}_{Q\bar{Q}}} \frac{\Pi^{0(1)}_{(Q\bar{Q})}(p_3)} {(p_{4}+ {p_5})^2}\right]v_{s^{\prime}}(p_5) .
\end{eqnarray}
\end{widetext}
\item The remaining two reduced amplitudes ${\cal A}_7^{\nu}$ and ${\cal A}_8^{\nu}$ for the case of $|(Q\bar{Q})[\left(1^3S_1^{\bf (8)}\right)]g\rangle$ production, as shown in Fig. \ref{QQQQoctet2}c, are
\begin{widetext}
\begin{eqnarray}
{\cal A}^{\nu}_7 &=&  {\bar u_s}(p_4){\Gamma^{\nu}_{Q\bar{Q}}}\frac{-\slashed{p}_{3} -\slashed{p}_5 + m_{Q}}{(p_{3} +p_5)^2 - m^2_{Q}}{\gamma_\rho} \textrm{Tr} \left[\frac{\Pi^{1}_{(Q\bar{Q})}(p_3)}{p_{3}^2}{\gamma^\rho}\right]v_{s^{\prime}}(p_5) , \\
{\cal A}^{\nu}_8 &=&  {\bar u_s}({p_4}){\gamma_\rho}\textrm{Tr} \left[\frac{\Pi^{1}_{(Q\bar{Q})}(p_3)}{p_{3}^2}{\gamma^\rho}\right]\frac{\slashed{p}_{3} +\slashed{p}_4 + m_{Q}}{(p_{3} +p_4)^2 - m^2_{Q}}{\Gamma^{\nu}_{Q\bar{Q}}} v_{s^{\prime}}(p_5) .
\end{eqnarray}
\end{widetext}
\end{itemize}

For the overall color-octet parameters, we have ${\cal C}_o=\frac{e^2 g_s^2}{\sin^{2}\theta_{w} (4\cos\theta_w)^2} \times \left({\cal C}^{(a)}_{ij},{\cal C}^{(b)}_{ij},{\cal C}^{(c)}_{ij}\right)$ respectively. Here, the color factors ${\cal C}^{(a),(b),(c)}_{ij}$, according to Figs. \ref{QQQQoctet2}a, \ref{QQQQoctet2}b and \ref{QQQQoctet2}c, are:
\begin{eqnarray}
&& {\cal C}^{(a)}_{ij}=\sum_{m,n}(T^{a})_{in}\times(\sqrt{2}T^d)_{nm}\times(T^{a})_{mj} , \\
&& {\cal C}^{(b)}_{ij}=\sum_{m,n}(T^{a})_{ij}\times(T^{a})_{mn}\times(\sqrt{2}T^d)_{nm} , \\
&& {\cal C}^{(c)}_{ij}=\sum_{m,n}(T^{a})_{ij}\times(T^{a})_{mn}\times(\sqrt{2}T^d)_{nm} ,
\end{eqnarray}
where $i,j=(1,2,3)$ are the color indices of the outgoing anti-quark $\bar{Q}$ and quark $Q$, and $m,n=(1,2,3)$ are those of the two constituent quarks $Q$ and $\bar{Q}$ in the heavy quarkonium. The superscript indices $(a)$, $(b)$ and $(c)$ of ${\cal C}_{ij}$ refer to the corresponding figures in Figs. \ref{QQQQoctet1} and \ref{QQQQoctet2}. After simplification, we obtain $${\cal C}^{(a)}_{ij}=-\frac{\sqrt 2}{6}T^d_{ij},\; {\cal C}^{(b)}_{ij}={\cal C}^{(c)}_{ij}=\frac{\sqrt 2}{2}T^d_{ij}.$$ Here $d=(1,\cdots,8)$ represents the color of the color-octet quarkonium.

\subsubsection{$e^+(p_2)e^-(p_1) \to Z^0 \to |(Q\bar{Q})[\left(1^1S_0^{\bf (8)}\right),\left(1^3S_1^{\bf (8)}\right)]g\rangle(p_3) + g(p_4)$}

\begin{figure}[ht]
\includegraphics[width=0.60\textwidth]{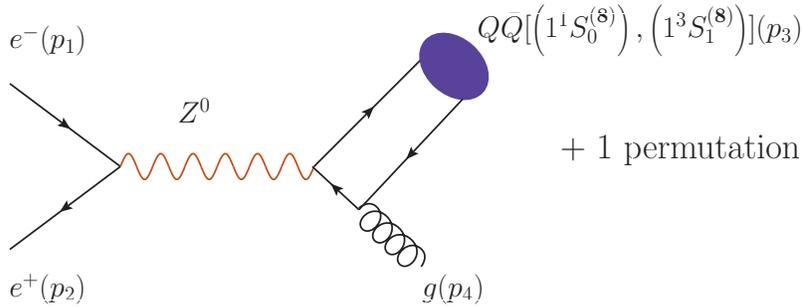}
\caption{Typical Feynman diagrams for $e^{+} e^{-}\rightarrow Z^0 \to |(Q\bar{Q})[\left(1^1S_0^{(\bf 8)}\right),\left(1^3S_1^{(\bf 8)}\right)]g\rangle+g$, where $Q$ stands for the heavy $c$ or $b$ quark.} \label{QQQQoctet3}
\end{figure}

Typical Feynman diagrams for the processes $e^+e^- \to Z^0 \to |(Q\bar{Q})[\left(1^1S_0^{\bf (8)}\right),\left(1^3S_1^{\bf (8)}\right)]g\rangle + g$ are presented in Fig. \ref{QQQQoctet3}. The reduced amplitude ${\cal A}_1^{\nu}$ takes the form
\begin{equation}
{\cal A}_1^{\nu} = \textrm{Tr}\left[\left( \Pi^{0(1)}_{(Q\bar{Q})}(p_3) \right){\Gamma^{\nu}_{Q\bar{Q}}}\frac{-\slashed{p}_{32}-\slashed{p}_4 + m_{Q}}{(p_{32} +p_4)^2 - m^2_{Q}}\slashed{\epsilon}(p_{4}) \right] ,
\end{equation}
and ${\cal A}_2^{\nu}$ can be obtained from ${\cal A}_1^{\nu}$ by permutation.

For the overall color-octet parameter, we have ${\cal C}_o=\frac{\sqrt 2}{2}\frac{e^2 g_s}{\sin^{2}\theta_{w} (4\cos\theta_w)^2}\times \delta_{ad}$. Here $ a,d=(1,\cdots,8)$ represent the color indices of the outgoing gluon and of the color-octet quarkonium, respectively.

\section{Numerical results}

\begin{table}[htb]
\begin{tabular}{|c| c|}
\hline
~~~Matrix elements~~~ & ~~~values~\cite{matrix element}~~~\\
\hline\hline
$\langle 0 |{\cal O}^{J/\psi}_1(^3S_1)|0 \rangle$~&~$1.2~\textrm{GeV}^3$~\\
\hline
$\langle 0 |{\cal O}^{\psi'}_1(^3S_1)|0 \rangle$~&~$7.6\times 10^{-1}~\textrm{GeV}^3$~\\
\hline
$\langle 0 |{\cal O}^{\Upsilon}_1(^3S_1)|0 \rangle$~&~$9.3~\textrm{GeV}^3$~\\
\hline
$\langle 0 |{\cal O}^{\Upsilon'}_1(^3S_1)|0 \rangle$~&~$4.6~\textrm{GeV}^3$~\\
\hline
$\langle 0 |{\cal O}^{\chi_{c1}}_1(^3P_1)|0 \rangle$~ &~$3.2\times10^{-1}~\textrm{GeV}^5$~\\
\hline
$\langle 0 |{\cal O}^{\chi_{b1}}_1(^3P_1)|0 \rangle$~&~$6.1~\textrm{GeV}^5$~\\
\hline\hline
$\langle 0 |{\cal O}^{J/\psi}_8(^3S_1)|0 \rangle$~&~$1.2\times 10^{-2}~\textrm{GeV}^3$~\\
\hline
$\langle 0 |{\cal O}^{\Upsilon}_8(^3S_1)|0 \rangle$~&~$9.5\times 10^{-3}~\textrm{GeV}^3$~\\
\hline
$\langle 0 |{\cal O}^{\chi_{c1}}_8(^3S_1)|0 \rangle$~&~$1.6\times 10^{-2}~\textrm{GeV}^3$~\\
\hline
$\langle 0 |{\cal O}^{\chi_{b1}}_8(^3S_1)|0 \rangle$~&~$4.3\times 10^{-2}~\textrm{GeV}^3$~\\
\hline
\end{tabular}
\caption{Input color-singlet and color-octet matrix elements. }
\label{me}
\end{table}

In doing numerical calculation, the heavy quark masses are taken as: $m_c=1.50^{+0.15}_{-0.15}$ GeV and $m_b=4.90^{+0.15}_{-0.15}$ GeV. The color-singlet and color-octet non-perturbative matrix elements, taken from Ref.\cite{matrix element}, are listed in Table \ref{me}. The $^1S_0$ matrix elements, not listed in the table, are derived adopting the relation
\begin{displaymath}
\langle 0|{\cal O}^{H}_{\bf [1,8]}(^3S_1)|0\rangle \simeq 3\langle 0|{\cal O}^{H}_{\bf [1,8]} (^1S_0)|0\rangle .
\end{displaymath}
The color-singlet matrix elements have been calculated using their relation to the wave functions at the origin~\cite{NRQCD1,NRQCD2}:
\begin{eqnarray}
\frac{\langle 0|{\cal O}_{\bf 1}^{\eta_c}|0 \rangle}{2N_c} &\simeq& \frac{\langle 0|{\cal O}_{\bf 1}^{J/\psi}|0 \rangle}{6N_c} = |\Psi_{1S}(0)|^2, \\
\frac{\langle 0|{\cal O}_{\bf 1}^{\eta'_c}|0 \rangle}{2N_c} &\simeq& \frac{\langle 0|{\cal O}_{\bf 1}^{\psi'}|0 \rangle}{6N_c} = |\Psi_{2S}(0)|^2
\end{eqnarray}
and
\begin{equation}
\frac{\langle 0|{\cal O}_{\bf 1}^{\chi_{c0}}|0 \rangle}{2N_c} \simeq \frac{\langle 0|{\cal O}_{\bf 1}^{\chi_{c1}}|0 \rangle}{6N_c} \simeq \frac{\langle 0|{\cal O}_{\bf 1}^{\chi_{c2}}|0 \rangle}{10N_c}= |\Psi'_{1P}(0)|^2 .
\end{equation}
For the production color-octet matrix elements, we adopt the relation~\cite{NRQCD1}
\begin{eqnarray}
\langle 0|{\cal O}_{\bf 8}^{\chi_{c2}}(^3S_1)|0\rangle&=&\frac{5}{3}\langle 0|{\cal O}_{\bf 8}^{h_c}(^1S_0)|0\rangle , \nonumber \\
\langle 0|{\cal O}_{\bf 8}^{\chi_{c1}}(^3S_1)|0\rangle&=&\frac{3}{3}\langle 0|{\cal O}_{\bf 8}^{h_c}(^1S_0)|0\rangle , \nonumber \\
\langle 0|{\cal O}_{\bf 8}^{\chi_{c0}}(^3S_1)|0\rangle&=&\frac{1}{3}\langle 0|{\cal O}_{\bf 8}^{h_c}(^1S_0)|0\rangle .
\end{eqnarray}
Here the heavy-quark spin symmetry~\cite{NRQCD1} has been implicitly adopted. That is, we will not distinguish the wavefunctions at the origin for the spin-singlet $(^1S_0)$ state and the spin-triplet $(^3S_1)$ state at the same energy level. The bottomonium color-singlet and color-octet matrix elements satisfy analogous relations~~\cite{matrix element}.

As for the renormalization scale, we take it to be $2m_c$ for charmonium production, and $2m_b$ for bottomonium production. Other input parameters are taken from the Particle Data Group~\cite{pdg}: $\Gamma_Z=2.4952$ GeV, $m_Z=91.1876$ GeV, ${\sin}^{2}{\theta _w}$=$0.2312$. By using the leading-order $\alpha_s$ running and $\alpha_s(m_Z)=0.1184$, we obtain $\alpha_s(2m_c)=0.237$ and $\alpha_s(2m_b)=0.175$.

\subsection{Properties of the color-singlet processes}

The color-singlet components provide the dominant contributions to the heavy quarkonium production processes. We will make a detailed discussion on the properties of the color-singlet processes, including their total and differential cross sections and the uncertainties related to a variation of the $e^+ e^-$ collision energy with respect to the $Z^0$-boson mass and to the knowledge of the heavy quark masses.

\subsubsection{Total and differential cross sections}

\begin{table}[h]
\begin{tabular}{|c|| c| c|}
\hline
 & \multicolumn{2}{c|}{Total cross sections}\\
\hline
~production channels~&~$Z^0$-propagator~&~$\gamma^*$-propagator~\\
\hline\hline
$e^+e^-\to\eta_c(1S) +c\bar{c}$~&~$1.76$~&~$4.09\times10^{-3}$\\
\hline
$e^+e^-\to\eta_c'(2S)+c\bar{c}$~&~$1.11$~&~$2.59\times10^{-3}$\\
\hline
$e^+e^-\to J/\psi(1S) +c\bar{c}$~&~$1.83$~&~$4.39\times10^{-3}$\\
\hline
$e^+e^-\to\psi'(2S)+c\bar{c}$~&~$1.16$~&~$2.78\times10^{-3}$\\
\hline
$e^+e^-\to J/\psi(1S)+gg$~&~$3.84\times10^{-2}$~&~$6.99\times10^{-4}$\\
\hline
$e^+e^-\to\psi'(2S)+gg$~&~$2.43\times10^{-2}$~&~$4.42\times10^{-4}$\\
\hline
$e^+e^-\to\eta_c(1S)+gg$~&~$1.72\times10^{-1}$~&~$0$\\
\hline
$e^+e^-\to\eta_c'(2S)+gg$~&~$1.09\times10^{-1}$~&~$0$\\
\hline
$e^+e^-\to h_c(1P) +c\bar{c}$~&~$2.34\times10^{-1}$~&~$5.60\times10^{-4}$\\
\hline
$e^+e^-\to\chi_{c0}(1P) +c\bar{c}$~&~$3.32\times10^{-1}$~&~$7.78\times10^{-4}$\\
\hline
$e^+e^-\to\chi_{c1}(1P) +c\bar{c}$~&~$3.66\times10^{-1}$~&~$8.37\times10^{-4}$\\
\hline
$e^+e^-\to\chi_{c2}(1P) +c\bar{c}$~&~$1.44\times10^{-1}$~&~$3.24\times10^{-4}$\\
\hline
\end{tabular}
\caption{Total cross section (in pb) for the color-singlet charmonium production. The channels are through $Z^0$ and $\gamma^*$ propagators, respectively, for $\sqrt{s} = m_Z$ and $m_c = 1.5$ GeV. }
\label{ccx1}
\end{table}

\begin{table}[h]
\begin{tabular}{|c ||c|| c|}
\hline
 & \multicolumn{2}{c|}{Total cross sections}\\
\hline
~production channels~&~$Z^0$-propagator~&~$\gamma^*$-propagator~\\
\hline\hline
$e^+e^-\to\eta_b(1S) +b\bar{b}$~&~$1.94\times10^{-1}$~&~$8.69\times10^{-5}$\\
\hline
$e^+e^-\to\eta_b'(2S) +b\bar{b}$~&~$9.60\times10^{-2}$~&~$4.30\times10^{-5}$\\
\hline
$e^+e^-\to\Upsilon(1S) +b\bar{b}$~&~$2.18\times10^{-1}$~&~$1.16\times10^{-4}$\\
\hline
$e^+e^-\to\Upsilon'(2S) +b\bar{b}$~&~$1.08\times10^{-1}$~&~$5.74\times10^{-5}$\\
\hline
$e^+e^-\to\Upsilon(1S) +gg$~&~$6.54\times10^{-2}$~&~$9.16\times10^{-5}$\\
\hline
$e^+e^-\to\Upsilon'(2S) +gg$~&~$3.24\times10^{-2}$~&~$4.53\times10^{-5}$\\
\hline
$e^+e^-\to\eta_b(1S) +gg$~&~$8.18\times10^{-2}$~&~$0$\\
\hline
$e^+e^-\to\eta_b'(2S) +gg$~&~$4.05\times10^{-2}$~&~$0$\\
\hline
$e^+e^-\to h_b(1P) +b\bar{b}$~&~$6.84\times10^{-3}$~&~$3.36\times10^{-6}$\\
\hline
$e^+e^-\to\chi_{b0}(1P) +b\bar{b}$~&~$1.09\times10^{-2}$~&~$5.18\times10^{-6}$\\
\hline
$e^+e^-\to\chi_{b1}(1P) +b\bar{b}$~&~$1.04\times10^{-2}$~&~$4.22\times10^{-6}$\\
\hline
$e^+e^-\to\chi_{b2}(1P) +b\bar{b}$~&~$4.30\times10^{-3}$~&~$1.67\times10^{-6}$\\
\hline
\end{tabular}
\caption{Total cross section (in pb) for the color-singlet bottomonium production. The channels are through $Z^0$ and $\gamma^*$ propagators, respectively, for $\sqrt{s} = m_Z$ and $m_b = 4.9$ GeV. }
\label{bbx1}
\end{table}

\begin{figure*}
\includegraphics[width=0.45\textwidth]{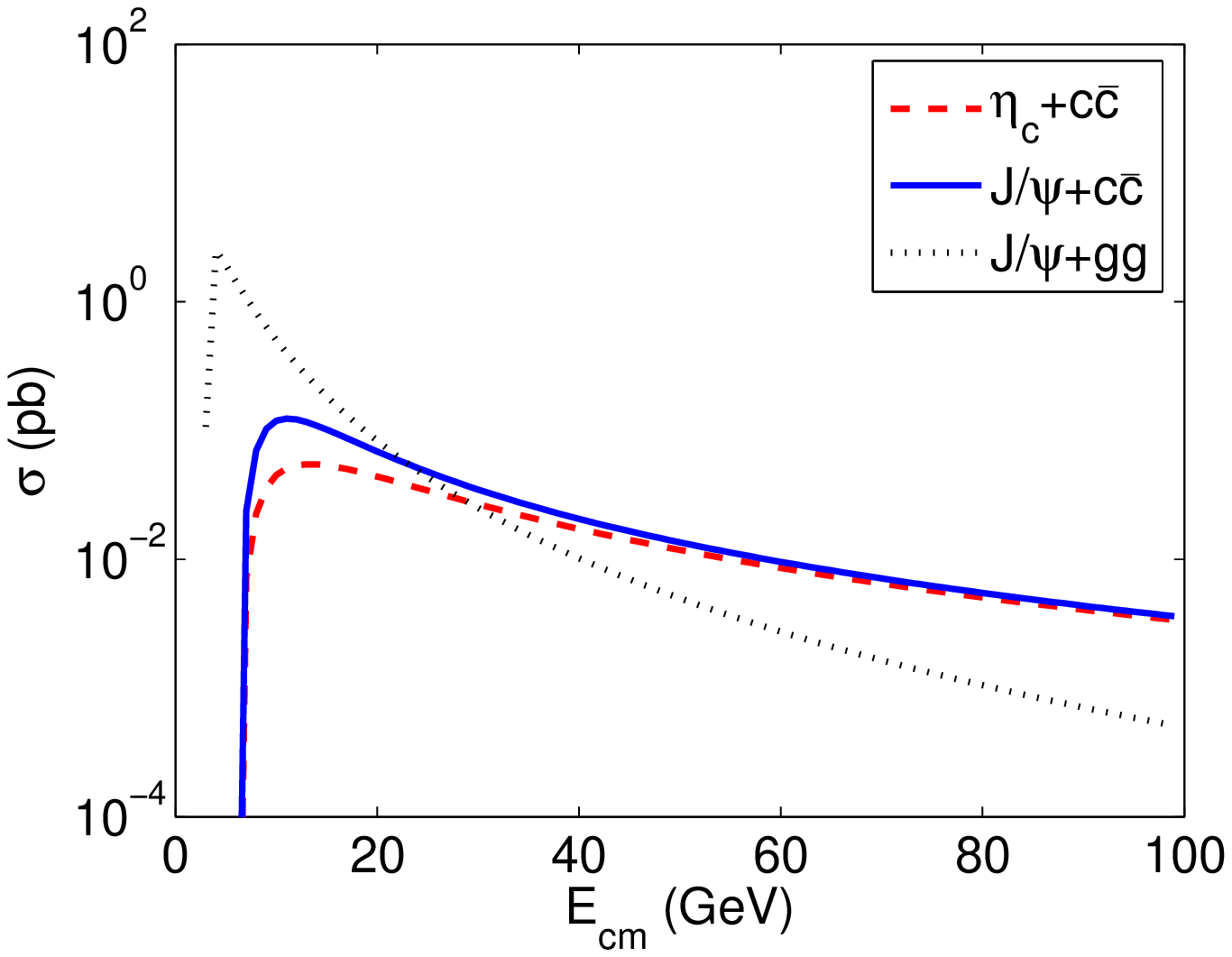}
\includegraphics[width=0.45\textwidth]{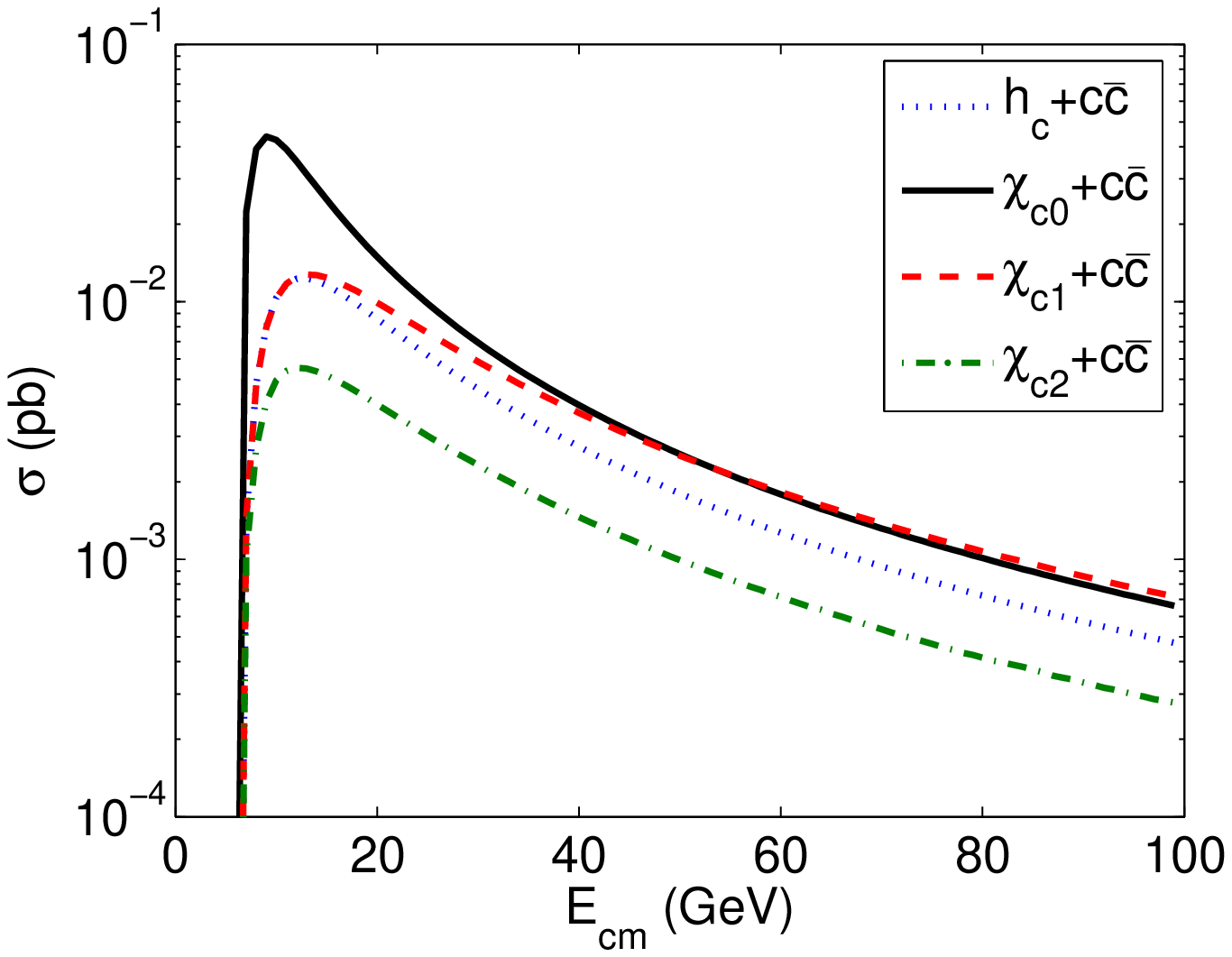}
\caption{Total cross sections of the channels $e^{+}+e^{-}\rightarrow \gamma^* \rightarrow |H_{c\bar{c}}\rangle+X$ for different $S$ wave and $P$ wave charmonium states versus the $e^+ e^-$ collision energy $E_{cm}=\sqrt{s}$.} \label{ccY}
\end{figure*}

\begin{figure*}
\includegraphics[width=0.45\textwidth]{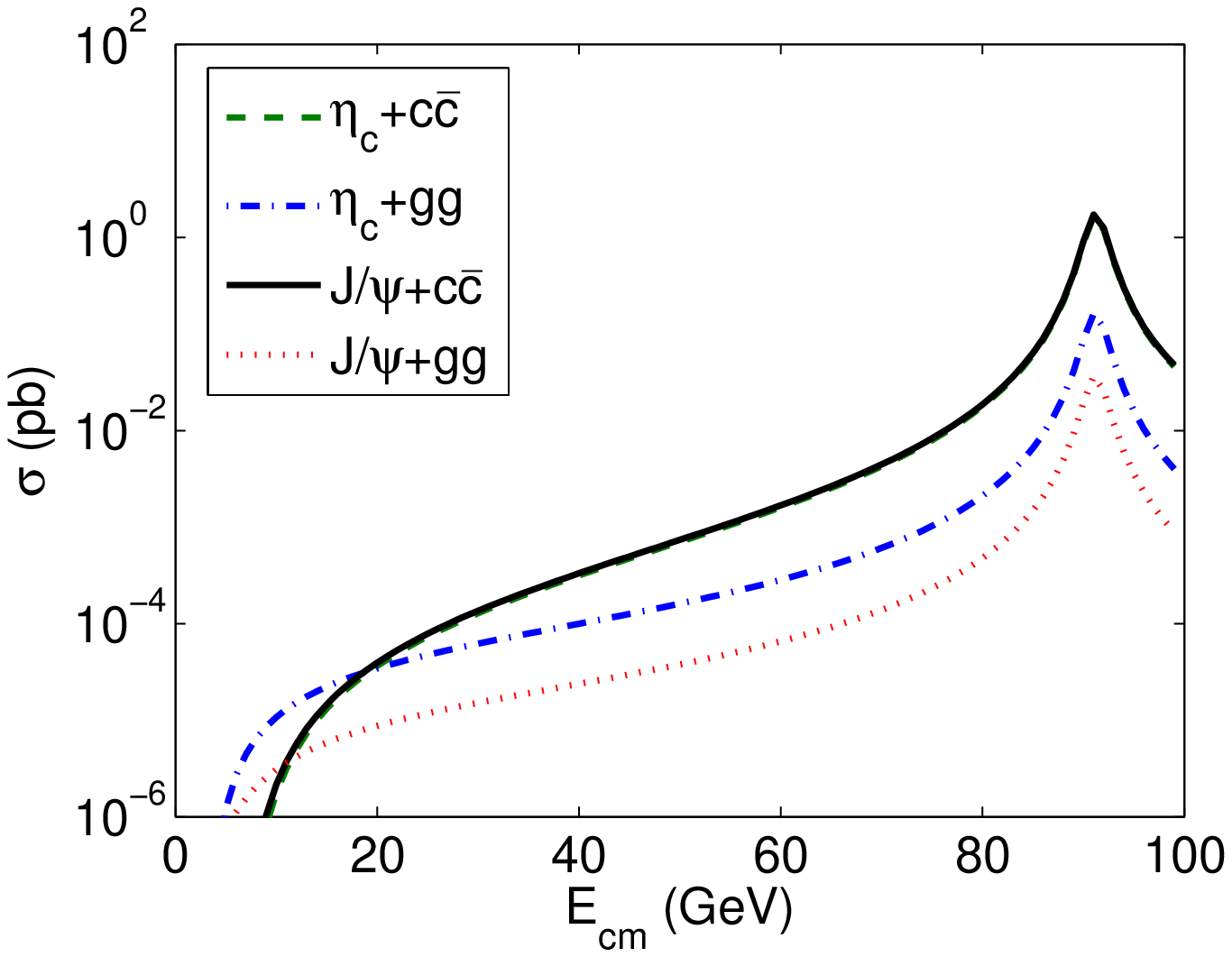}
\includegraphics[width=0.45\textwidth]{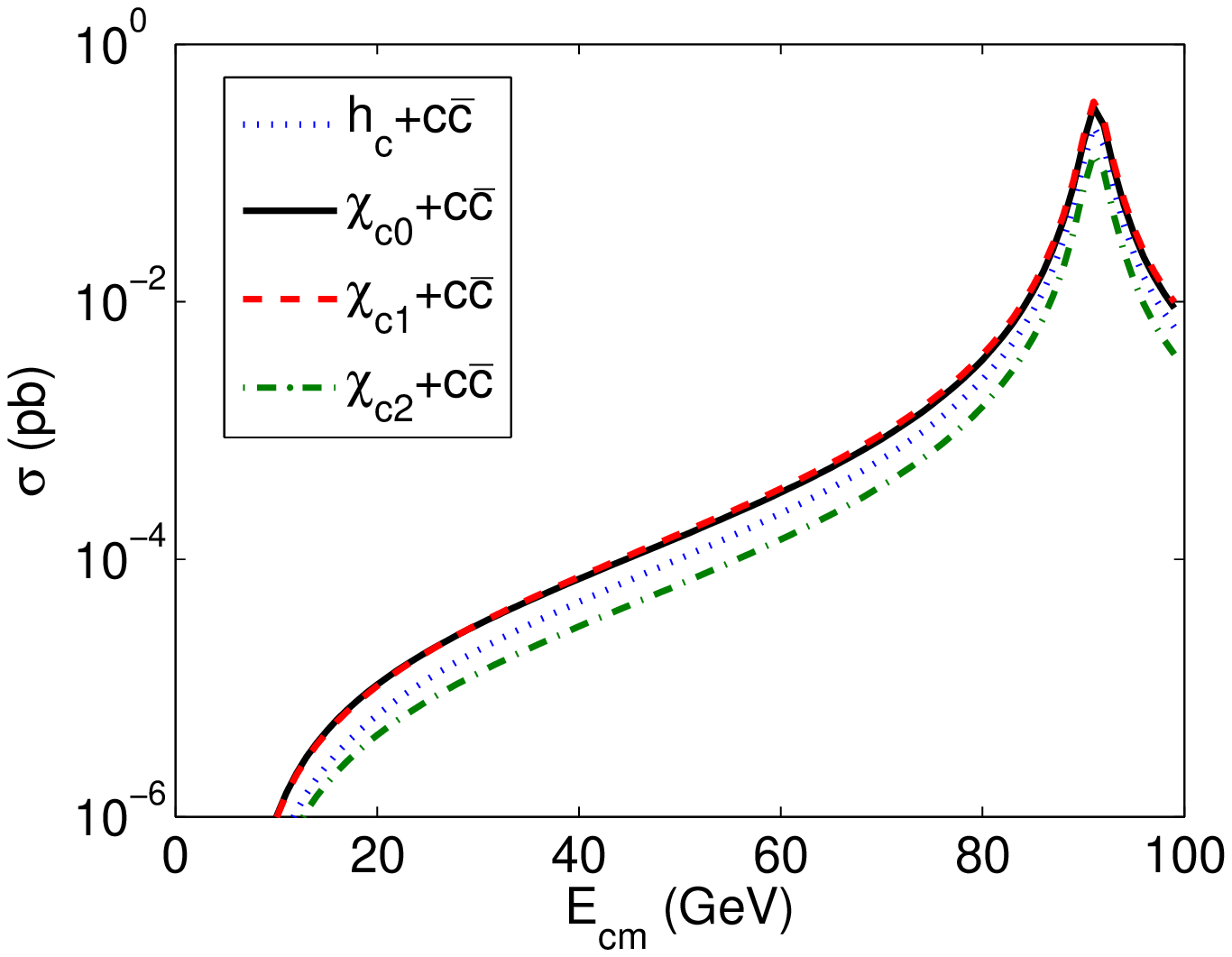}
\caption{Total cross sections of the channels $e^{+}+e^{-}\rightarrow Z^0 \rightarrow |H_{c\bar{c}}\rangle+X$ for different $S$ wave and $P$ wave charmonium states versus the $e^+ e^-$ collision energy $E_{cm}=\sqrt{s}$. The two curves for $\eta_c+c\bar{c}$ and $J/\psi+c\bar{c}$ almost coincide with each other, and the two curves for $\chi_{c0}+c\bar{c}$ and $\chi_{c1}+c\bar{c}$ also almost coincide with each other. } \label{ccZ}
\end{figure*}

\begin{figure*}
\includegraphics[width=0.45\textwidth]{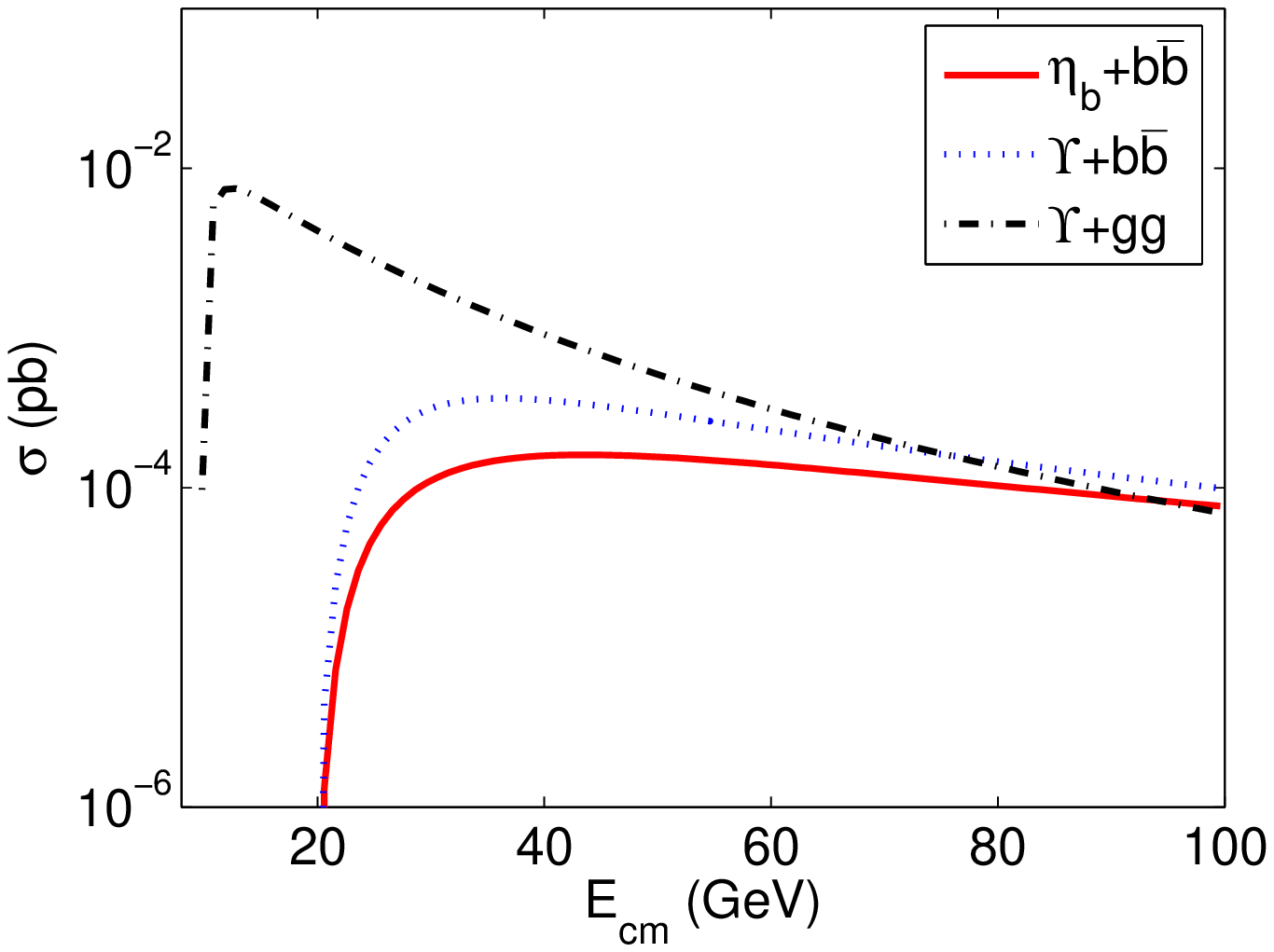}
\includegraphics[width=0.45\textwidth]{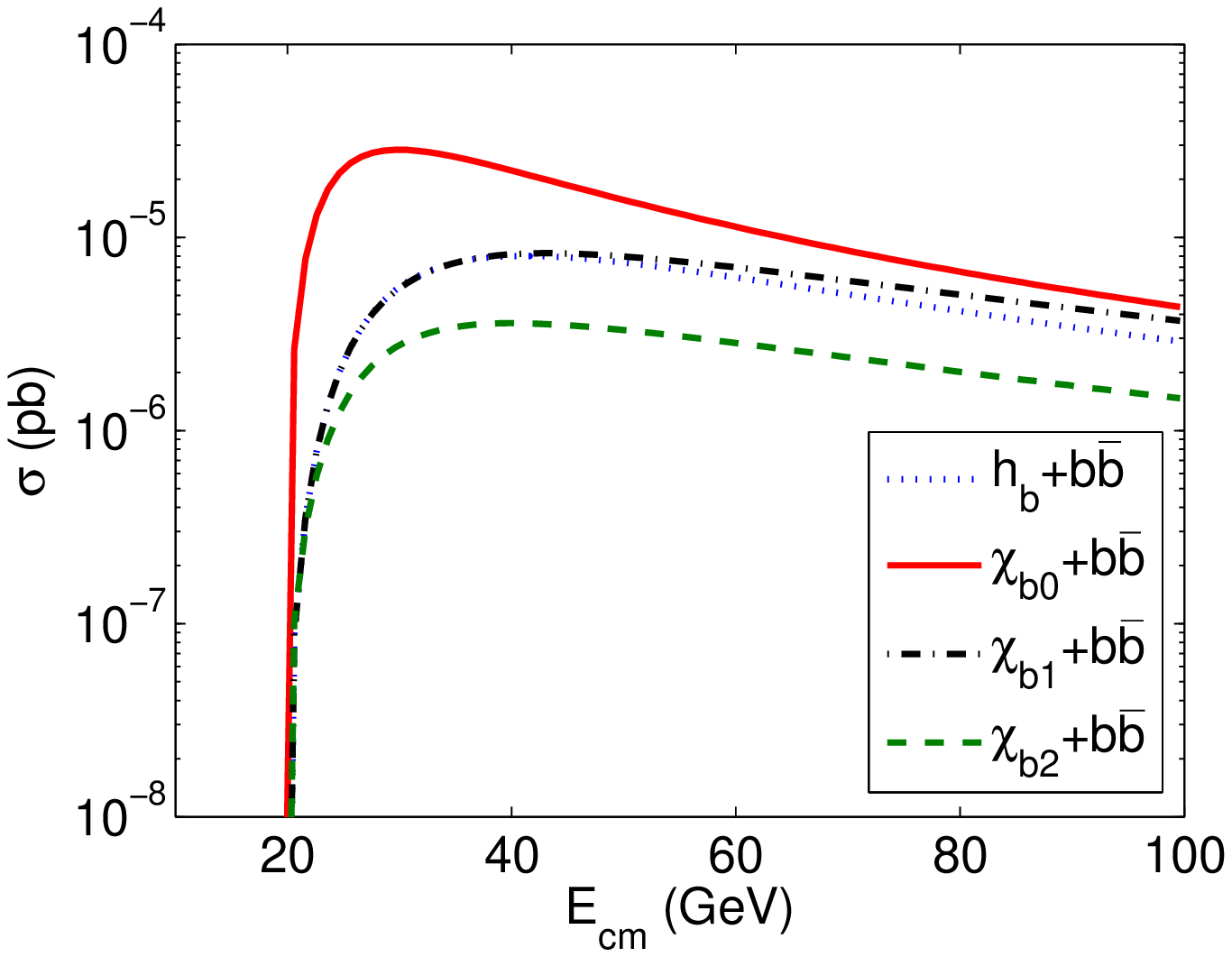}
\caption{Total cross sections of the channels $e^{+}+e^{-}\rightarrow \gamma^* \rightarrow |H_{b\bar{b}}\rangle+X$ for different $S$ wave and $P$ wave bottomonium states versus the $e^+ e^-$ collision energy $E_{cm}=\sqrt{s}$.} \label{bbY}
\end{figure*}

\begin{figure*}
\includegraphics[width=0.45\textwidth]{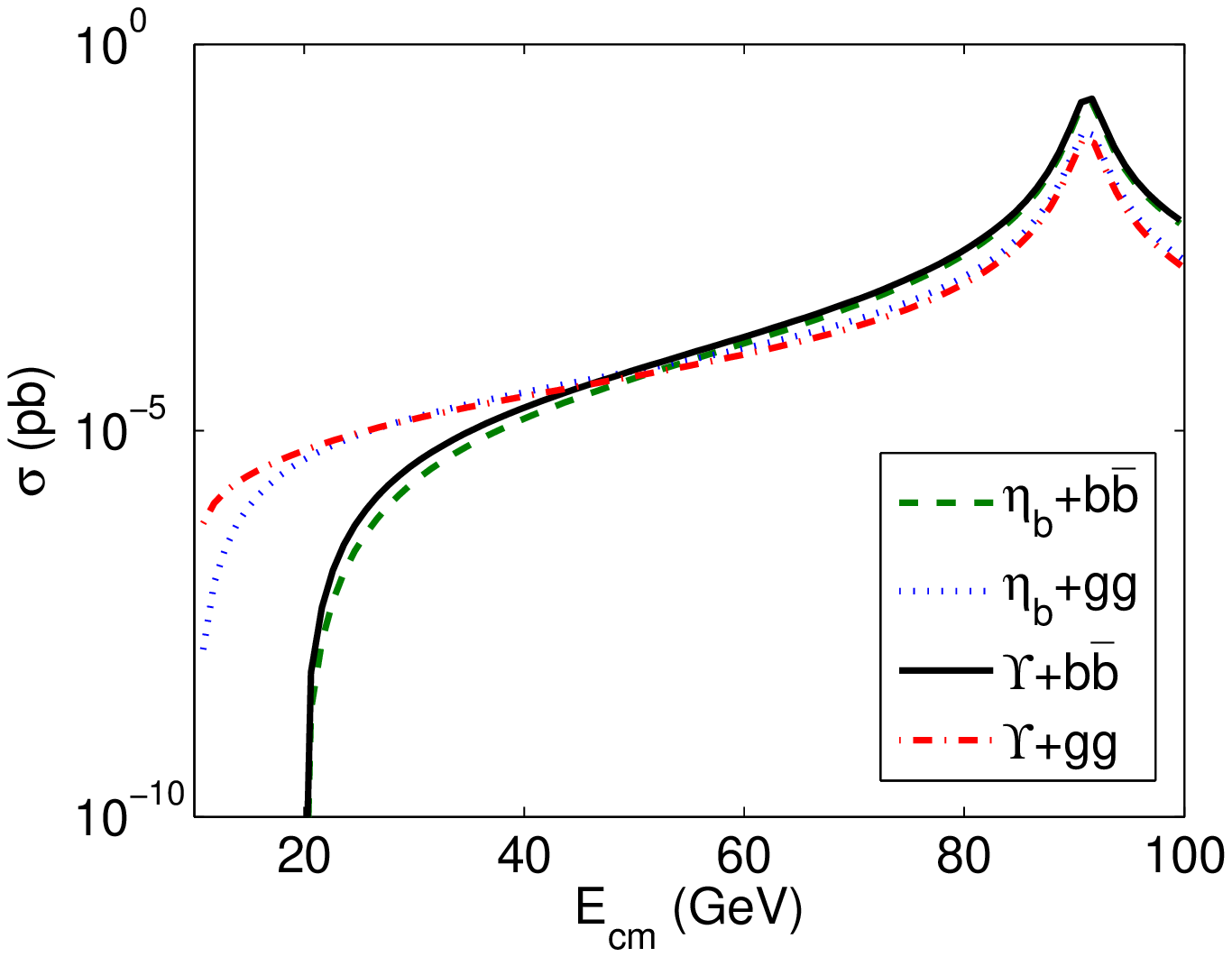}
\includegraphics[width=0.45\textwidth]{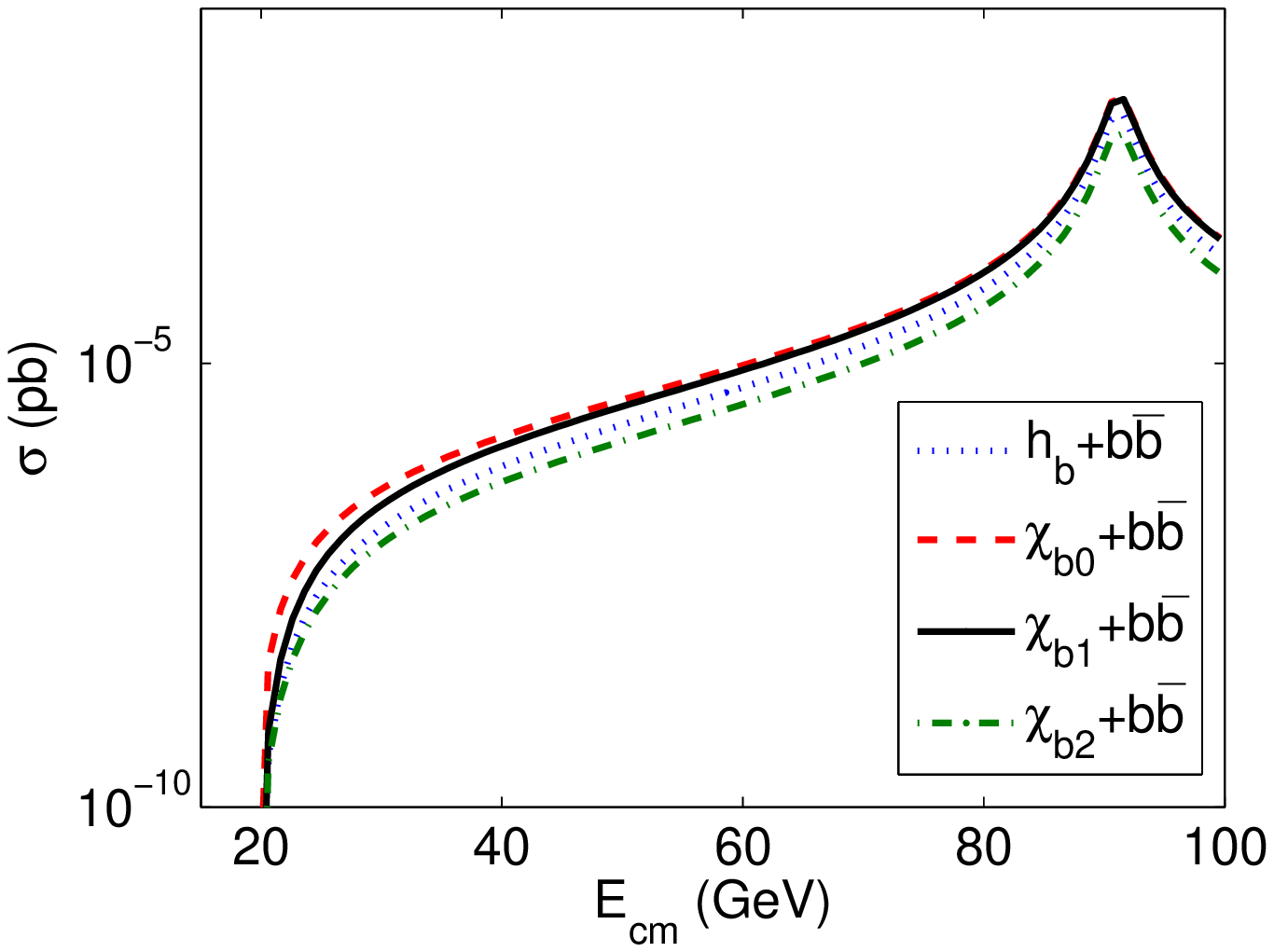}
\caption{Total cross sections of the channels $e^{+}+e^{-}\rightarrow Z^0 \rightarrow |H_{b\bar{b}}\rangle+X$ for different $S$ wave and $P$ wave bottomonium states versus the $e^+ e^-$ collision energy $E_{cm}=\sqrt{s}$. The two curves for $\eta_b+b\bar{b}$ and $\Upsilon+b\bar{b}$ almost coincide with each other, and the two curves for $\chi_{b0}+b\bar{b}$ and $\chi_{b1}+b\bar{b}$ also almost coincide with each other.} \label{bbZ}
\end{figure*}

Total cross sections for the charmonium and bottomonium productions at the $Z^0$ peak ($\sqrt{s}=m_Z$) are presented in Tables \ref{ccx1} and \ref{bbx1}. By adding all the $1S$-wave charmonium or bottomonium states together, we obtain $$\sigma(e^+e^- \to |H_{c\bar{c}}\rangle(1S)+c\bar{c})\simeq 3.59\; {\rm pb}$$ and $$\sigma(e^+e^- \to |H_{b\bar{b}}\rangle(1S)+b\bar{b})\simeq 0.41 \;{\rm pb}.$$ By adding all the $1P$-wave charmonium or bottomonium states together, we obtain $$\sigma(e^+e^- \to |H_{c\bar{c}}\rangle(1P)+c\bar{c})\simeq 1.08\; {\rm pb}$$ and $$\sigma(e^+e^- \to |H_{b\bar{b}}\rangle(1P)+b\bar{b})\simeq 3.25\times 10^{-2} \;{\rm pb}.$$ By adding all the $2S$-wave charmonium or bottomonium states together, we obtain $$\sigma(e^+e^- \to |H_{c\bar{c}}\rangle(2S)+c\bar{c})\simeq 2.27\; {\rm pb}$$ and $$\sigma(e^+e^- \to |H_{b\bar{b}}\rangle(2S)+b\bar{b})\simeq 0.20 \;{\rm pb}.$$ Then, in addition to the $1S$-level quarkonium states, both the $2S$-level and $1P$-level quarkonium states can provide sizable contributions to the production channel $e^+e^- \to |H_{Q\bar{Q}}\rangle +Q\bar{Q}$. More explicitly, to show the relative importance of those channels via the $Z^0$ propagator, we define two type of ratios
    \begin{equation}
     R_{H_{c\bar{c}}+X} = \frac{\sigma_{e^+e^-\rightarrow Z^0 \rightarrow |H_{c\bar{c}}\rangle+X}} {\sigma_{e^+e^-\rightarrow Z^0 \rightarrow J/\psi + c\bar{c}}}
    \end{equation}
    and
    \begin{equation}
     R_{H_{b\bar{b}}+X} = \frac{\sigma_{e^+e^-\rightarrow Z^0 \rightarrow |H_{b\bar{b}}\rangle+X}} {\sigma_{e^+e^-\rightarrow Z^0 \rightarrow \Upsilon + b\bar{b}}} .
    \end{equation}
    We obtain
    \begin{eqnarray*}
      R_{\eta_c+gg}  &=& 9.4 \% \;,\; R_{\eta_c'+gg} = 6.0 \% \;,\; R_{J/\psi+gg} = 2.1 \% \;, \nonumber\\
      R_{\psi'+gg} &=& 1.3 \% \;,\; R_{\eta_c+c\bar{c}} = 96 \% \;,\; R_{\eta_c'+c\bar{c}} = 61 \% \;, \nonumber\\
      R_{\psi'+c\bar{c}} &=& 63 \% \;,\; R_{h_c+c\bar{c}} = 13 \% \;,\; R_{\chi_{c0}+c\bar{c}} = 18 \% \;,\nonumber\\
      R_{\chi_{c1}+c\bar{c}} &=& 20 \% \;,\; R_{\chi_{c2}+c\bar{c}} = 7.8\%
    \end{eqnarray*}
    and
    \begin{eqnarray*}
      R_{\eta_b+gg}  &=& 38 \% \;,\; R_{\eta_b'+gg} = 19 \% \;,\; R_{\Upsilon+gg} = 30 \% \;,\nonumber\\
      R_{\Upsilon'+gg} &=& 15 \% \;,\; R_{\eta_b+b\bar{b}} = 89 \% \;,\; R_{\eta_b'+b\bar{b}} = 44 \% \;, \\
      R_{\Upsilon'+b\bar{b}} &=& 50 \% \;,\; R_{h_b+b\bar{b}}  = 3.1 \% \;,\; R_{\chi_{b0}+b\bar{b}} = 5.0 \% \;,\nonumber\\
      R_{\chi_{b1}+b\bar{b}} &=& 4.8 \% \;,\; R_{\chi_{b2}+b\bar{b}} = 2.0 \% .
    \end{eqnarray*}

For the $e^+e^-$ collision energy $E_{cm}=\sqrt{s}=m_Z$, the cross sections for the channels via the $\gamma^*$ propagator are much smaller than the same channels via the $Z^0$ propagator. For example, we have $$\frac{\sigma_{e^+e^- \to \gamma^* \to J/\psi+c\bar{c}}}{\sigma_{e^+e^- \to Z^0 \to J/\psi+c\bar{c}}}=2.4\times10^{-3}$$ and $$\frac{\sigma_{e^+e^- \to \gamma^* \to \Upsilon+b\bar{b}}}{\sigma_{e^+e^- \to Z^0 \to \Upsilon+b\bar{b}}}=5.0\times10^{-4} . $$ To show this point clearly, we present the total cross sections for the color-singlet channels versus the collision energy $E_{cm}$ in Figs. \ref{ccY}, \ref{ccZ}, \ref{bbY} and \ref{bbZ}. In the low collision energy region, the production cross sections are dominated by the cases via the $\gamma^*$ propagator, whose main contributions are around $6-20$ GeV for charmonium production and $10-40$ GeV for bottomonium production. The production via $Z^0$ propagator is small in the low energy region, but has a peak value at $E_{cm}=m_{Z}$ due to the $Z^0$-boson resonance effect.

For the charmonium production channels, from Fig. \ref{ccZ} and Fig. \ref{bbZ}, it is found that the total cross sections for $e^+e^- \to Z^0 \to J/\psi+c\bar{c}$ and $e^+e^- \to Z^0 \to \eta_c+c\bar{c}$, and the total cross sections for $e^+e^- \to Z^0 \to \chi_{c0}+c\bar{c}$ and $e^+e^- \to Z^0 \to \chi_{c1}+c\bar{c}$ are almost coincident with each other, especially for larger collision energy. The conditions for the bottomonium cases are similar. This shows that there is approximate ``spin degeneracy" for the production channels $e^{+} e^{-}\to Z^{0}\to |H_{Q\bar{Q}}\rangle + Q\bar{Q}$. Quantitatively, when the collision energy $E_{cm}=m_Z$, we have
\begin{eqnarray}
    \frac{\sigma_{e^+e^- \to Z^0 \to \eta_c+c\bar{c}}}{\sigma_{e^+e^- \to Z^0 \to J/\psi+c\bar{c}}} &\simeq& 96\% ,\label{csc1} \\
    \frac{\sigma_{e^+e^- \to Z^0 \to \chi_{c0}+c\bar{c}}}{\sigma_{e^+e^- \to Z^0 \to \chi_{c1}+c\bar{c}}} &\simeq& 91\%  ,\label{csc2} \\
    \frac{\sigma_{e^+e^- \to Z^0 \to \eta_b+b\bar{b}}}{\sigma_{e^+e^- \to Z^0 \to \Upsilon+b\bar{b}}} &\simeq& 89\%  ,\label{csc3}  \\
    \frac{\sigma_{e^+e^- \to Z^0 \to \chi_{b1}+b\bar{b}}}{\sigma_{e^+e^- \to Z^0 \to \chi_{b0}+b\bar{b}}} &\simeq& 95\%  .\label{csc4}
\end{eqnarray}

\begin{figure*}
\includegraphics[width=0.45\textwidth]{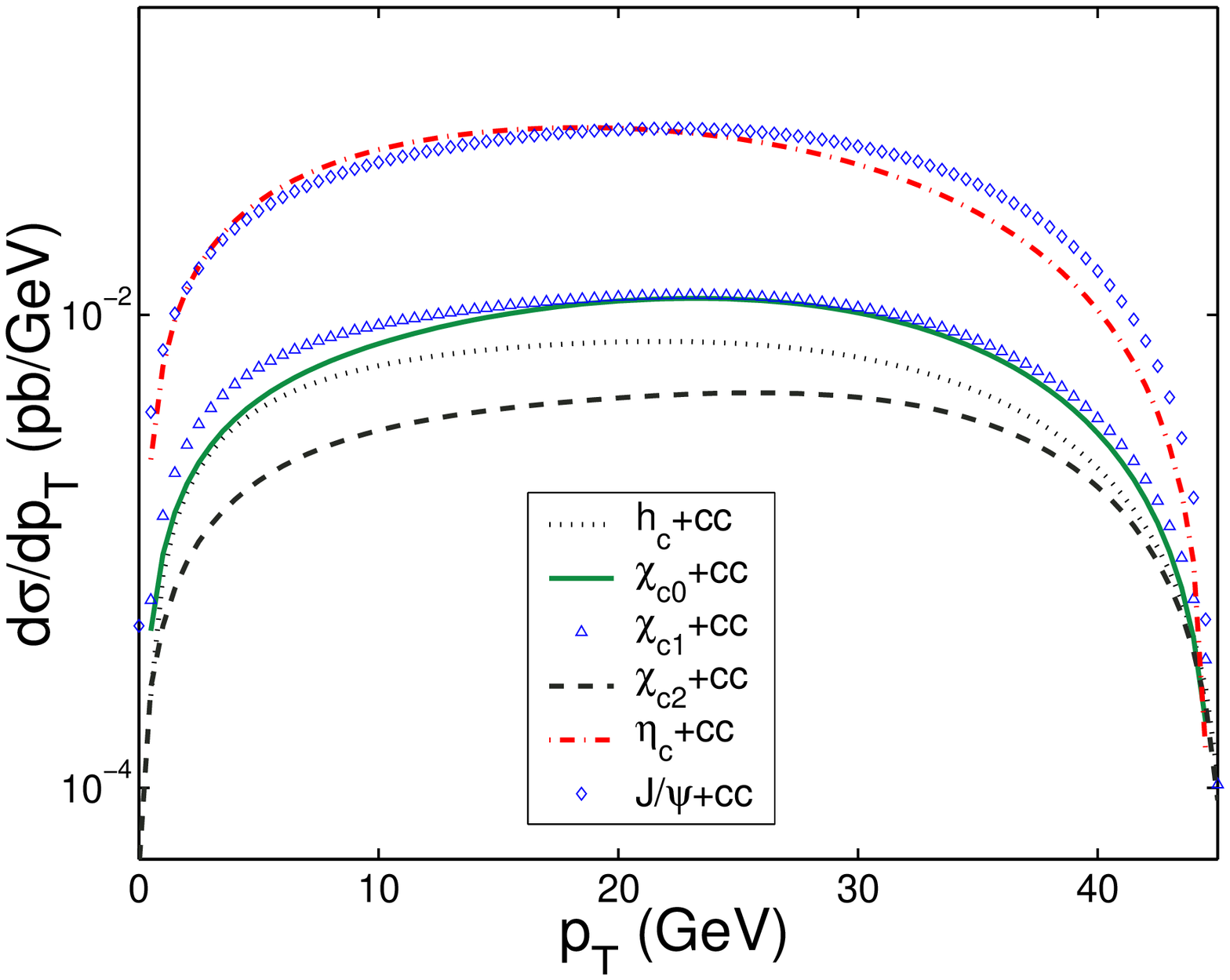}
\includegraphics[width=0.45\textwidth]{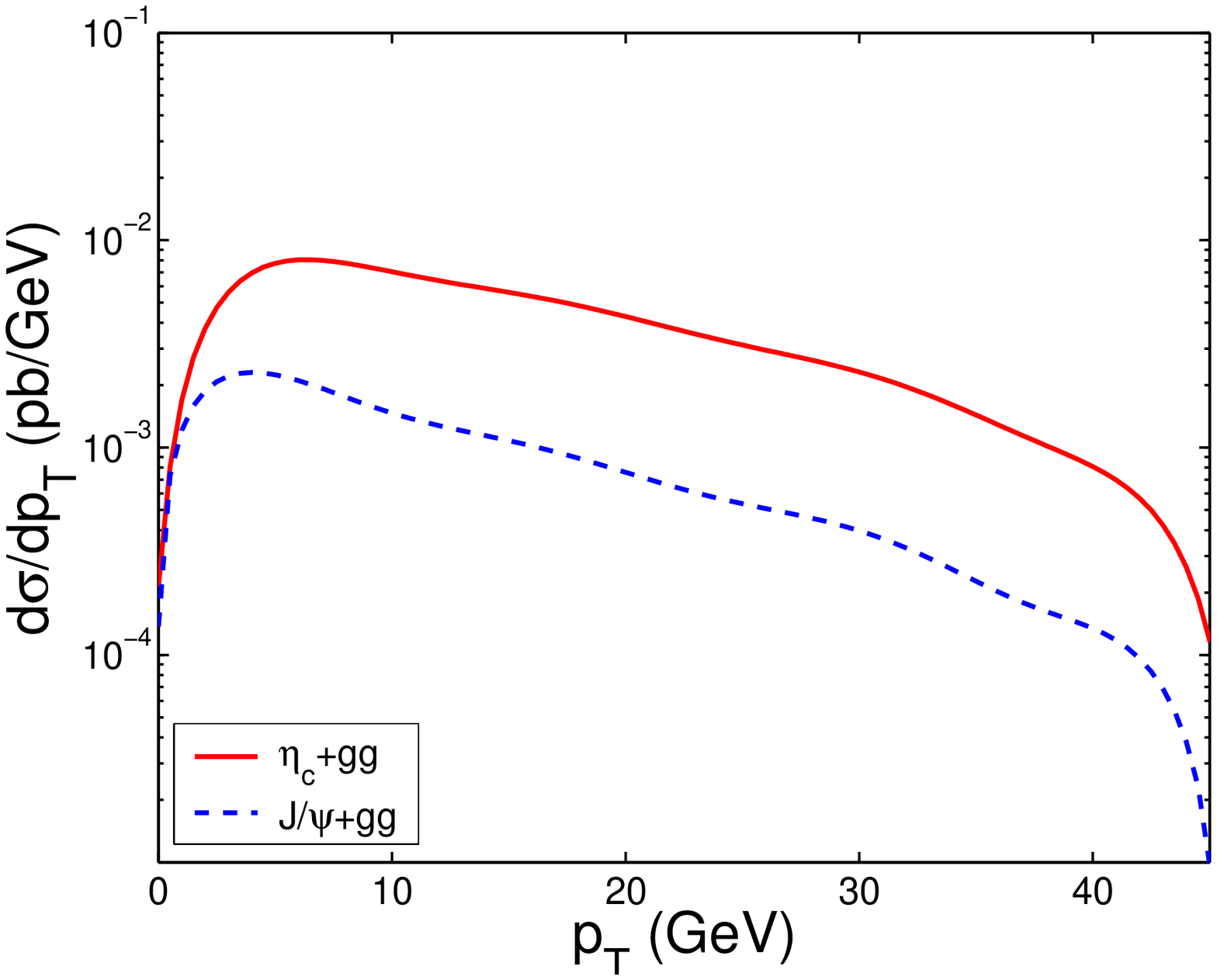}
\caption{The charmonium $p_T$ distributions for the production processes $e^+e^- \to Z^0 \to |H_{c\bar{c}}\rangle+X$ at the $e^+ e^-$ collision energy $E_{cm}=m_Z$. The left panel is for $X=c\bar{c}$, the right one is for $X=gg$. } \label{ptccZ}
\end{figure*}

\begin{figure*}
\includegraphics[width=0.45\textwidth]{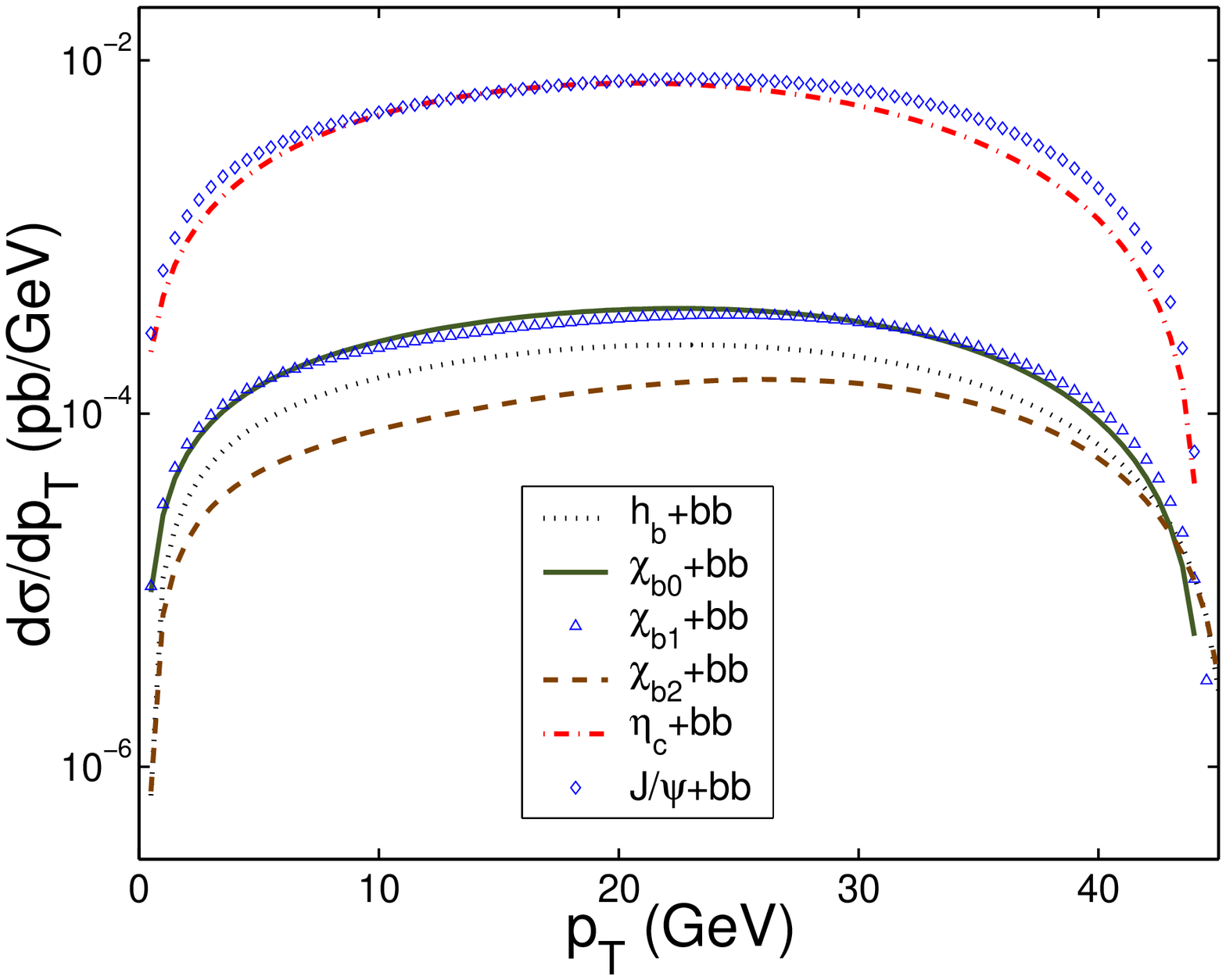}
\includegraphics[width=0.45\textwidth]{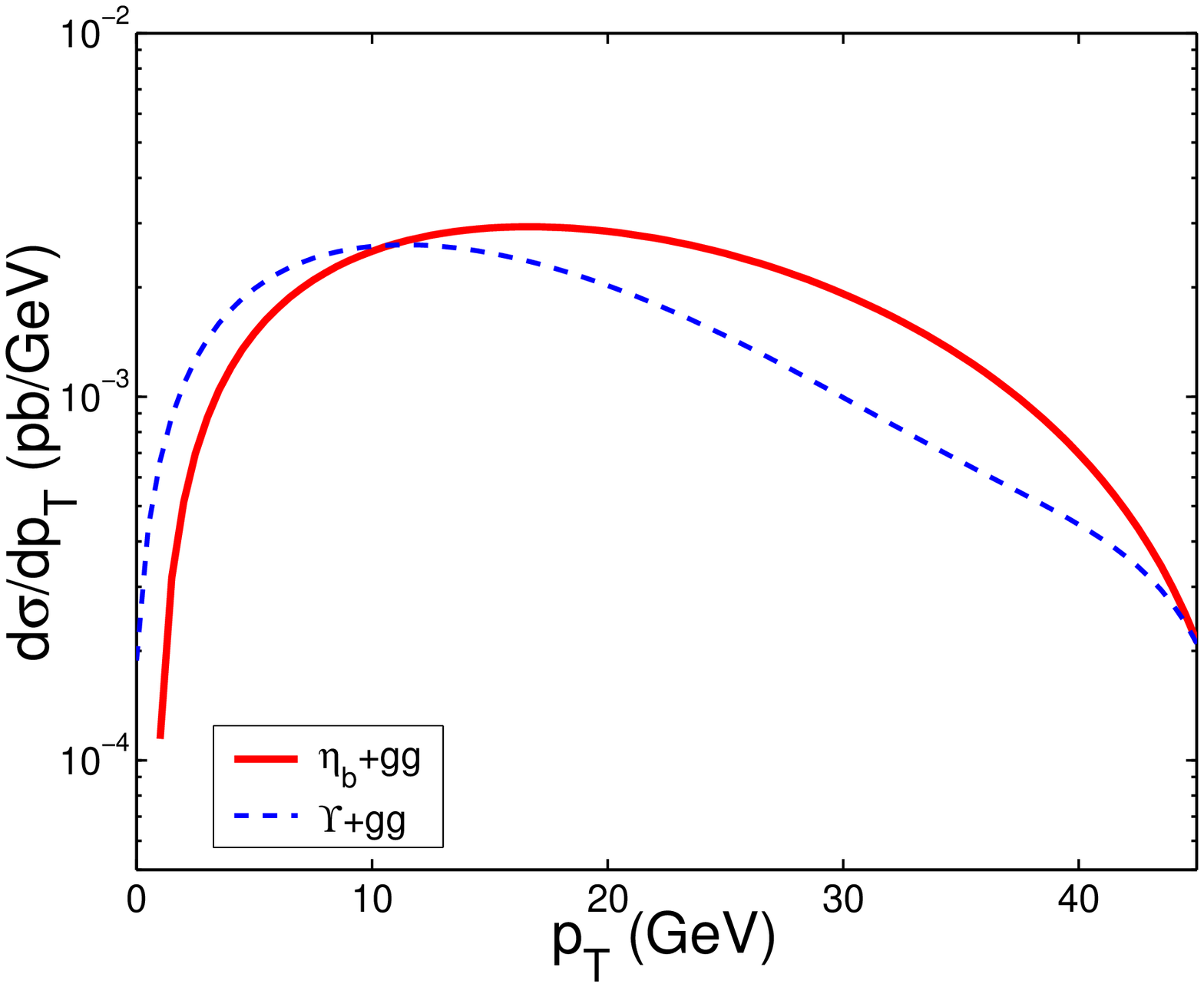}
\caption{The bottomonium $p_T$ distributions for the production processes $e^+e^- \to Z^0 \to |H_{b\bar{b}}\rangle+X$ at the $e^+ e^-$ collision energy $E_{cm}=m_Z$. The left panel is for $X=b\bar{b}$, the right one is for $X=gg$. } \label{ptbbZ}
\end{figure*}

\begin{figure*}
\includegraphics[width=0.45\textwidth]{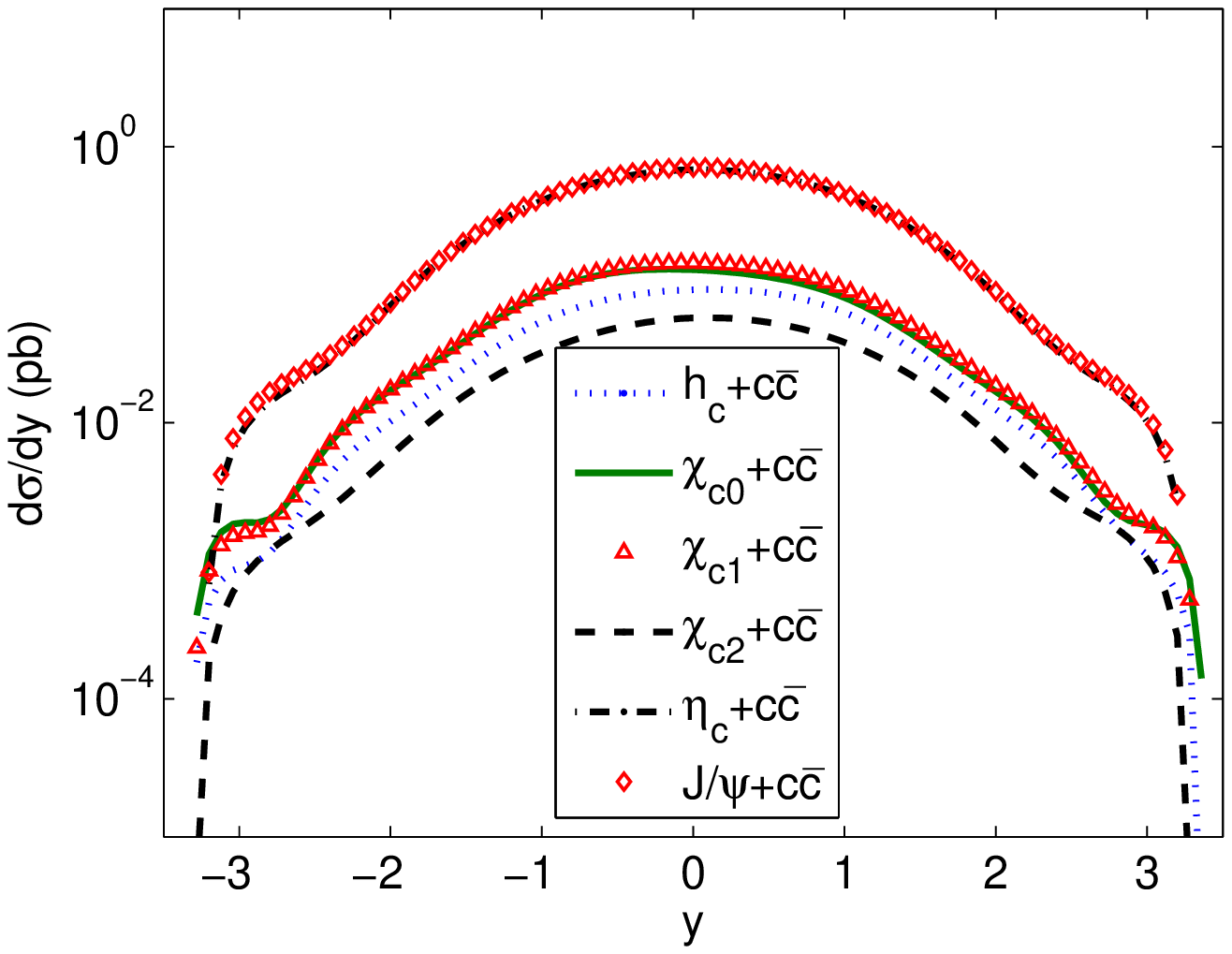}
\includegraphics[width=0.45\textwidth]{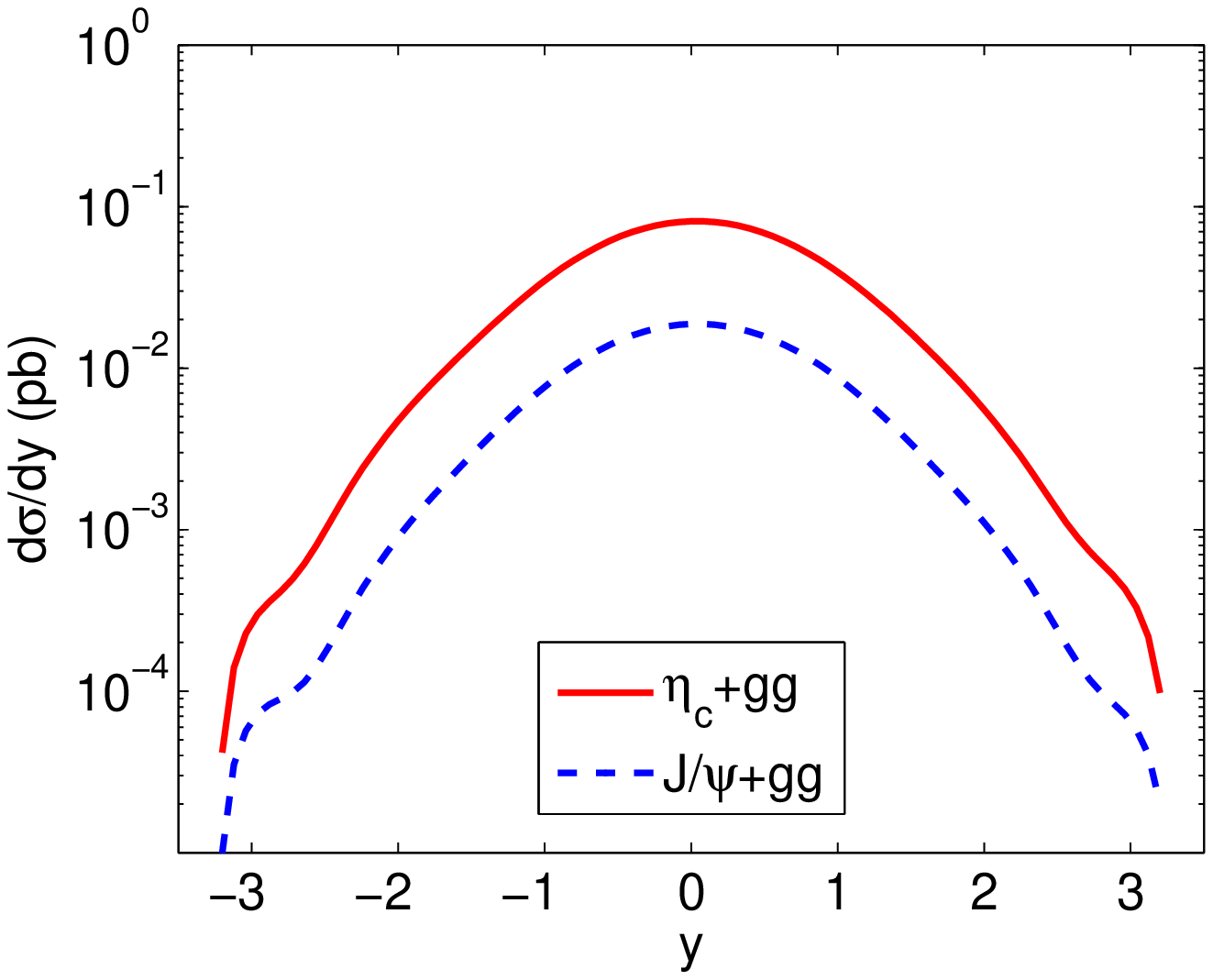}
\caption{The charmonium $y$ distributions for the production processes $e^+e^- \to Z^0 \to |H_{c\bar{c}}\rangle+X$ at the $e^+ e^-$ collision energy $E_{cm}=m_Z$. The left panel is for $X=c\bar{c}$, the right one is for $X=gg$. } \label{dyccZ}
\end{figure*}

\begin{figure*}
\includegraphics[width=0.45\textwidth]{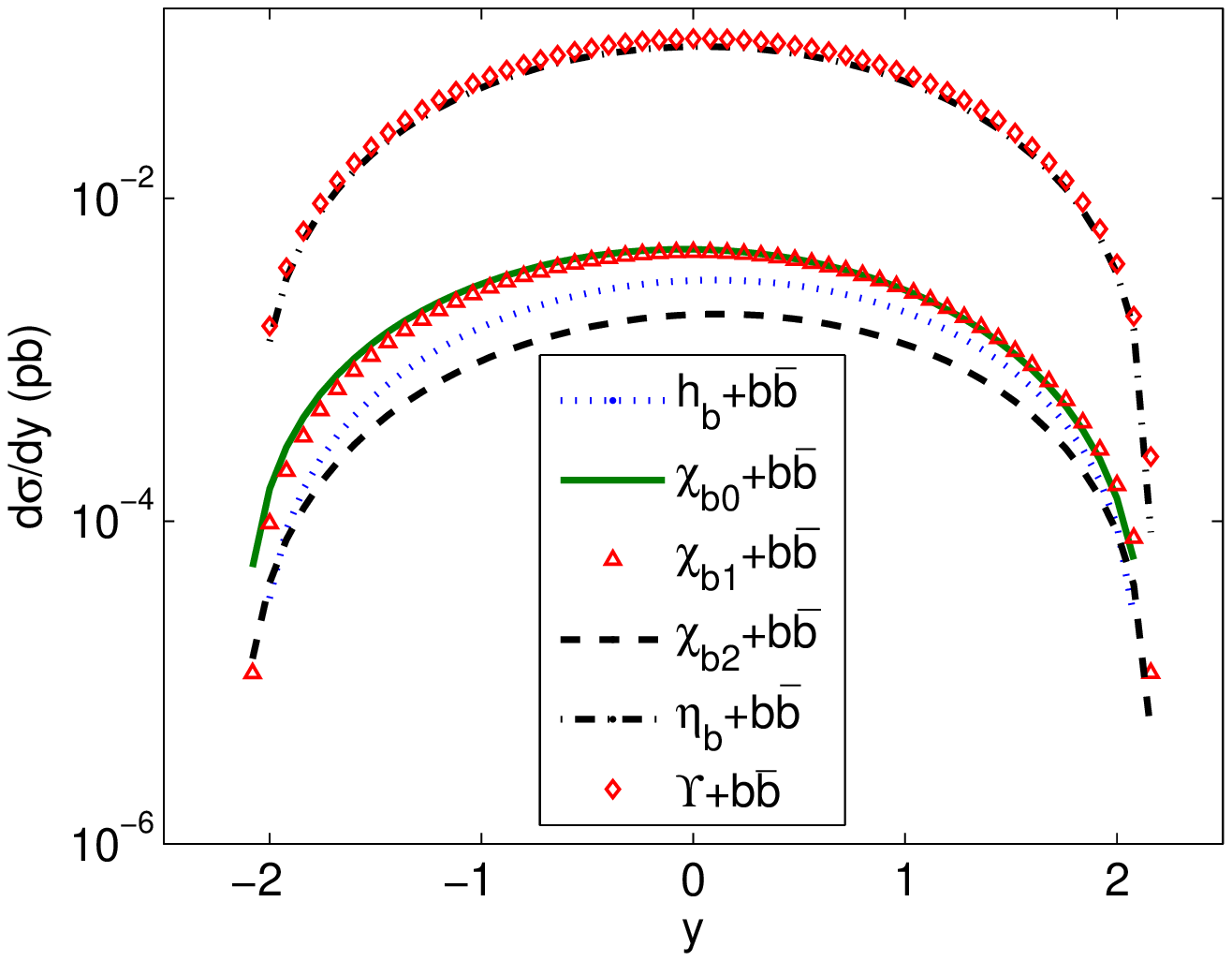}
\includegraphics[width=0.45\textwidth]{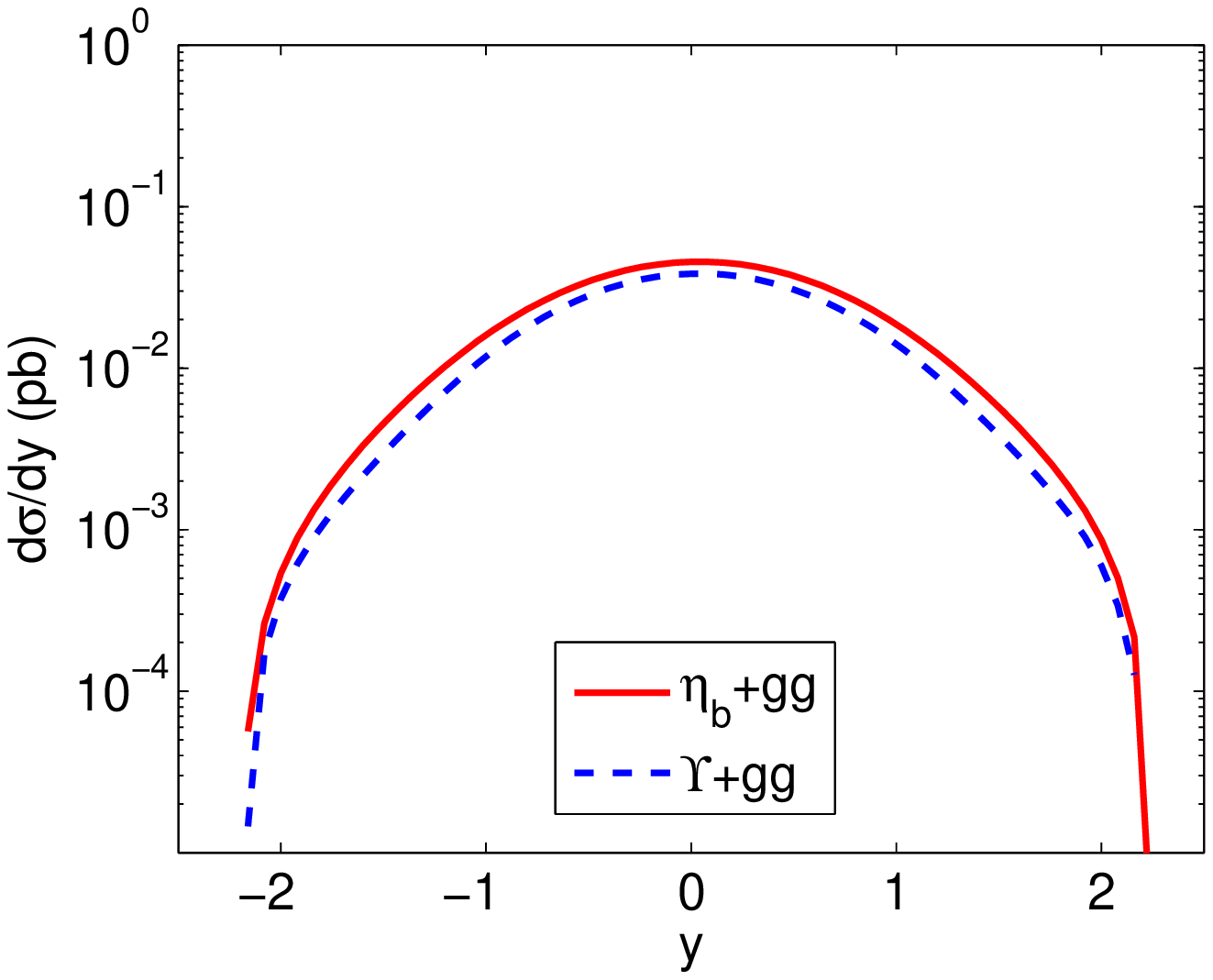}
\caption{The bottomonium $y$ distributions for the production processes $e^+e^- \to Z^0 \to |H_{b\bar{b}}\rangle+X$ at the $e^+ e^-$ collision energy $E_{cm}=m_Z$. The left panel is for $X=b\bar{b}$, the right one is for $X=gg$. } \label{dybbZ}
\end{figure*}

We present the heavy quarkonium transverse momentum $(p_T)$ distributions and rapidity ($y$) distributions for $E_{cm}=m_Z$ in Figs. \ref{ptccZ}, \ref{ptbbZ}, \ref{dyccZ} and \ref{dybbZ}. These figures show that both for the $p_T$ distributions and for the $y$ distributions, the two curves for $e^+e^- \to Z^0 \to J/\psi+c\bar{c}$ and $e^+e^- \to Z^0 \to \eta_c+c\bar{c}$, and the two curves for $e^+e^- \to Z^0 \to \chi_{c0}+c\bar{c}$ and $e^+e^- \to Z^0 \to \chi_{c1}+c\bar{c}$ are very close to each other. The curves for the bottomonium production are similar.

As a cross check / explanation for the approximate ``spin degeneracy", we adopt the fragmentation approach. Using the fragmentation approach the most important/dominant higher-order effects can be included by using the Dokshitzer-Gribov-Lipatov-Altarelli-Parisi evolution equation, and the fragmentation approach gives a reasonable approximation to the full tree-level calculation as long as the transverse momentum of the produced heavy hadron is large enough~\cite{frag1,frag2,frag3,frag4,frag5,frag6,frag7,frag8,frag9,frag10,frag11,frag12,frag13}. Moveover, it has been argued that the fragmentation approach may be extended to lower $p_T$ region under the FONLL scheme~\cite{fonll1,fonll2}.

\begin{figure}
\includegraphics[width=0.60\textwidth]{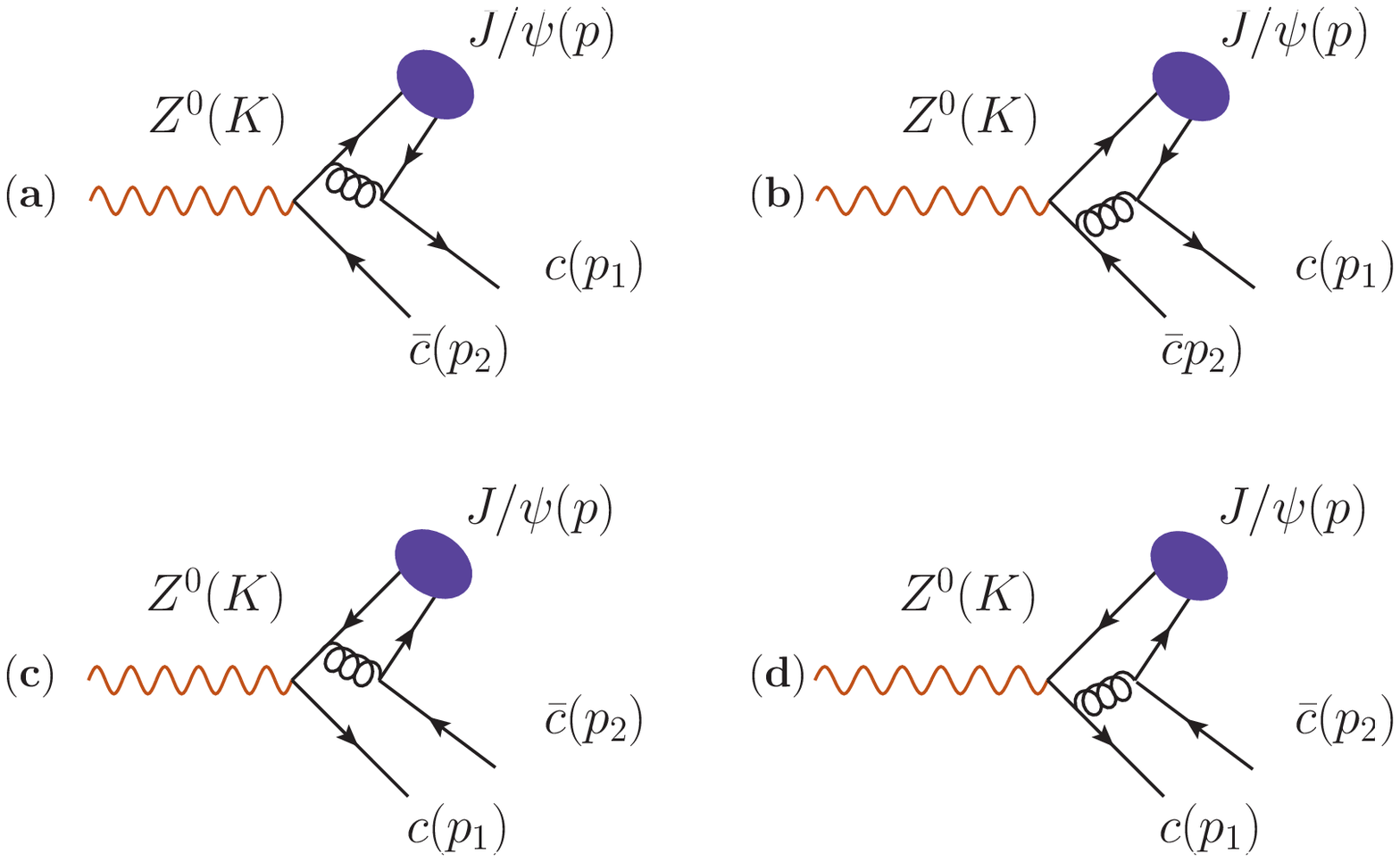}
\caption{Feynman diagrams for calculating the total fragmentation probabilities for $c\to J/\psi$ and $c\to\eta_c$. } \label{ZQQ}
\end{figure}

In the large $p_T$ region, the probabilities of $Z^0\to J/\psi$ and $Z^0\to\eta_c$ are proportional to each other and dominated by the fragmentations $c\to J/\psi$ and $c\to\eta_c$, respectively. To calculate them, we need to deal with the Feynman diagrams shown in Fig. \ref{ZQQ} for the $J/\psi$ case. With the choice of the axial gauge, the amplitudes of Figs. \ref{ZQQ}b and \ref{ZQQ}d are suppressed by a factor $m_c/m_Z$ in comparison to those of Figs. \ref{ZQQ}a and \ref{ZQQ}c and they are, therefore, neglected in deriving the fragmentation probabilities for $c\to J/\psi$ and $c\to\eta_c$. Following the same procedure as shown by Ref.\cite{frag11}, one can obtain the total fragmentation probabilities:
\begin{widetext}
\begin{eqnarray}
\int^{1}_{0}{\rm d} z D_{c \rightarrow J/\psi}(z,3m_c) &=& \frac{256}{27} \alpha_s(3m_c)^{2} \times\frac{|\Psi_{1S}(0)|^2}{M_{J/\psi}^{3}} \left(\frac{1189}{30}-57\ln 2\right) , \label{frag1} \\
\int^{1}_{0}{\rm d} z D_{c \rightarrow \eta_c}(z,3m_c) &=& \frac{256}{27} \alpha_s(3m_c)^{2} \times\frac{|\Psi_{1S}(0)|^2}{M_{\eta_c}^{3}} \left(\frac{773}{30}-37\ln 2\right) . \label{frag2}
\end{eqnarray}
\end{widetext}

Then, the ratio of the two fragmentation probabilities is $$\frac{\int^{1}_{0}{\rm d}z D_{c\rightarrow J/\psi}(z,3m_c)}{\int^{1}_{0}{\rm d}z D_{c\rightarrow \eta_c}(z,3m_c)} \simeq 97\%.$$ This is consistent with Eq.(\ref{csc1}) and confirms that there is spin-degeneracy for $e^+e^-$ at the high collision energy. The total fragmentation probabilities for $b\to\Upsilon$ and $b\to\eta_b$ can be obtained from these expressions by replacing $m_c$ with $m_b$. The probabilities for $P$-wave production can be derived in a similar way, obtaining the results found in Ref.\cite{frag13}, which we confirm with our calculations.

\begin{widetext}
\begin{center}
\begin{table}
\begin{tabular}{|c|| c| c| c|}
\hline
~Channel~&~$p_T>5$ GeV~&~$p_T>10$ GeV~&~$p_T>15$ GeV~\\
\hline\hline
$\sigma_{J/\psi(1S) +c\bar{c}}$~&~$1.75(96\%)$~&~$1.55(85\%)$~&~$1.29(70\%)$~\\
\hline
$\sigma_{\psi'(2S)+c\bar{c}}$~&~$1.11(96\%)$~&~$9.81 \times 10^{-1}(85\%)$~&~$8.17 \times 10^{-1}(70\%)$~\\
\hline
$\sigma_{J/\psi(1S)+gg}$~&~$2.89\times 10^{-2}(75\%)$~&~$1.96 \times 10^{-2}(51\%)$~&~$1.30 \times 10^{-2}(34\%)$~\\
\hline
$\sigma_{\psi'(2S)+gg}$~&~$1.83\times 10^{-2}(75\%)$~&~$1.24 \times 10^{-2}(51\%)$~&~$8.23 \times 10^{-3}(34\%)$~\\
\hline\hline
$\sigma_{\eta_c(1S) +c\bar{c}}$~&~$1.68(95\%)$~&~$1.46(83\%)$~&~$1.17(66\%)$~\\
\hline
$\sigma_{\eta_c'(2S)+c\bar{c}}$~&~$1.06(95\%)$~&~$9.24 \times 10^{-1}(83\%)$~&~$7.41 \times 10^{-1}(66\%)$~\\
\hline
$\sigma_{\eta_c(1S)+gg}$~&~$1.47\times 10^{-1}(85\%)$~&~$1.08 \times 10^{-1}(63\%)$~&~$7.51 \times 10^{-2}(44\%)$~\\
\hline
$\sigma_{\eta_c'(2S)+gg}$~&~$9.31\times 10^{-2}(85\%)$~&~$6.84 \times 10^{-2}(63\%)$~&~$4.75 \times 10^{-2}(44\%)$~\\
\hline\hline
$\sigma_{h_c(1P) +c\bar{c}}$~&~$2.22\times 10^{-1}(95\%)$~&~$1.96 \times 10^{-1}(84\%)$~&~$1.61 \times 10^{-1}(69\%)$~\\
\hline
$\sigma_{\chi_{c0}(1P) +c\bar{c}}$~&~$3.17\times 10^{-1}(95\%)$~&~$2.86 \times 10^{-1}(86\%)$~&~$2.42 \times 10^{-1}(73\%)$~\\
\hline
$\sigma_{\chi_{c1}(1P) +c\bar{c}}$~&~$3.46\times 10^{-1}(94\%)$~&~$3.08 \times 10^{-1}(84\%)$~&~$2.56 \times 10^{-1}(70\%)$~\\
\hline
$\sigma_{\chi_{c2}(1P) +c\bar{c}}$~&~$1.37\times 10^{-1}(95\%)$~&~$1.24 \times 10^{-1}(86\%)$~&~$1.05 \times 10^{-1}(73\%)$~\\
\hline
\end{tabular}
\caption{Cross section (in pb) for the color-singlet charmonium production with different $p_{T}$ cuts. The channels are through $Z^0$ propagator for $\sqrt{s} = m_Z$ and $m_c$=1.5 GeV. The percentages in the parentheses represent the ratios between the cross section with and without $p_{T}$ cut.}
\label{ptccZcs}
\end{table}
\end{center}

\begin{center}
\begin{table}
\begin{tabular}{|c|| c| c| c|}
\hline
~Channel~&~$p_T>5$ GeV~&~$p_T>10$ GeV~&~$p_T>15$ GeV~\\
\hline\hline
$\sigma_{\Upsilon(1S) +b\bar{b}}$~&~$2.10 \times 10^{-1}(96\%)$~&~$1.89\times 10^{-1}(87\%)$~&~$1.59\times 10^{-1}(73\%)$~\\
\hline
$\sigma_{\Upsilon'(2S)+b\bar{b}}$~&~$1.04\times 10^{-1}(96\%)$~&~$9.35 \times 10^{-2}(87\%)$~&~$7.86\times 10^{-2}(73\%)$~\\
\hline
$\sigma_{\Upsilon(1S)+gg}$~&~$5.90\times 10^{-2}(90\%)$~&~$4.63 \times 10^{-2}(71\%)$~&~$3.37 \times 10^{-2}(51\%)$~\\
\hline
$\sigma_{\Upsilon'(2S)+gg}$~&~$2.92\times 10^{-2}(90\%)$~&~$2.29 \times 10^{-2}(71\%)$~&~$1.67 \times 10^{-2}(51\%)$~\\
\hline\hline
$\sigma_{\eta_b(1S) +b\bar{b}}$~&~$1.88\times 10^{-1}(97\%)$~&~$1.68\times 10^{-1}(87\%)$~&~$1.38\times 10^{-1}(71\%)$~\\
\hline
$\sigma_{\eta_b'(2S)+b\bar{b}}$~&~$9.30\times 10^{-2}(97\%)$~&~$8.31\times 10^{-2}(87\%)$~&~$6.83\times 10^{-2}(71\%)$~\\
\hline
$\sigma_{\eta_b(1S)+gg}$~&~$7.85\times 10^{-2}(96\%)$~&~$6.80 \times 10^{-2}(83\%)$~&~$5.35 \times 10^{-2}(65\%)$~\\
\hline
$\sigma_{\eta_b'(2S)+gg}$~&~$3.88\times 10^{-2}(96\%)$~&~$3.36 \times 10^{-2}(83\%)$~&~$2.65 \times 10^{-2}(65\%)$~\\
\hline\hline
$\sigma_{h_b(1P) +b\bar{b}}$~&~$6.60\times 10^{-3}(96\%)$~&~$6.00 \times 10^{-3}(88\%)$~&~$5.00 \times 10^{-3}(73\%)$~\\
\hline
$\sigma_{\chi_{b0}(1P) +b\bar{b}}$~&~$1.05\times 10^{-2}(96\%)$~&~$9.40 \times 10^{-3}(86\%)$~&~$7.85 \times 10^{-3}(72\%)$~\\
\hline
$\sigma_{\chi_{b1}(1P) +b\bar{b}}$~&~$9.95\times 10^{-3}(96\%)$~&~$8.95 \times 10^{-3}(86\%)$~&~$7.60 \times 10^{-3}(73\%)$~\\
\hline
$\sigma_{\chi_{b2}(1P) +b\bar{b}}$~&~$4.16\times 10^{-3}(97\%)$~&~$3.84 \times 10^{-3}(89\%)$~&~$3.33 \times 10^{-3}(77\%)$~\\
\hline
\end{tabular}
\caption{Cross section (in pb) for the color-singlet bottomonium production with different $p_{T}$ cuts. The channels are through $Z^0$ propagator for $\sqrt{s} = m_Z$ and $m_b$=4.9 GeV. The percentages in the parentheses represent the ratios between the cross section with and without $p_{T}$ cut. }
\label{ptbbZcs}
\end{table}
\end{center}

\begin{center}
\begin{table}
\begin{tabular}{|c|| c| c| c|}
\hline
~Channel~&~$|y|<0.5$~&~$|y|<1.0$~&~$|y|<1.5$~\\
\hline\hline
$\sigma_{J/\psi(1S) +c\bar{c}}$~&~$7.00\times 10^{-1}(38\%)$~&~$1.27(69\%)$~&~$1.60(87\%)$~\\
\hline
$\sigma_{\psi'(2S)+c\bar{c}}$~&~$4.43\times 10^{-1}(38\%)$~&~$8.04\times 10^{-1}(69\%)$~&~$1.01(87\%)$~\\
\hline
$\sigma_{J/\psi(1S)+gg}$~&~$1.82\times 10^{-2}(47\%)$~&~$3.02\times 10^{-2}(79\%)$~&~$3.58 \times 10^{-2}(93\%)$~\\
\hline
$\sigma_{\psi'(2S)+gg}$~&~$1.15\times 10^{-2}(47\%)$~&~$1.91\times 10^{-2}(79\%)$~&~$2.27 \times 10^{-2}(93\%)$~\\
\hline
\hline
$\sigma_{\eta_c(1S) +c\bar{c}}$~&~$6.71\times 10^{-1}(38\%)$~&~$1.23(70\%)$~&~$1.55(88\%)$~\\
\hline
$\sigma_{\eta_c'(2S)+c\bar{c}}$~&~$4.25\times 10^{-1}(38\%)$~&~$7.79\times 10^{-1}(70\%)$~&~$9.81\times 10^{-1}(88\%)$~\\
\hline
$\sigma_{\eta_c(1S)+gg}$~&~$7.88\times 10^{-2}(46\%)$~&~$1.32\times 10^{-1}(77\%)$~&~$1.58 \times 10^{-1}(92\%)$~\\
\hline
$\sigma_{\eta_c'(2S)+gg}$~&~$4.99\times 10^{-2}(46\%)$~&~$8.36\times 10^{-2}(77\%)$~&~$1.00 \times 10^{-1}(92\%)$~\\
\hline
\hline
$\sigma_{h_c(1P) +c\bar{c}}$~&~$8.97\times 10^{-2}(38\%)$~&~$1.63\times 10^{-1}(70\%)$~&~$2.05 \times 10^{-1}(88\%)$~\\
\hline
$\sigma_{\chi_{c0}(1P) +c\bar{c}}$~&~$1.25\times 10^{-1}(38\%)$~&~$2.28\times 10^{-1}(68\%)$~&~$2.88 \times 10^{-1}(86\%)$~\\
\hline
$\sigma_{\chi_{c1}(1P) +c\bar{c}}$~&~$1.39\times 10^{-1}(38\%)$~&~$2.53\times 10^{-1}(69\%)$~&~$3.20 \times 10^{-1}(87\%)$~\\
\hline
$\sigma_{\chi_{c2}(1P) +c\bar{c}}$~&~$5.48\times 10^{-2}(38\%)$~&~$9.97\times 10^{-2}(69\%)$~&~$1.25 \times 10^{-1}(87\%)$~\\
\hline
\end{tabular}
\caption{Cross section (in pb) for the color-singlet charmonium production with different $y$ cuts. The channels are through $Z^0$ propagator for $\sqrt{s} = m_Z$ and $m_c$=1.5 GeV. The percentages in the parentheses represent the ratios between the cross sections with and without rapidity cut.}
\label{dyccZcs}
\end{table}
\end{center}

\begin{center}
\begin{table}
\begin{tabular}{|c|| c| c| c|}
\hline
~Channel~&~$|y|<0.5$~&~$|y|<1.0$~&~$|y|<1.5$~\\
\hline\hline
$\sigma_{\Upsilon(1S) +b\bar{b}}$~&~$9.40\times 10^{-2}(43\%)$~&~$1.67\times 10^{-1}(77\%)$~&~$2.06\times 10^{-1}(94\%)$~\\
\hline
$\sigma_{\Upsilon'(2S)+b\bar{b}}$~&~$4.65\times 10^{-2}(43\%)$~&~$8.26\times 10^{-2}(77\%)$~&~$1.02\times 10^{-1}(94\%)$~\\
\hline
$\sigma_{\Upsilon(1S)+gg}$~&~$3.54\times 10^{-2}(54\%)$~&~$5.63\times 10^{-2}(86\%)$~&~$6.38 \times 10^{-2}(98\%)$~\\
\hline
$\sigma_{\Upsilon'(2S)+gg}$~&~$1.75\times 10^{-2}(54\%)$~&~$2.78\times 10^{-2}(86\%)$~&~$3.16 \times 10^{-2}(98\%)$~\\
\hline
\hline
$\sigma_{\eta_b(1S) +b\bar{b}}$~&~$8.26\times 10^{-2}(43\%)$~&~$1.49\times 10^{-1}(77\%)$~&~$1.84\times 10^{-1}(95\%)$~\\
\hline
$\sigma_{\eta_b'(2S)+b\bar{b}}$~&~$4.09\times 10^{-2}(43\%)$~&~$7.37\times 10^{-2}(77\%)$~&~$9.10\times 10^{-2}(95\%)$~\\
\hline
$\sigma_{\eta_b(1S)+gg}$~&~$4.23\times 10^{-2}(52\%)$~&~$6.87\times 10^{-2}(84\%)$~&~$7.92 \times 10^{-2}(97\%)$~\\
\hline
$\sigma_{\eta_b'(2S)+gg}$~&~$2.09\times 10^{-2}(52\%)$~&~$3.40\times 10^{-2}(84\%)$~&~$3.92 \times 10^{-2}(97\%)$~\\
\hline
\hline
$\sigma_{h_b(1P) +b\bar{b}}$~&~$3.01\times 10^{-3}(44\%)$~&~$5.29\times 10^{-3}(77\%)$~&~$6.48 \times 10^{-3}(95\%)$~\\
\hline
$\sigma_{\chi_{b0}(1P) +b\bar{b}}$~&~$4.60\times 10^{-3}(42\%)$~&~$8.26\times 10^{-3}(76\%)$~&~$1.02 \times 10^{-2}(94\%)$~\\
\hline
$\sigma_{\chi_{b1}(1P) +b\bar{b}}$~&~$4.47\times 10^{-3}(43\%)$~&~$7.93\times 10^{-3}(76\%)$~&~$9.77 \times 10^{-3}(94\%)$~\\
\hline
$\sigma_{\chi_{b2}(1P) +b\bar{b}}$~&~$1.84\times 10^{-3}(43\%)$~&~$3.27\times 10^{-3}(76\%)$~&~$4.02 \times 10^{-3}(93\%)$~\\
\hline
\end{tabular}
\caption{Cross section (in pb) for the color-singlet bottomonium production with different $y$ cuts. The channels are through $Z^0$ propagator for $\sqrt{s} = m_Z$ and $m_b$=4.9 GeV. The percentages in the parentheses represent the ratios between the cross sections with and without rapidity cut.}
\label{dybbZcs}
\end{table}
\end{center}
\end{widetext}

In real experiments, due to the limited detector sensitivity and acceptance, events with small $p_T$ and/or large $|y|$ may not be detected. To account for this effect, we report in Tables \ref{ptccZcs}, \ref{ptbbZcs}, \ref{dyccZcs} and \ref{dybbZcs} the cross sections with different cuts on the variables $p_T$ and $|y|$.

\subsubsection{Uncertainties from the determinations of $E_{cm}$ and $m_Q$}

For the leading-order calculation, the uncertainty sources include the bound-state matrix elements, the renormalization scale, the quark masses $m_b$ and $m_c$. The conventional scale setting assigns the typical momentum flow of the process as the renormalization scale, e.g. $2m_c$ for charmonium production and $2m_b$ for bottomonium production. This rough assignment of scale and its range leads to an important systematic error in the present theoretical estimations. In the literature, the principle of maximum conformality (PMC)~\cite{pmc} provides a feasible way to derive precise QCD predictions. The main idea of PMC is to sum all the non-conformal $\beta$-terms in the perturbative expansion into the running coupling. The remaining terms are then identical to that of a conformal theory. The PMC estimation is then scheme independent, and the remaining scale dependence is greatly suppressed. In the present framework, the matrix elements and the strong coupling constant $\alpha_s$ emerge as overall factors and their uncertainties can be conveniently discussed, so we will not discuss their uncertainties in the present paper.

As shown in the last subsection, if the collision energy $E_{cm}$ is around $m_{Z}$, the total cross sections for the channels via the $\gamma^*$ propagator are much smaller than those of the channels via the $Z^0$ propagator. In the present subsection we address the production via $Z^0$ propagator, i.e. the channel $e^{+}e^{-} \rightarrow Z^{0}\rightarrow |H_{Q\bar{Q}}\rangle+X$. For clarity, when we discuss the uncertainty due to a given parameter we fix all others to their central values.

\begin{table}[t]
\begin{tabular}{|c||c|c|c|}
\hline
~~~Channel~~~& ~$E_{cm}=97\% m_Z$~ & ~$E_{cm}=103\%m_Z$~ & ~$\left(^{R_-}_{R_+}\right)$~ \\
\hline\hline
$\sigma_{\eta_c+c\bar{c}}$&$2.91\times10^{-1}$ &$3.14\times10^{-1}$ & $\left(^{17\%}_{18\%}\right)$ \\
\hline
$\sigma_{\eta'_c+c\bar{c}}$&$1.84\times10^{-1}$&$1.99\times10^{-1}$ & $\left(^{17\%}_{18\%}\right)$\\
\hline
$\sigma_{J/\psi+c\bar{c}}$ & $3.02\times10^{-1}$ & $3.26\times10^{-1}$ & $\left(^{17\%}_{18\%}\right)$\\
\hline
$\sigma_{\psi'+c\bar{c}}$ & $1.91\times10^{-1}$ & $2.06\times10^{-1}$ & $\left(^{17\%}_{18\%}\right)$ \\
\hline
$\sigma_{J/\psi+gg}$&$6.66\times10^{-3}$& $6.57\times10^{-3}$ & $\left(^{17\%}_{17\%}\right)$\\
\hline
$\sigma_{\psi'+gg}$&$4.22\times10^{-3}$& $4.16\times10^{-3}$ & $\left(^{17\%}_{17\%}\right)$ \\
\hline
$\sigma_{\eta_c+gg}$ & $2.97\times10^{-2}$ & $2.94\times10^{-2}$ & $\left(^{17\%}_{17\%}\right)$\\
\hline
$\sigma_{\eta'_c+gg}$ & $1.88\times10^{-2}$ & $1.86\times10^{-2}$ & $\left(^{17\%}_{17\%}\right)$\\
\hline
$\sigma_{h_c+c\bar{c}}$ & $3.88\times10^{-2}$ & $4.18\times10^{-2}$ & $\left(^{17\%}_{18\%}\right)$\\
\hline
$\sigma_{\chi_{c0}+c\bar{c}}$ & $5.51\times10^{-2}$ & $5.92\times10^{-2}$ & $\left(^{17\%}_{18\%}\right)$\\
\hline
$\sigma_{\chi_{c1}+c\bar{c}}$ & $6.08\times10^{-2}$ & $6.55\times10^{-2}$ & $\left(^{17\%}_{18\%}\right)$\\
\hline
$\sigma_{\chi_{c2}+c\bar{c}}$ & $2.40\times10^{-2}$ & $2.59\times10^{-2}$ & $\left(^{17\%}_{18\%}\right)$\\
\hline
\end{tabular}
\caption{Total cross sections (in pb) for the production channels $e^{+}e^{-} \rightarrow Z^{0}\rightarrow |H_{c\bar{c}}\rangle+X$, with varying values of $E_{cm}$. The ratios $R_{\mp}$ in the last column show how the cross sections are changed with varying values of $E_{cm}$. } \label{Ecm uncertainty1}
\end{table}

\begin{table}[t]
\begin{tabular}{|c||c|c|c|}
\hline
~~~Channel~~~&~$E_{cm}=97\% m_Z$~ & ~$E_{cm}=103\%m_Z$~ & ~$\left(^{R_-}_{R_+}\right)$~ \\
\hline\hline
$\sigma_{\eta_b+b\bar{b}}$ & $3.17\times10^{-2}$ & $3.53\times10^{-2}$ & $\left(^{16\%}_{18\%}\right)$ \\
\hline
$\sigma_{\eta_b'+b\bar{b}}$ & $1.57\times10^{-2}$ & $1.75\times10^{-2}$ & $\left(^{16\%}_{18\%}\right)$ \\
\hline
$\sigma_{\Upsilon+b\bar{b}}$ & $3.57\times10^{-2}$ & $3.95\times10^{-2}$ & $\left(^{16\%}_{18\%}\right)$ \\
\hline
$\sigma_{\Upsilon'+b\bar{b}}$ & $1.77\times10^{-2}$ & $1.95\times10^{-2}$ & $\left(^{16\%}_{18\%}\right)$ \\
\hline
$\sigma_{\Upsilon+gg}$ & $1.12\times10^{-2}$ & $1.13\times10^{-2}$ & $\left(^{17\%}_{17\%}\right)$\\
\hline
$\sigma_{\Upsilon'+gg}$ & $5.54\times10^{-3}$ & $5.59\times10^{-3}$ & $\left(^{17\%}_{17\%}\right)$\\
\hline
$\sigma_{\eta_b+gg}$ & $1.40\times10^{-2}$ & $1.42\times10^{-2}$ & $\left(^{17\%}_{17\%}\right)$\\
\hline
$\sigma_{\eta_b'+gg}$ & $6.93\times10^{-3}$ & $7.03\times10^{-3}$ & $\left(^{17\%}_{17\%}\right)$\\
\hline
$\sigma_{h_b+b\bar{b}}$ & $1.13\times10^{-3}$ & $1.24\times10^{-3}$ & $\left(^{16\%}_{18\%}\right)$\\
\hline
$\sigma_{\chi_{b0}+b\bar{b}}$ & $1.80\times10^{-3}$ & $1.95\times10^{-3}$ & $\left(^{16\%}_{18\%}\right)$\\
\hline
$\sigma_{\chi_{b1}+b\bar{b}}$ & $1.71\times10^{-3}$ & $1.88\times10^{-3}$ & $\left(^{16\%}_{18\%}\right)$\\
\hline
$\sigma_{\chi_{b2}+b\bar{b}}$ & $7.06\times10^{-4}$ & $7.74\times10^{-4}$ & $\left(^{16\%}_{18\%}\right)$\\
\hline
\end{tabular}
\caption{Total cross sections (in pb) for the production channels $e^{+}e^{-} \rightarrow Z^{0}\rightarrow |H_{b\bar{b}}\rangle+X$, with varying values of $E_{cm}$. The ratios $R_{\mp}$ in the last column show how the cross sections are changed with varying values of $E_{cm}$. } \label{Ecm uncertainty2}
\end{table}

To show the sensitivity of the total cross sections to the collision energy around the $Z^0$ peak, we calculate the total cross sections by taking $E_{cm}=(1 \pm 3\%) m_Z$. Our results are presented in Tables \ref{Ecm uncertainty1} and \ref{Ecm uncertainty2}, where we define two ratios $$R_-=\frac{\sigma(E_{cm}=97\%m_Z)}{\sigma(E_{cm}=m_Z)}$$ and $$R_+=\frac{\sigma(E_{cm}=103\%m_Z)}{\sigma(E_{cm}=m_Z)}$$ to show how the cross sections are changed with varying values of $E_{cm}$. For example, by varying $E_{cm}$ within this range, the total cross sections for the production of $J/\psi$ and $\Upsilon$ drop to $16-18\%$ of their peak values.

\begin{table}[tb]
\begin{tabular}{|c ||c |c ||c|}
\hline
~$m_c$ GeV~ & ~$1.35$ GeV~ & ~$1.65$ GeV~ & ~uncertainty~ \\
\hline\hline
$\sigma_{\eta_c+c\bar{c}}$~&~$2.44$~&~$1.31$~&~${}^{+0.680}_{-0.450}$ \\
\hline
$\sigma_{\eta_c'+c\bar{c}}$~&~$1.54$~&~$8.29\times10^{-1}$~ &~${}^{+0.430}_{-0.281}$ \\
\hline
$\sigma_{J/\psi+c\bar{c}}$~&~$2.53$~&~$1.36$~ &~${}^{+0.700}_{-0.470}$ \\
\hline
$\sigma_{\psi'+c\bar{c}}$~&~$1.60$~&~$8.61\times10^{-1}$~&~${}^{+0.440}_{-0.299}$ \\
\hline
$\sigma_{J/\psi+gg}$~&~$4.55\times10^{-2}$~&~$3.29\times10^{-2}$~&~${}^{+0.007}_{-0.006}$ \\
\hline
$\sigma_{\psi'+gg}$~&~$2.88\times10^{-2}$~&~$2.08\times10^{-2}$~&~${}^{+0.005}_{-0.004}$ \\
\hline
$\sigma_{\eta_c+gg}$~&~$2.04\times10^{-1}$~&~$1.46\times10^{-1}$~&~${}^{+0.032}_{-0.026}$ \\
\hline
$\sigma_{\eta_c'+gg}$~&~$1.29\times10^{-1}$~&~$9.24\times10^{-2}$~&~${}^{+0.020}_{-0.017}$ \\
\hline
$\sigma_{h_c+c\bar{c}}$~&~$3.99\times10^{-1}$~&~ $1.45\times10^{-1}$~&~${}^{+0.165}_{-0.089}$ \\
\hline
$\sigma_{\chi_{c0}+c\bar{c}}$~&~$5.64\times10^{-1}$~&~$2.04\times10^{-1}$~&~${}^{+0.232}_{-0.128}$ \\
\hline
$\sigma_{\chi_{c1}+c\bar{c}}$~&~$6.25\times10^{-1}$~&~$2.26\times10^{-1}$~&~${}^{+0.259}_{-0.140}$ \\
\hline
$\sigma_{\chi_{c2}+c\bar{c}}$~&~$2.44\times10^{-1}$~&~$8.90\times10^{-2}$~&~${}^{+0.100}_{-0.055}$ \\
\hline
\end{tabular}
\caption{Total cross sections (in pb) for the channels $e^{+}e^{-} \rightarrow Z^{0} \rightarrow |H_{c\bar{c}}\rangle +X$ with varying values of $m_c$. The uncertainties in the last column are the deviations from the central values corresponding to $m_c = 1.5$ GeV. }
\label{ccx2}
\end{table}

\begin{table}[tb]
\begin{tabular}{|c ||c |c ||c|}
\hline
$m_b(\textrm{GeV})$~&~$4.75\rm{GeV}$~&~$5.05\rm{GeV}$~&~uncertainty~\\
\hline
\hline
$\sigma_{\eta_b+b\bar{b}}$~&~$2.17\times10^{-1}$~&~ $1.74\times10^{-1} $~&~${}^{+0.023}_{-0.020}$ \\
\hline
$\sigma_{\eta_b'+b\bar{b}}$~&~$1.07\times10^{-1}$~&~$8.61\times10^{-2}$~&~${}^{+0.011}_{-0.010}$ \\
\hline
$\sigma_{\Upsilon+b\bar{b}}$~&~$2.43\times10^{-1}$~&~$1.96\times10^{-1}$~&~${}^{+0.025}_{-0.022}$ \\
\hline
$\sigma_{\Upsilon'+b\bar{b}}$~&~$1.20\times10^{-1}$~&~$9.70\times10^{-2}$~&~${}^{+0.012}_{-0.011}$ \\
\hline
$\sigma_{\Upsilon+gg}$~&~$6.95\times10^{-2}$~&~$6.17\times10^{-2}$~ &~${}^{+0.004}_{-0.003}$ \\
\hline
$\sigma_{\Upsilon'+gg}$~&~$3.44\times10^{-2}$~&~$3.05\times10^{-2}$~ &~${}^{+0.002}_{-0.002}$ \\
\hline
$\sigma_{\eta_b+gg}$~&~$8.72\times10^{-2}$~&~$7.69\times10^{-2}$~ &~${}^{+0.005}_{-0.005}$ \\
\hline
$\sigma_{\eta_b'+gg}$~&~$4.32\times10^{-2}$~&~$3.81\times10^{-2}$~ &~${}^{+0.003}_{-0.002}$ \\
\hline
$\sigma_{h_b+b\bar{b}}$~&~$8.12\times10^{-3}$~&~$5.83\times10^{-3}$~&~${}^{+0.001}_{-0.001}$ \\
\hline
$\sigma_{\chi_{b0}+b\bar{b}}$~&~$1.28\times10^{-2}$~&~$9.31\times10^{-3}$~&~${}^{+0.002}_{-0.002}$ \\
\hline
$\sigma_{\chi_{b1}+b\bar{b}}$~&~$1.23\times10^{-2}$~&~$8.87\times10^{-3}$~&~${}^{+0.002}_{-0.002}$ \\
\hline
$\sigma_{\chi_{b2}+b\bar{b}}$~&~$5.09\times10^{-3}$~&~$3.67\times10^{-3}$~&~${}^{+0.001}_{-0.001}$ \\
\hline
\end{tabular}
\caption{Total cross sections (in pb) for the channels $e^{+}e^{-} \rightarrow Z^{0} \rightarrow |H_{b\bar{b}}\rangle +X$ with varying values of $m_b$ GeV. The uncertainties in the last column are the deviations from the central values corresponding to $m_b = 4.9$ GeV. }
\label{bbx2}
\end{table}

Next, we discuss the uncertainties from the heavy quark masses by varying $m_c= 1.50 \pm 0.15$ GeV and $m_b= 4.9 \pm 0.15$ GeV respectively. Our results for  $e^{+}e^{-} \rightarrow Z^{0} \rightarrow |H_{Q\bar{Q}}\rangle +X$ are presented in Tables \ref{ccx2} and \ref{bbx2}. The tables show that
\begin{itemize}
\item for charmonium production, with $X=c\bar{c}$, the uncertainties
     associated with the variation $m_c=1.50\pm0.15$ GeV are $\sim40\%$ for the $S$-wave case and $70\%$ for the $P$-wave case; for $X = gg$, the uncertainty are $\sim 15\%-19\%$.
\item for bottomonium production, with $X=b\bar{b}$, the uncertainties
     associated with the variation $m_b=4.90\pm0.15$ GeV are $\sim11\%$ for the $S$-wave case and $15\%-18\%$ for the $P$-wave case; for $X = gg$, the uncertainties are $\sim6\%$.
\end{itemize}

\subsection{Properties of the color-octet processes}

\subsubsection{Total cross sections}

\begin{table*}[htb]
\begin{tabular}{|c|| c| c| c| c| c|}
\hline
~~$J/\psi$ or $\Upsilon$ production channels~~ &~~$\sigma_{(a)}$~~ &~~$\sigma_{(b)}$~~ &~~$\sigma_{(c)}$~~&~~$\sigma_{\rm tot}$~~ \\
\hline\hline
$e^+e^- \to Z^0 \to |(c\bar{c})[(1^1S_0^{\bf (8)})]g\rangle + c\bar{c} \to J/\psi+c\bar{c}$ ~&~$1.65\times10^{-3}$~&~$2.65\times10^{-5}$~&~$\sim$~&~$1.69\times10^{-3}$~\\
\hline
$e^+e^- \to Z^0 \to |(c\bar{c})[(1^3S_1^{\bf (8)})]g\rangle + c\bar{c} \to J/\psi+c\bar{c}$ ~&~$1.71 \times10^{-3}$~&~$1.80\times10^{-4}$~&~$8.92\times10^{-1}$~&~ $9.11\times10^{-1}$~ \\
\hline
$e^+e^- \to Z^0 \to |(c\bar{c})[(1^1S_0^{\bf (8)})]g\rangle + g\to J/\psi+g$ ~&~$\sim$~&~$\sim$~&~$\sim$~&~$4.87\times10^{-4}$~ \\
\hline
$e^+e^- \to Z^0 \to |(c\bar{c})[(1^3S_1^{\bf (8)})]g\rangle + g\to J/\psi+g$ ~&~$\sim$~&~$\sim$~&~$\sim$~&~$3.31\times10^{-3}$~ \\
\hline\hline
$e^+e^- \to Z^0 \to |(b\bar{b})[(1^1S_0^{\bf (8)})]g\rangle + b\bar{b}\to \Upsilon+b\bar{b}$ ~&~$1.86\times10^{-5}$~&~$5.22\times10^{-6}$~&~$\sim$~ &~$2.76\times10^{-5}$~ \\
\hline
$e^+e^- \to Z^0 \to |(b\bar{b})[(1^3S_1^{\bf (8)})]g\rangle + b\bar{b}\to \Upsilon+b\bar{b}$ ~&~$2.09 \times10^{-5}$~&~$1.13\times10^{-5}$~&~$4.05\times10^{-3}$~ &~$4.23\times10^{-3}$~ \\
\hline
$e^+e^- \to Z^0 \to |(b\bar{b})[(1^1S_0^{\bf (8)})]g\rangle + g\to \Upsilon+g$ ~&~$\sim$~ &~$\sim$~&~$\sim$~&~$2.81\times10^{-4}$~ \\
\hline
$e^+e^- \to Z^0 \to |(b\bar{b})[(1^3S_1^{\bf (8)})]g\rangle + g\to \Upsilon+g$ ~&~$\sim$~ &~$\sim$~&~$\sim$~&~$5.93\times10^{-4}$~ \\
\hline
\end{tabular}
\caption{Total cross sections (in pb) for the color-octet heavy quarkonium production via the $e^+ e^-$ annihilation at $\sqrt{s}=m_Z$, where the subscripts $(a)$, $(b)$ and $(c)$ refer to the Feynman diagrams shown in the corresponding panels of Figs.\ref{QQQQoctet1} and \ref{QQQQoctet2} respectively. The symbol $\sigma_{\rm tot}$ refers to the sum of $\sigma_{(a)}$, $\sigma_{(b)}$, $\sigma_{(c)}$ and their interference terms for the channel $e^+e^- \to Z^0 \to |(Q\bar{Q})[\left(1^1S_0^{(\bf 8)}\right),\left(1^3S_1^{(\bf 8)}\right)]g\rangle + Q\bar{Q}$, and to the cross section of Fig.\ref{QQQQoctet3} for the channel $e^+e^- \to Z^0 \to |(Q\bar{Q})[\left(1^1S_0^{(\bf 8)}\right),\left(1^3S_1^{(\bf 8)}\right)]g\rangle + g$.} \label{co}
\end{table*}

\begin{table*}[htb]
\begin{tabular}{|c|| c| c| c| c| c|}
\hline
~~$h_{c}(h_b)$ or $\chi_{cJ}(\chi_{bJ})$ production channels~~ &~~$\sigma_{(a)}$~~ &~~$\sigma_{(b)}$~~ &~~$\sigma_{(c)}$~~&~~$\sigma_{\rm tot}$~~ \\
\hline\hline
$e^+e^- \to Z^0 \to |(c\bar{c})[(1^1S_0^{\bf (8)})]g\rangle + c\bar{c}\to h_c+c\bar{c}$ ~&~$6.59\times10^{-3}$~&~$1.06\times10^{-4}$~&~$\sim$~&~$6.78\times10^{-3}$~\\
\hline
$e^+e^- \to Z^0 \to |(c\bar{c})[(1^3S_1^{\bf (8)})]g\rangle + c\bar{c}\to \chi_{cJ}+c\bar{c}$ ~&~$6.85 \times10^{-3}$~&~$7.21\times10^{-4}$~&~$3.56$~&~ $3.65$~ \\
\hline
$e^+e^- \to Z^0 \to |(c\bar{c})[(1^1S_0^{\bf (8)})]g\rangle + g \to h_c+g$ ~&~$\sim$~&~$\sim$~&~$\sim$~&~$1.95\times10^{-3}$~ \\
\hline
$e^+e^- \to Z^0 \to |(c\bar{c})[(1^3S_1^{\bf (8)})]g\rangle + g \to \chi_{cJ}+g$ ~&~$\sim$~&~$\sim$~&~$\sim$~&~$1.32\times10^{-2}$~ \\
\hline\hline
$e^+e^- \to Z^0 \to |(b\bar{b})[(1^1S_0^{\bf (8)})]g\rangle + b\bar{b} \to h_b+b\bar{b}$ ~&~$2.53\times10^{-4}$~&~$7.09\times10^{-5}$~&~$\sim$~ &~$3.75\times10^{-4}$~ \\
\hline
$e^+e^- \to Z^0 \to |(b\bar{b})[(1^3S_1^{\bf (8)})]g\rangle + b\bar{b} \to \chi_{bJ}+b\bar{b}$ ~&~$2.09 \times10^{-4}$~&~$1.53\times10^{-4}$~&~$4.26\times10^{-2}$~ &~$4.44\times10^{-2}$~ \\
\hline
$e^+e^- \to Z^0 \to |(b\bar{b})[(1^1S_0^{\bf (8)})]g\rangle + g \to h_b+g$ ~&~$\sim$~ &~$\sim$~&~$\sim$~&~$3.81\times10^{-3}$~ \\
\hline
$e^+e^- \to Z^0 \to |(b\bar{b})[(1^3S_1^{\bf (8)})]g\rangle + g \to \chi_{bJ}+g$ ~&~$\sim$~ &~$\sim$~&~$\sim$~&~$8.05\times10^{-3}$~ \\
\hline
\end{tabular}
\caption{Total cross sections (in pb) for the color-octet heavy quarkonium production via the $e^+ e^-$ annihilation at $\sqrt{s}=m_Z$, where the subscripts $(a)$, $(b)$ and $(c)$ refer to the Feynman figures shown in corresponding panel of Figs.\ref{QQQQoctet1} and \ref{QQQQoctet2} respectively. The symbol $\sigma_{\rm tot}$ refers to the sum of $\sigma_{(a)}$, $\sigma_{(b)}$, $\sigma_{(c)}$ and their interference terms for the channel $e^+e^- \to Z^0 \to |(Q\bar{Q})[\left(1^1S_0^{(\bf 8)}\right),\left(1^3S_1^{(\bf 8)}\right)]g\rangle + Q\bar{Q}$, and to the cross section of Fig.\ref{QQQQoctet3} for the channel $e^+e^- \to Z^0 \to |(Q\bar{Q})[\left(1^1S_0^{(\bf 8)}\right),\left(1^3S_1^{(\bf 8)}\right)]g\rangle + g$. Here the contributions to $^3P_0, ^3P_1, ^3P_2$ from the color-octet component $(Q\bar{Q})[(1^3S_1^{\bf (8)})]g\rangle$ have been summed up.} \label{co1}
\end{table*}

As discussed in the introduction, we will consider the following sizeable color-octet processes
\begin{eqnarray}
e^+e^- \to Z^0 &\to& |(Q\bar{Q})[\left(1^1S_0^{\bf (8)}\right), \left(1^3S_1^{\bf (8)}\right)]g\rangle + Q\bar{Q} \nonumber \\
&\to& \psi_Q+Q\bar{Q} \ , \nonumber \\
e^+e^- \to Z^0 &\to& |(Q\bar{Q})[\left(1^1S_0^{\bf (8)}\right)]g\rangle + Q\bar{Q} \to h_Q+Q\bar{Q}  \ , \nonumber \\
e^+e^- \to Z^0 &\to& |(Q\bar{Q})[\left(1^3S_1^{\bf (8)}\right)]g\rangle + Q\bar{Q} \to \chi_{QJ}+Q\bar{Q}\nonumber
\end{eqnarray}
and
\begin{eqnarray}
e^+e^- \to Z^0 &\to& |(Q\bar{Q})[\left(1^1S_0^{\bf (8)}\right), \left(1^3S_1^{\bf (8)}\right)]g\rangle + g \ , \nonumber \\
&\to& \psi_Q+g\nonumber \\
e^+e^- \to Z^0 &\to& |(Q\bar{Q})[\left(1^1S_0^{\bf (8)}\right)]g\rangle + g \to h_Q+g \ , \nonumber \\
e^+e^- \to Z^0 &\to& |(Q\bar{Q})[\left(1^3S_1^{\bf (8)}\right)]g\rangle + g \to \chi_{QJ}+g . \nonumber
\end{eqnarray}
It is noted that the total cross sections for the color-octet channels versus the collision energy have similar shapes to those in Figs. \ref{ccY}, \ref{ccZ}, \ref{bbY} and \ref{bbZ}. We present their total cross sections at the $e^+ e^-$ collision energy $E_{cm}=\sqrt{s}=m_Z$ in Tables \ref{co} and \ref{co1}. For the channels with $X=c\bar{c}$ or $b\bar{b}$, the results in these tables, compared to those in Tables \ref{ccx1} and \ref{bbx1}, imply
\begin{widetext}
\begin{eqnarray}
\frac{\sigma(e^+e^- \to Z^0 \to |(c\bar{c})[(1^1S_0^{\bf (8)})]g\rangle + c\bar{c} \to J/\psi+c\bar{c})} {\sigma(e^+e^- \to Z^0 \to |(c\bar{c})[(1^3S_1^{\bf (1)})] \rangle + c\bar{c} \to J/\psi+c\bar{c})} = 0.1\% , \nonumber \\
\frac{\sigma(e^+e^- \to Z^0 \to |(c\bar{c})[(1^3S_1^{\bf (8)})]g\rangle + c\bar{c} \to J/\psi+c\bar{c})} {\sigma(e^+e^- \to Z^0 \to |(c\bar{c})[(1^3S_1^{\bf (1)})] \rangle + c\bar{c} \to J/\psi+c\bar{c})} = 50\% , \nonumber \\
\frac{\sigma(e^+e^- \to Z^0 \to |(b\bar{b})[(1^1S_0^{\bf (8)})]g\rangle + b\bar{b} \to \Upsilon+b\bar{b})} {\sigma(e^+e^- \to Z^0 \to |(b\bar{b})[(1^3S_1^{\bf (1)})] \rangle + b\bar{b} \to \Upsilon+ b\bar{b})}= 0.01\% , \nonumber \\
\frac{\sigma(e^+e^- \to Z^0 \to |(b\bar{b})[(1^3S_1^{\bf (8)})]g\rangle + b\bar{b} \to \Upsilon+ b\bar{b})} {\sigma(e^+e^- \to Z^0 \to |(b\bar{b})[(1^3S_1^{\bf (1)})] \rangle + b\bar{b} \to \Upsilon +b\bar{b})} = 1.9\% \nonumber \\
\frac{\sigma(e^+e^- \to Z^0 \to |(c\bar{c})[(1^1S_0^{\bf (8)})]g\rangle + c\bar{c}\to h_c +c\bar{c})} {\sigma(e^+e^- \to Z^0 \to |(c\bar{c})[(1^1P_1^{\bf (1)})] \rangle + c\bar{c}\to h_c +c\bar{c})} = 2.9\% , \nonumber \\
\frac{\sigma(e^+e^- \to Z^0 \to |(c\bar{c})[(1^3S_1^{\bf (8)})]g\rangle +c\bar{c} \to \chi_{cJ} +c\bar{c})} {\sigma(e^+e^- \to Z^0 \to |(c\bar{c})[(1^3P_J^{\bf (1)})] \rangle + c\bar{c} \to \chi_{cJ}+ c\bar{c})}= 433\% , \nonumber \\
\frac{\sigma(e^+e^- \to Z^0 \to |(b\bar{b})[(1^1S_0^{\bf (8)})]g\rangle + b\bar{b} \to h_b +b\bar{b})} {\sigma(e^+e^- \to Z^0 \to |(b\bar{b})[(1^1P_1^{\bf (1)})] \rangle + b\bar{b} \to h_b + b\bar{b})} = 5.5\% , \nonumber \\
\frac{\sigma(e^+e^- \to Z^0 \to |(b\bar{b})[(1^3S_1^{\bf (8)})]g\rangle + b\bar{b} \to \chi_{bJ} +b\bar{b})} {\sigma(e^+e^- \to Z^0 \to |(b\bar{b})[(1^3P_J^{\bf (1)})] \rangle + b\bar{b} \to \chi_{bJ} + b\bar{b})} = 173\% . \nonumber
\end{eqnarray}
\end{widetext}
The analogous ratios for the cases $X = g$ have similar values.

We now discuss the relative importance of different channels for quarkonium production at the super $Z$ factory with $E_{cm} = m_Z$ and at a $B$ factory with $E_{cm} = 10.6$ GeV.
\begin{itemize}
\item At the super $Z$ factory, the channel $e^+e^- \to Z^0 \to |(Q\bar{Q})[\left(1^3S_1^{\bf (8)}\right)]g\rangle + Q\bar{Q}$, with the topology shown in Fig. \ref{QQQQoctet2}c, provides the dominant contribution with respect to other octet channels. A similar enhancement is not expected at the $B$ factory, where quarkonium production is dominated by the $\gamma*$ propagator and, for example,
    \begin{widetext}
    \begin{eqnarray}
    \frac{\sigma(e^+e^- \to \gamma^* \to |(c\bar{c})[(1^3S_1^{\bf (8)})]g\rangle + c\bar{c}\to J/\psi+c\bar{c})} {\sigma(e^+e^- \to \gamma^* \to |(c\bar{c})[(1^3S_1^{\bf (1)})] \rangle + c\bar{c}\to J/\psi+c\bar{c})} \sim 3.1\% , \nonumber \\
    \frac{\sigma(e^+e^- \to \gamma^* \to |(c\bar{c})[(1^3S_1^{\bf (8)})]g\rangle + c\bar{c}\to \chi_{cJ}+c\bar{c})} {\sigma(e^+e^- \to \gamma^* \to |(c\bar{c})[(1^3P_J^{\bf (1)})] \rangle + c\bar{c}\to \chi_{cJ}+c\bar{c})} \sim 19\% . \nonumber
    \end{eqnarray}
    \end{widetext}

\item At the super Z factory, the channels $e^+e^- \to Z^0 \to |(Q\bar{Q})[(1^1S_0^{\bf (8)}), (1^3S_1^{\bf (8)})]g\rangle + g$ are less important than $e^+e^- \to Z^0 \to |(Q\bar{Q})[(1^1S_0^{\bf (8)}), (1^3S_1^{\bf (8)})]g\rangle + Q\bar{Q}$ by at least an order of magnitude for the production of $\psi_Q$, $h_Q$ and $\chi_{QJ}$, while at the B factory $e^+e^- \to \gamma^* \to |(Q\bar{Q})[1^1S_0^{\bf (8)}]g\rangle + g$ gives a significant contribution:

    \begin{widetext}
    \begin{eqnarray}
    \frac{\sigma(e^+e^- \to \gamma^* \to |(c\bar{c})[(1^1S_0^{\bf (8)})]g\rangle + g \to J/\psi+g)} {\sigma(e^+e^- \to \gamma^* \to |(c\bar{c})[(1^3S_1^{\bf (1)})] \rangle + c\bar{c}\to J/\psi+c\bar{c})}\sim 30\% . \nonumber
    \end{eqnarray}
    \end{widetext}
\end{itemize}

\subsubsection{Differential cross sections for the octet channels}

\begin{figure*}[htb]
\includegraphics[width=0.45\textwidth]{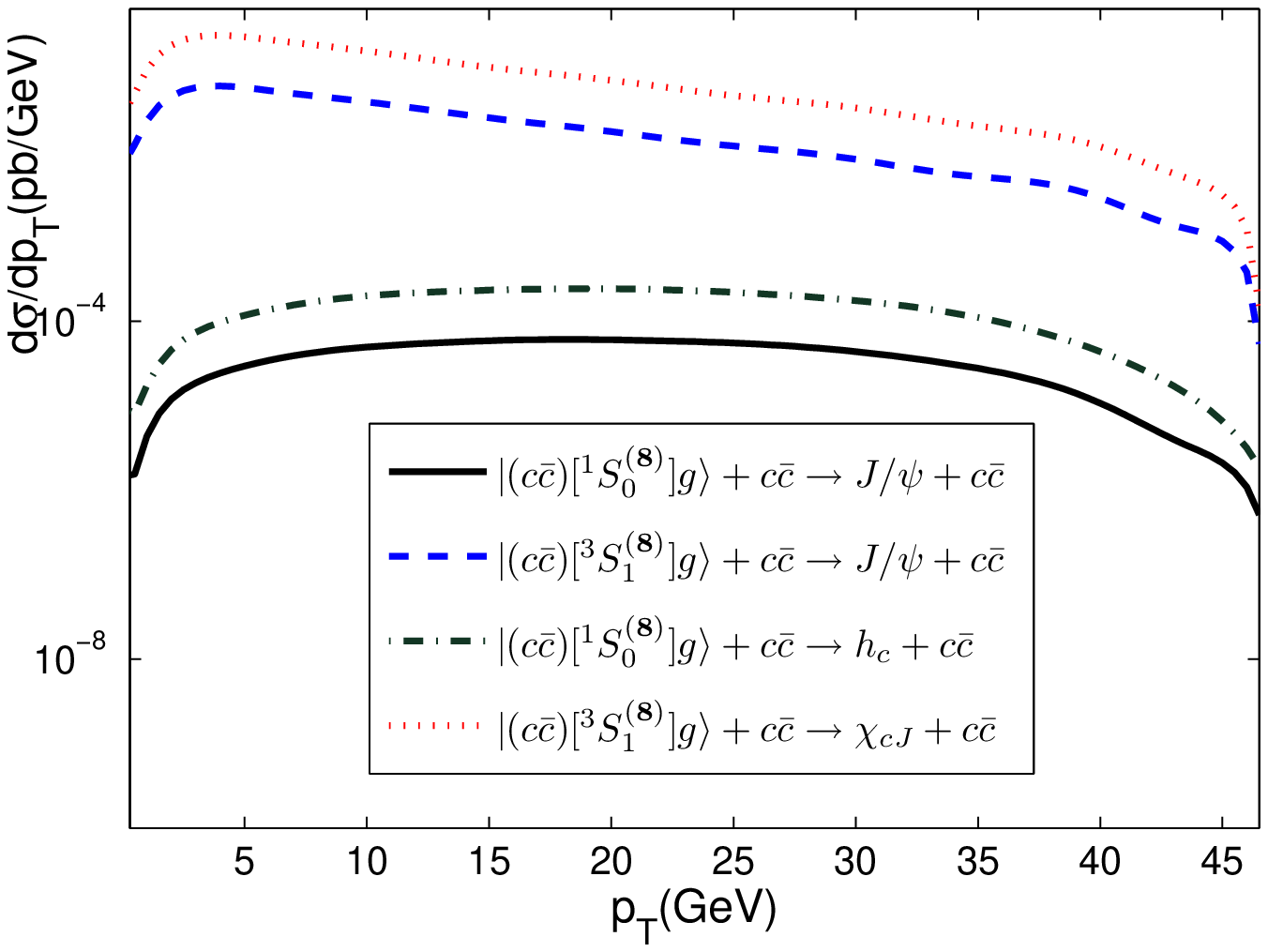}
\includegraphics[width=0.45\textwidth]{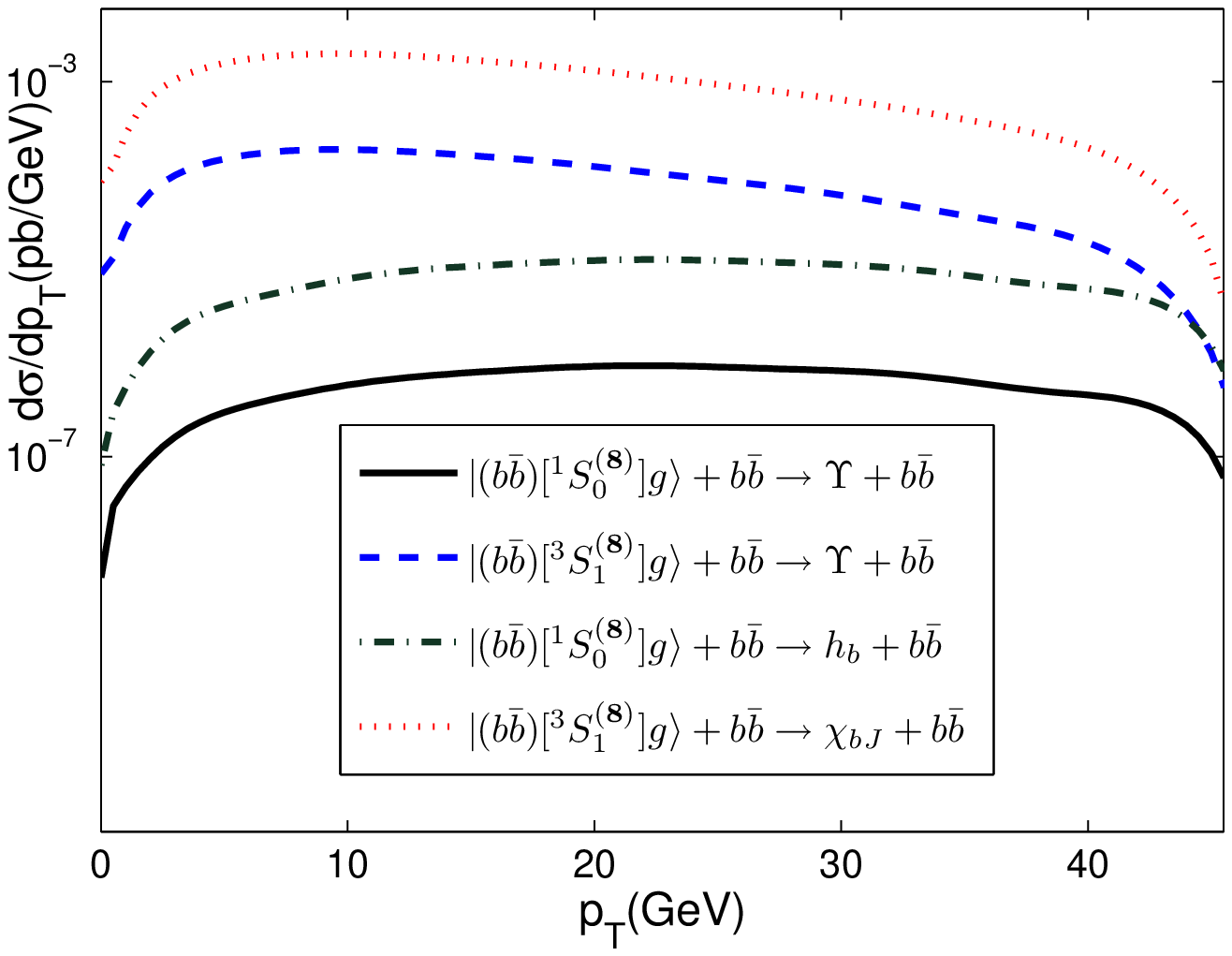}
\caption{The charmonium (left) and bottomonium (right) $p_T$ distributions for the production processes $e^+e^- \to Z^0 \to |(Q\bar{Q})[\left(1^1S_0^{\bf (8)}\right), \left(1^3S_1^{\bf (8)}\right)]g\rangle + Q\bar{Q} \to (\psi_Q, h_Q, \chi_{QJ})+Q\bar{Q}$ at the collision energy $E_{cm} = m_Z$. } \label{ptco}
\end{figure*}

\begin{figure*}[htb]
\includegraphics[width=0.45\textwidth]{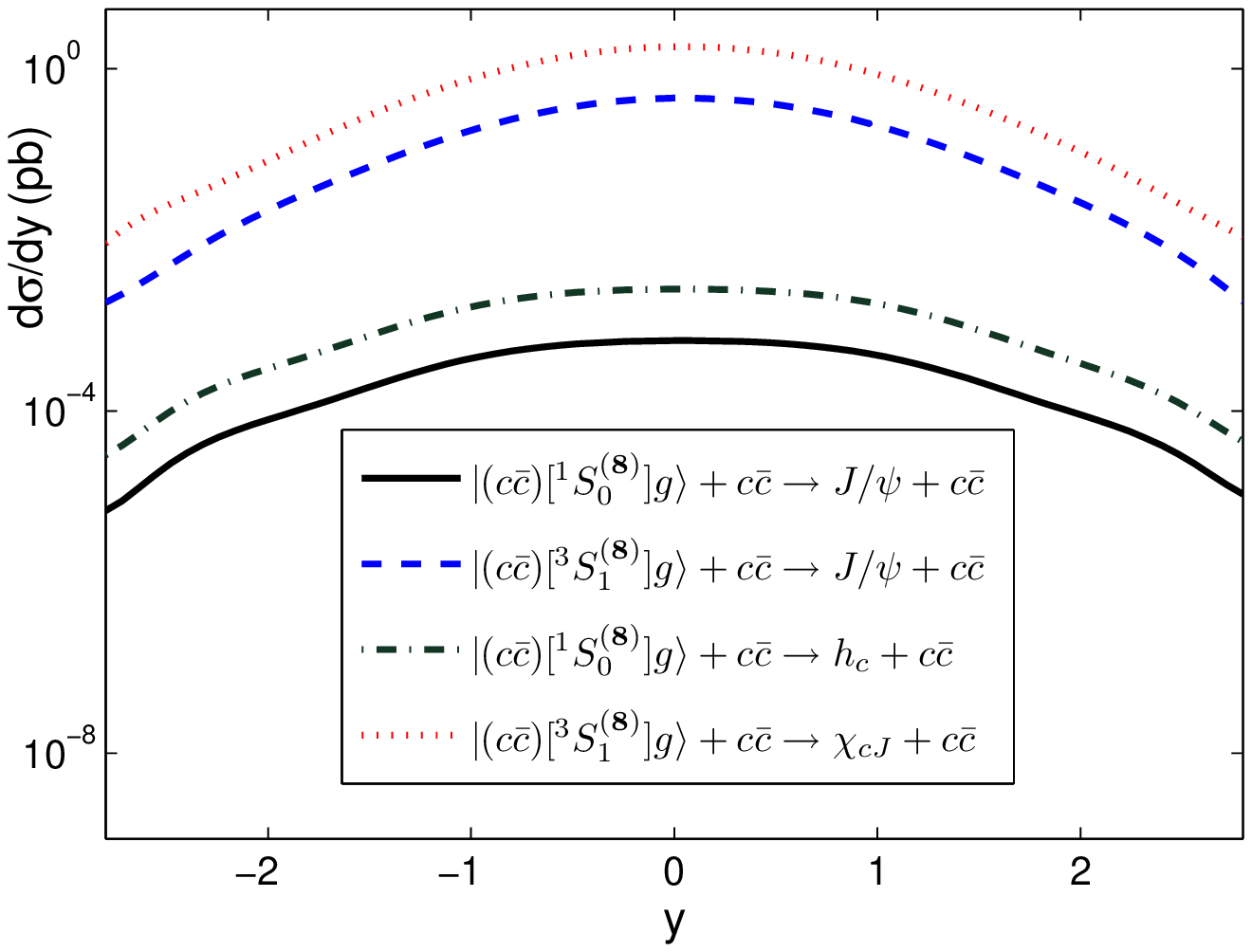}
\includegraphics[width=0.45\textwidth]{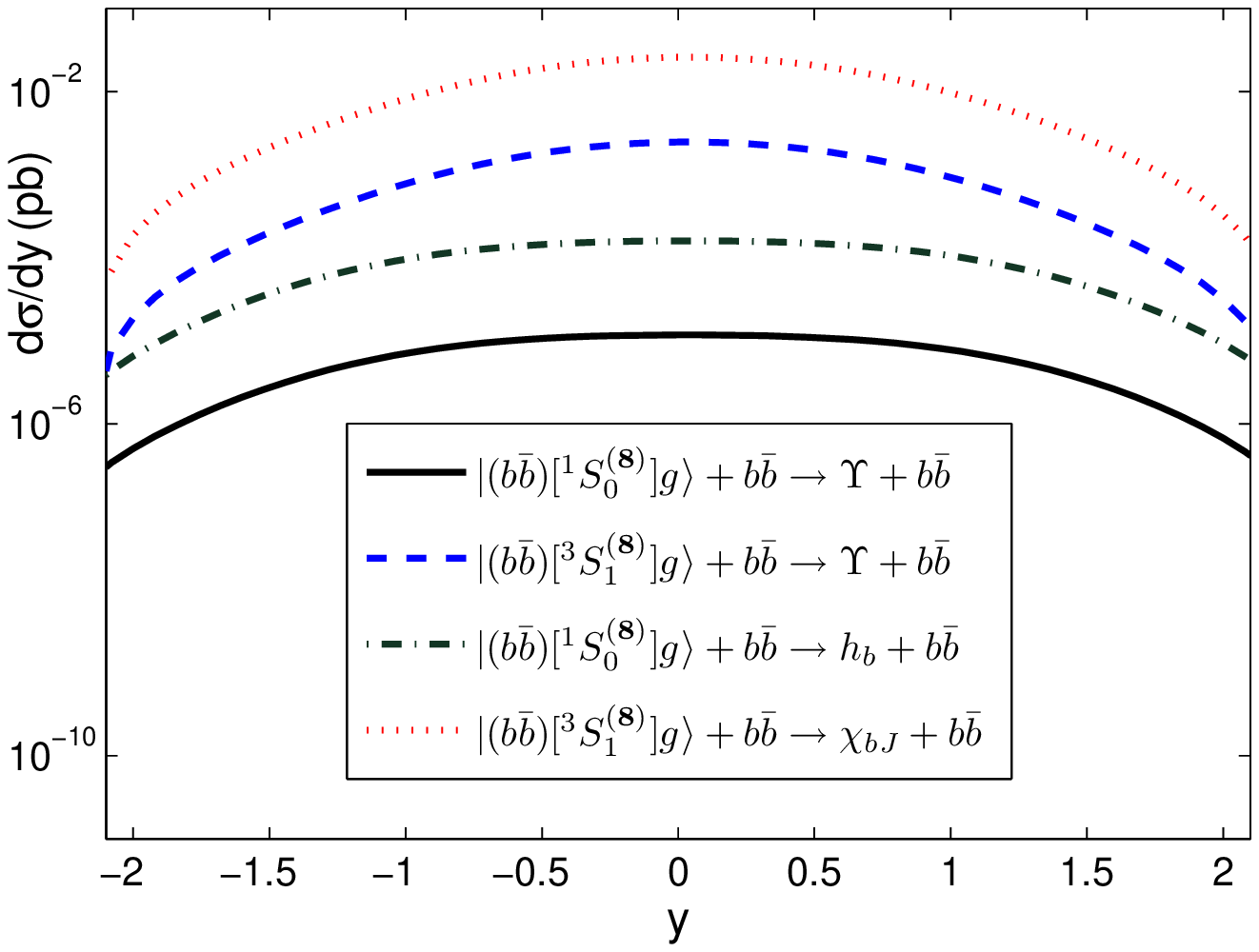}
\caption{The charmonium (left) and bottomonium (right) $y$ distributions for the production processes $e^+e^- \to Z^0 \to |(Q\bar{Q})[\left(1^1S_0^{\bf (8)}\right), \left(1^3S_1^{\bf (8)}\right)]g\rangle + Q\bar{Q} \to (\psi_Q, h_Q, \chi_{QJ})+Q\bar{Q}$ at the collision energy $E_{cm} = m_Z$. } \label{dyco}
\end{figure*}

\begin{table*}[htb]
\begin{tabular}{|c|| c| c| c|}
\hline
~Channel~&~$p_T>5$ GeV~&~$p_T>10$ GeV~&~$p_T>15$ GeV~\\
\hline\hline
$\sigma(e^+e^- \to Z^0 \to |(c\bar{c})[(1^1S_0^{\bf (8)})]g\rangle + c\bar{c}\to J/\psi+c\bar{c})$~&~$1.62\times 10^{-3}(96\%)$~&~$1.41\times 10^{-3}(83\%)$~&~$1.13\times 10^{-3}(67\%)$~\\
\hline
$\sigma(e^+e^- \to Z^0 \to |(c\bar{c})[(1^3S_1^{\bf (8)})]g\rangle + c\bar{c}\to J/\psi+c\bar{c})$~&~$6.80\times 10^{-1}(75\%)$~&~$4.41 \times 10^{-1}(48\%)$~&~$2.85 \times 10^{-1}(31\%)$~\\
\hline
$\sigma(e^+e^- \to Z^0 \to |(c\bar{c})[(1^1S_0^{\bf (8)})]g\rangle + c\bar{c}\to h_c+c\bar{c})$~&~$6.50\times 10^{-3}(96\%)$~&~$5.65 \times 10^{-3}(83\%)$~&~$4.54 \times 10^{-3}(67\%)$~\\
\hline
$\sigma(e^+e^- \to Z^0 \to |(c\bar{c})[(1^3S_1^{\bf (8)})]g\rangle + c\bar{c}\to \chi_{cJ}+c\bar{c})$~&~$2.72(75\%)$~&~$1.76(48\%)$~&~$1.14(31\%)$~\\
\hline\hline
$\sigma(e^+e^- \to Z^0 \to |(b\bar{b})[(1^1S_0^{\bf (8)})]g\rangle + b\bar{b}\to \Upsilon+b\bar{b})$~&~$2.69\times 10^{-5}(97\%)$~&~$2.46\times 10^{-5}(89\%)$~&~$2.11\times 10^{-5}(76\%)$~\\
\hline
$\sigma(e^+e^- \to Z^0 \to |(b\bar{b})[(1^3S_1^{\bf (8)})]g\rangle + b\bar{b}\to \Upsilon+b\bar{b})$~&~$3.80\times 10^{-3}(90\%)$~&~$2.90\times 10^{-3}
(69\%)$~&~$2.02\times 10^{-3}(48\%)$~\\
\hline
$\sigma(e^+e^- \to Z^0 \to |(b\bar{b})[(1^1S_0^{\bf (8)})]g\rangle + b\bar{b}\to h_b+b\bar{b})$~&~$3.66\times 10^{-4}(97\%)$~&~$3.34\times 10^{-4}(89\%)$~&~$2.87\times 10^{-4}(76\%)$~\\
\hline
$\sigma(e^+e^- \to Z^0 \to |(b\bar{b})[(1^3S_1^{\bf (8)})]g\rangle + b\bar{b}\to \chi_{bJ}+b\bar{b})$~&~$3.99\times 10^{-2}(90\%)$~&~$3.05\times 10^{-2}
(69\%)$~&~$2.12\times 10^{-2}(48\%)$~\\
\hline
\end{tabular}
\caption{Total cross sections (in pb) for the octet channels of charmonium and bottomonium production via $Z^0$ propagator at the super $Z$ factory with different $p_{T}$ cuts, with $m_c$=1.5 GeV and $m_b$=4.9 GeV. The percentages in the parentheses represent the ratios between the cross sections with and without $p_T$ cut.} \label{tabptco}
\end{table*}

\begin{table*}[htb]
\begin{tabular}{|c|| c| c| c|}
\hline
~Channel~&~$|y|<0.5$~&~$|y|<1.0$~&~$|y|<1.5$~\\
\hline\hline
$\sigma(e^+e^- \to Z^0 \to |(c\bar{c})[(1^1S_0^{\bf (8)})]g\rangle + c\bar{c}\to J/\psi+c\bar{c})$~&~$6.45\times 10^{-4}(38\%)$~&~$1.18\times 10^{-3}(70\%)$~&~$1.49\times 10^{-3}(88\%)$~\\
\hline
$\sigma(e^+e^- \to Z^0 \to |(c\bar{c})[(1^3S_1^{\bf (8)})]g\rangle + c\bar{c}\to J/\psi+c\bar{c})$~&~$4.26\times 10^{-1}(47\%)$~&~$7.14 \times 10^{-1}(78\%)$~&~$8.48 \times 10^{-1}(93\%)$~\\
\hline
$\sigma(e^+e^- \to Z^0 \to |(c\bar{c})[(1^1S_0^{\bf (8)})]g\rangle + c\bar{c}\to h_c+c\bar{c})$~&~$2.59\times 10^{-3}(38\%)$~&~$4.73\times 10^{-3}(70\%)$~&~$5.98\times 10^{-3}(88\%)$~\\
\hline
$\sigma(e^+e^- \to Z^0 \to |(c\bar{c})[(1^3S_1^{\bf (8)})]g\rangle + c\bar{c}\to \chi_{cJ}+c\bar{c})$~&~$1.71(47\%)$~&~$2.86(78\%)$~&~$3.40(93\%)$~\\
\hline\hline
$\sigma(e^+e^- \to Z^0 \to |(b\bar{b})[(1^1S_0^{\bf (8)})]g\rangle + b\bar{b}\to \Upsilon+b\bar{b})$~&~$1.14\times 10^{-5}(41\%)$~&~$2.07\times 10^{-5}(75\%)$~&~$2.58\times 10^{-5}(93\%)$~\\
\hline
$\sigma(e^+e^- \to Z^0 \to |(b\bar{b})[(1^3S_1^{\bf (8)})]g\rangle + b\bar{b}\to \Upsilon+b\bar{b})$~&~$2.27\times 10^{-3}(54\%)$~&~$3.63\times 10^{-3} (86\%)$~&~$4.13\times 10^{-3}(98\%)$~\\
\hline
$\sigma(e^+e^- \to Z^0 \to |(b\bar{b})[(1^1S_0^{\bf (8)})]g\rangle + b\bar{b}\to h_b+b\bar{b})$~&~$1.55\times 10^{-4}(41\%)$~&~$2.81\times 10^{-4}(75\%)$~&~$3.51\times 10^{-4}(93\%)$~\\
\hline
$\sigma(e^+e^- \to Z^0 \to |(b\bar{b})[(1^3S_1^{\bf (8)})]g\rangle + b\bar{b}\to \chi_{bJ}+b\bar{b})$~&~$2.38\times 10^{-2}(54\%)$~&~$3.81\times 10^{-2} (86\%)$~&~$4.33\times 10^{-2}(98\%)$~\\
\hline
\end{tabular}
\caption{Total cross sections (in pb) for the octet channels of charmonium and bottomonium production via $Z^0$ propagator at the super $Z$ factory with different $y$ cuts, with $m_c$=1.5 GeV and $m_b$=4.9 GeV. The percentages in the parentheses represent the ratios between the cross sections with and without rapidity cut.} \label{tabdyco}
\end{table*}

To better illustrate the relative importance of different production channels, we present the $p_T$ and $y$ distributions for the octet production channels of $J/\psi$, $\Upsilon$, $h_c$, $h_b$, $\chi_{cJ}$ and $\chi_{bJ}$ in Figs. \ref{ptco} and \ref{dyco}. Furthermore, the total cross sections with different $p_T$ and $y$ cuts are reported in Tables \ref{tabptco} and \ref{tabdyco}. The results show that the channel $e^+e^- \to Z^0 \to |(Q\bar{Q})[(1^3S_1^{\bf (8)})]g\rangle + Q\bar{Q}$ will lead to a peak in the low $p_{T}$ region at about $5$ GeV for charmonium and $10$ GeV for bottomonium.

\begin{figure*}[htb]
\includegraphics[width=0.45\textwidth]{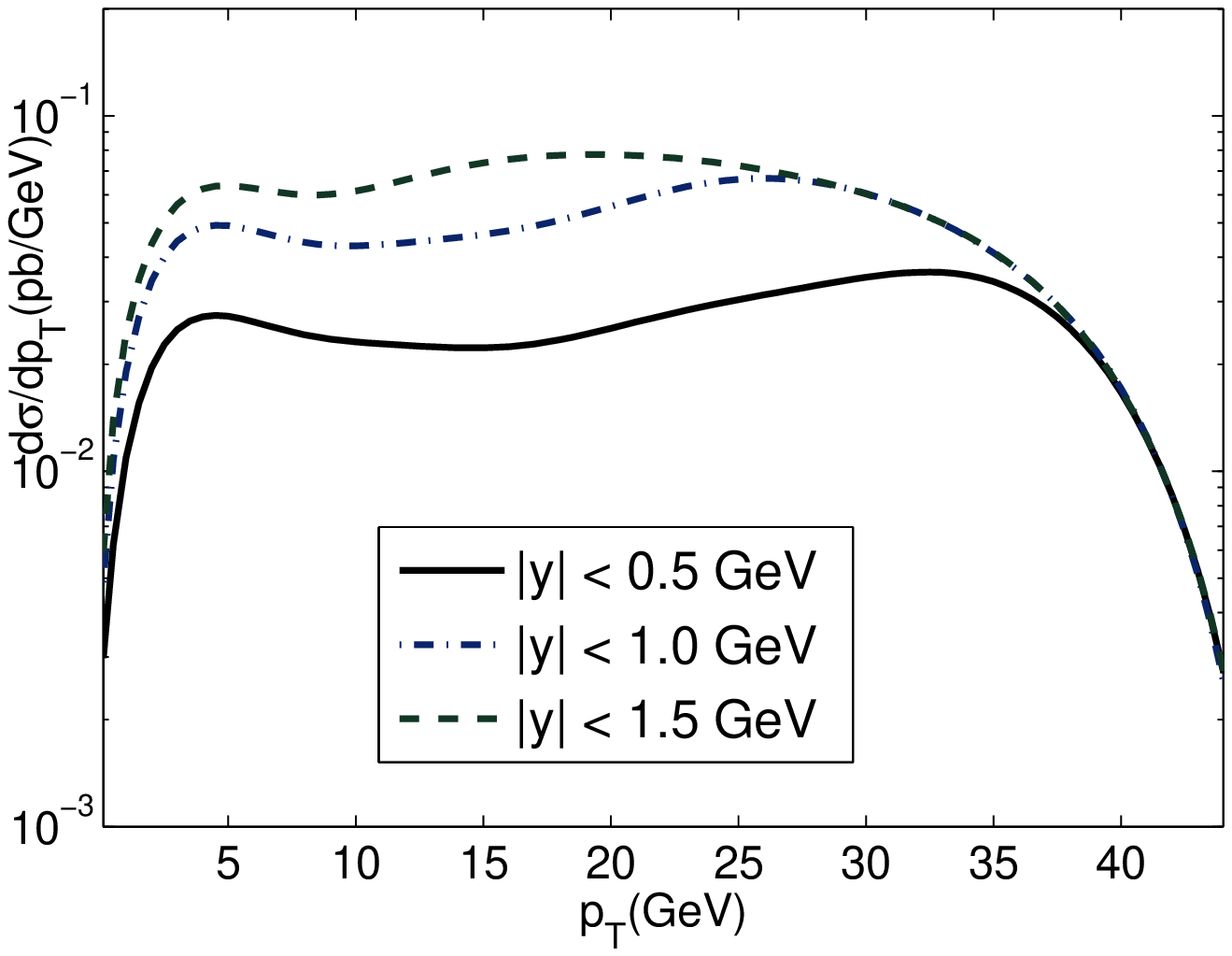}
\includegraphics[width=0.45\textwidth]{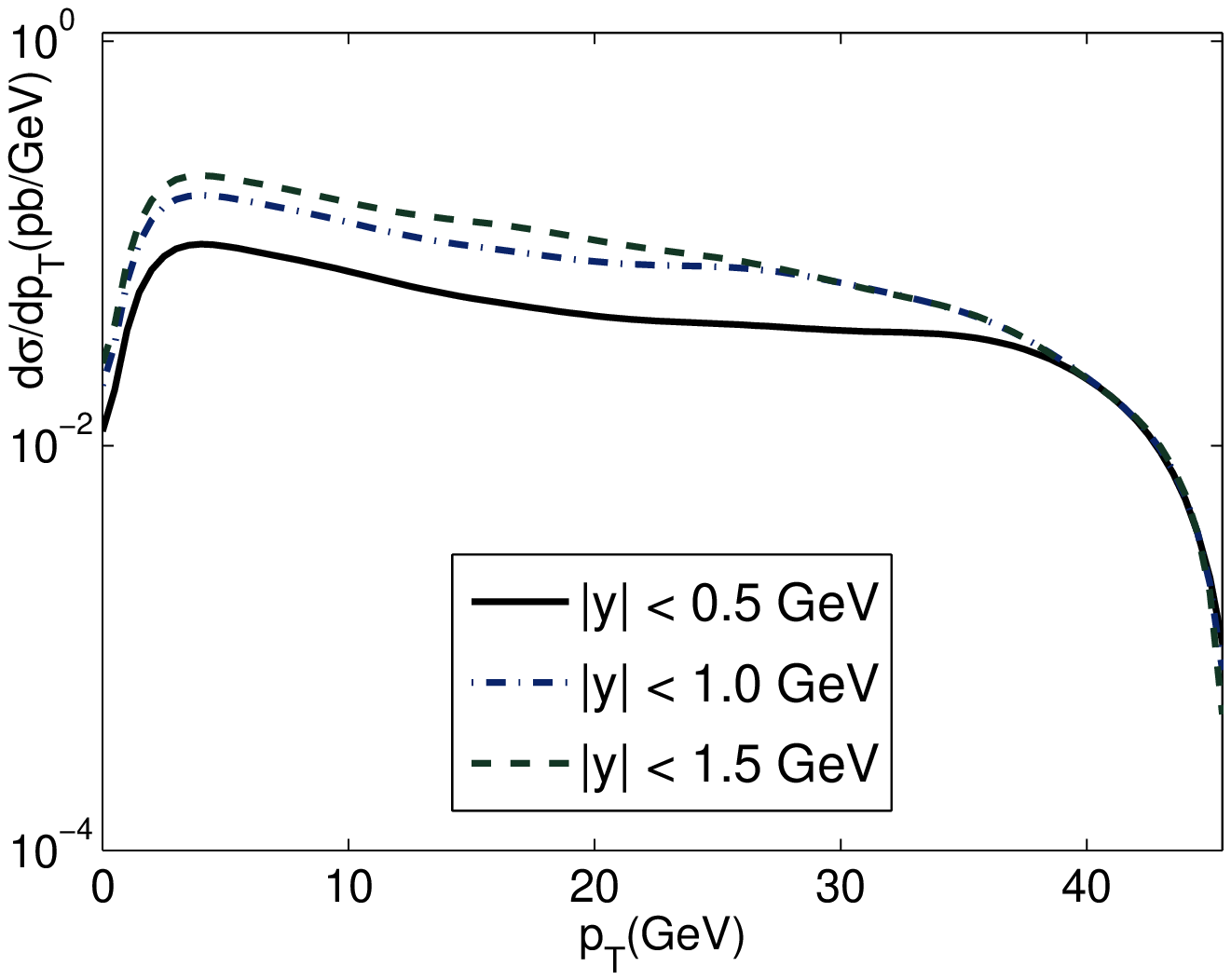}
\caption{The charmonium $p_T$ distributions (left: $J/\psi$, right: $P$-wave charmonium) with various $y$ cuts for $E_{cm}=m_Z$ and $m_c$ = 1.5 GeV. } \label{dyptcutcc1}
\end{figure*}

\begin{figure*}[htb]
\includegraphics[width=0.45\textwidth]{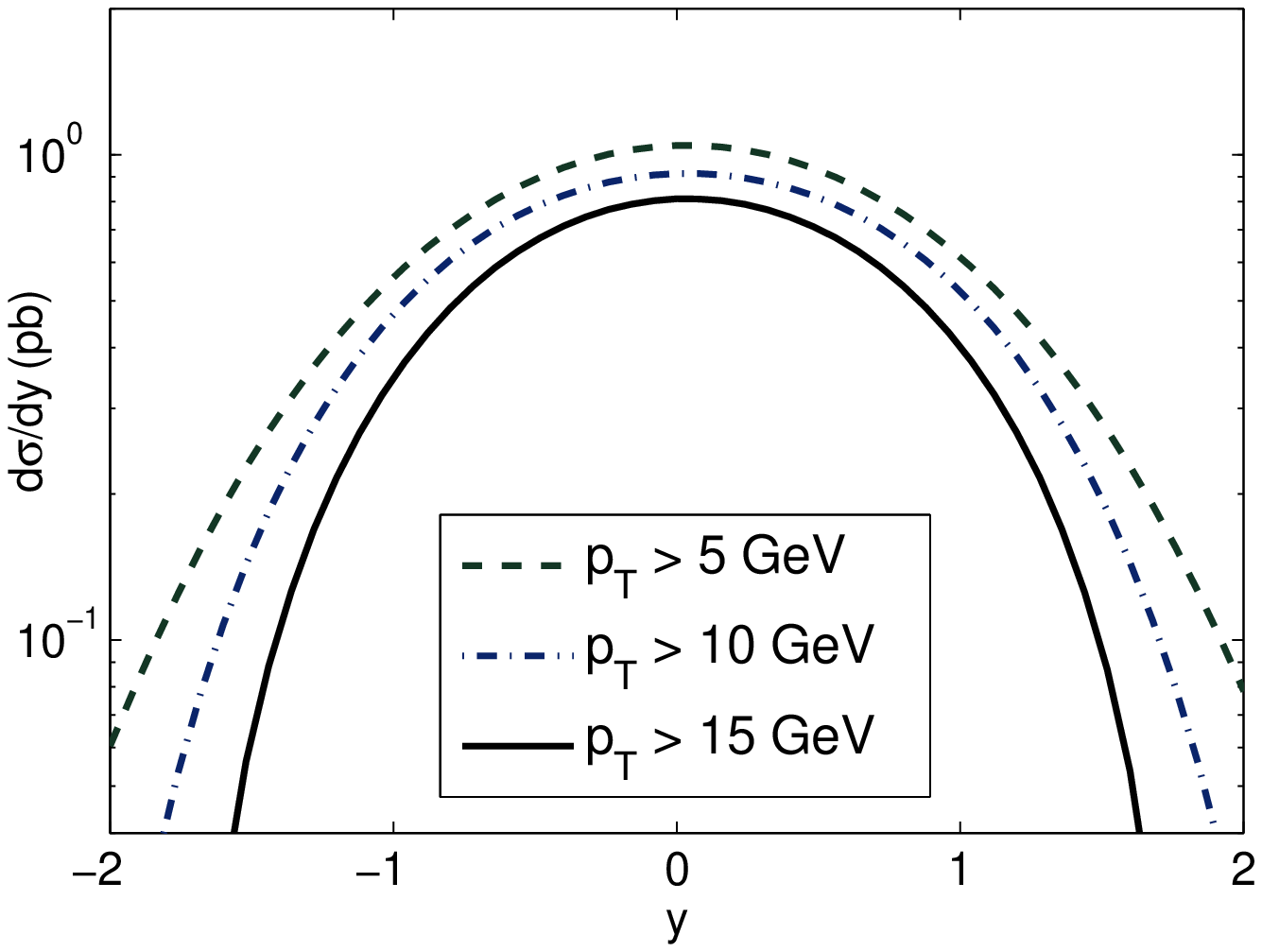}
\includegraphics[width=0.45\textwidth]{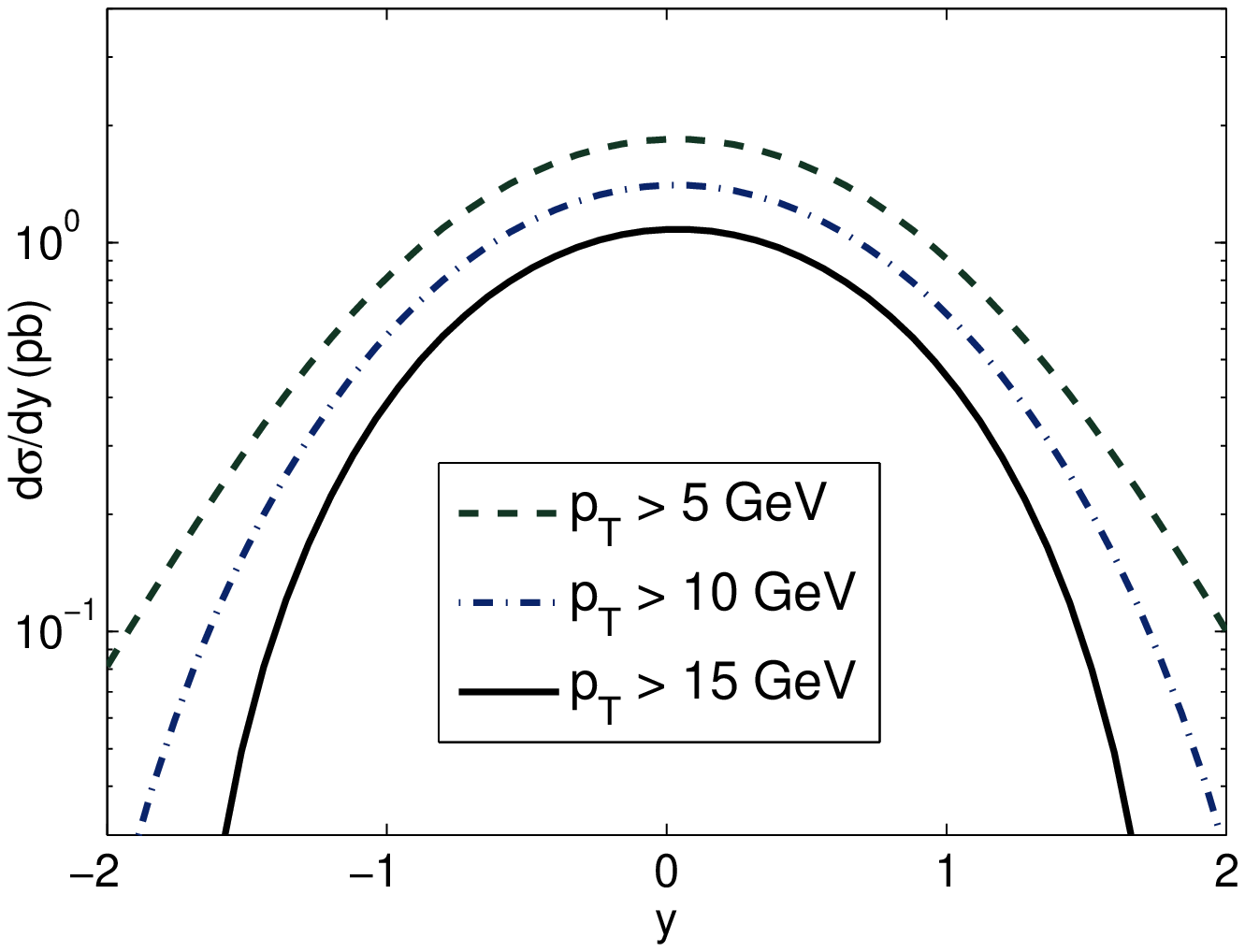}
\caption{The charmonium $y$ distributions (left: $J/\psi$, right: $P$-wave charmonium) with various $p_T$ cuts for $E_{cm}=m_Z$ and $m_c$ = 1.5 GeV. } \label{dyptcutcc2}
\end{figure*}

\begin{figure*}[htb]
\includegraphics[width=0.45\textwidth]{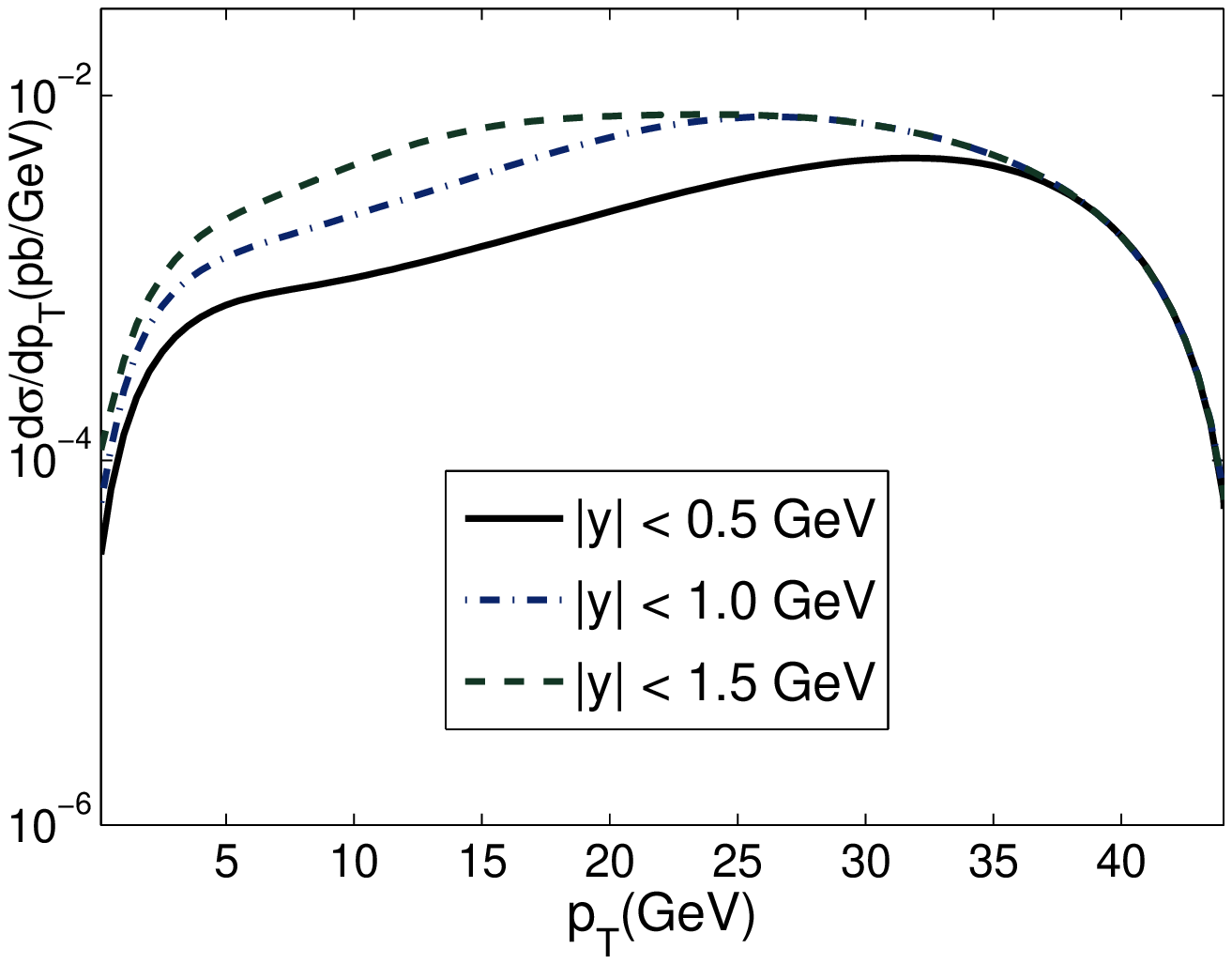}
\includegraphics[width=0.45\textwidth]{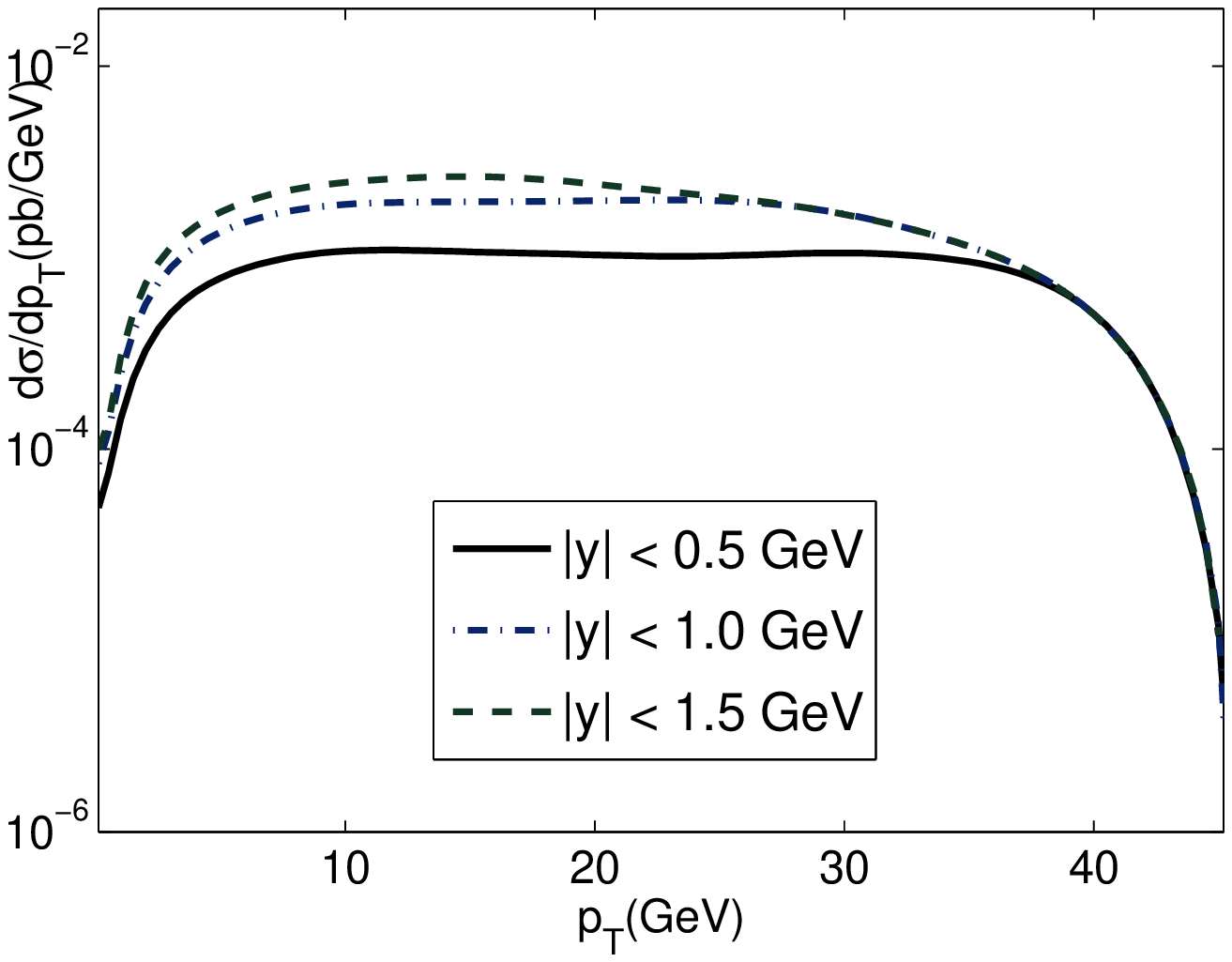}
\caption{The bottomonium $p_T$ distributions (left: $\Upsilon$, right: $P$-wave bottomonium) with various $y$ cuts for $E_{cm}=m_Z$ and $m_b$ = 4.9 GeV. } \label{dyptcutbb1}
\end{figure*}

\begin{figure*}[htb]
\includegraphics[width=0.45\textwidth]{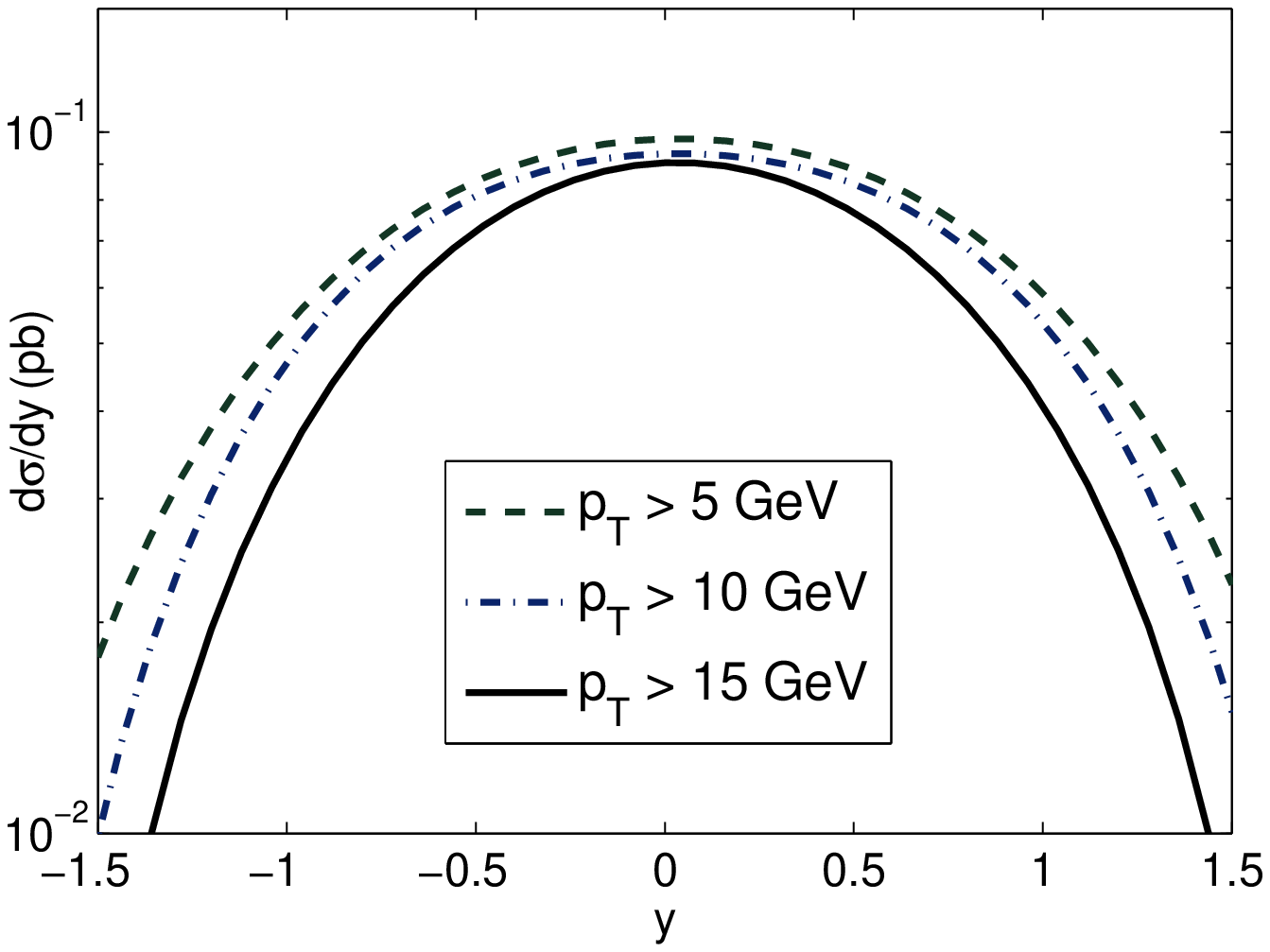}
\includegraphics[width=0.45\textwidth]{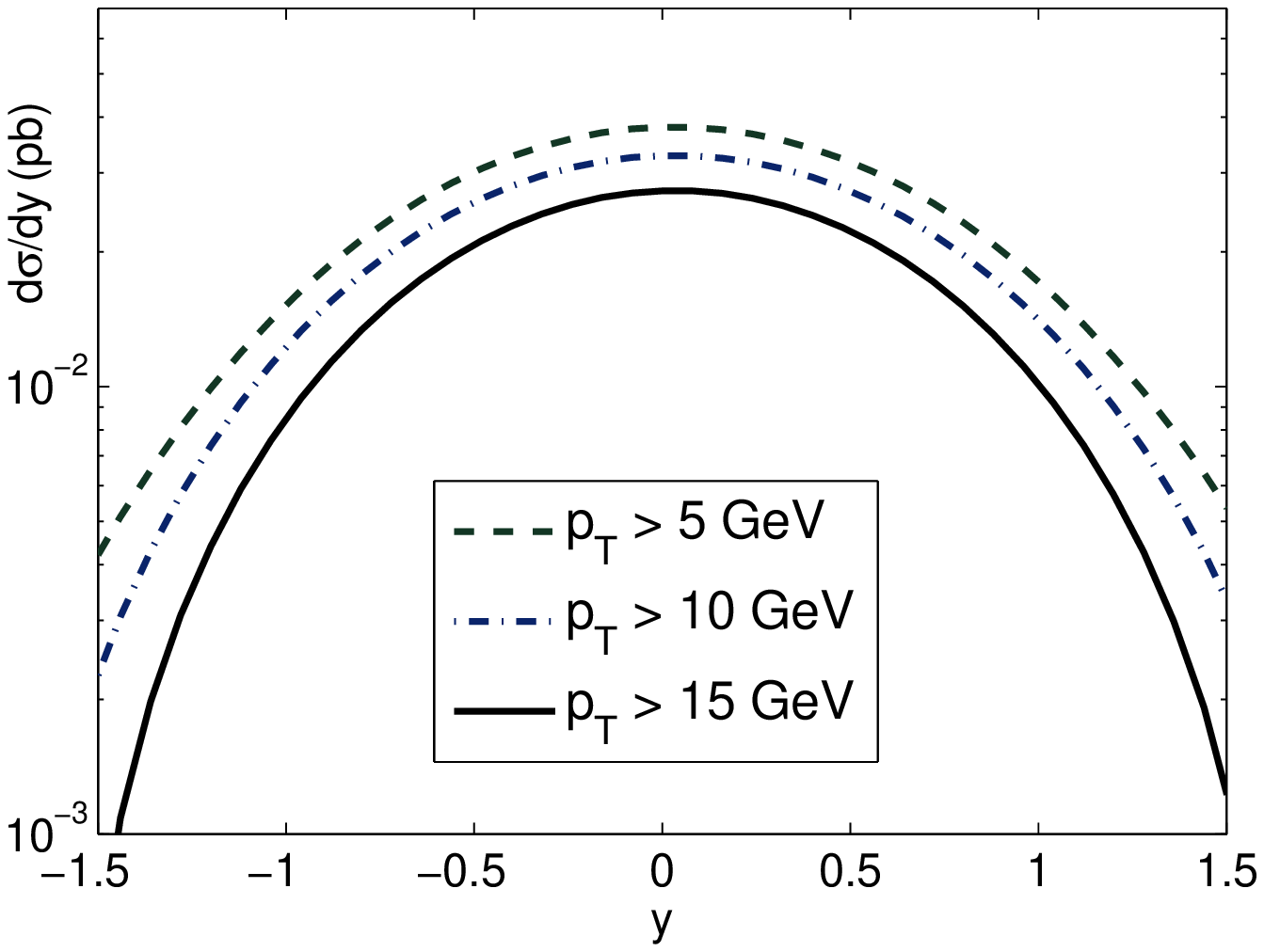}
\caption{The bottomonium $y$ distributions (left: $\Upsilon$, right: $P$-wave bottomonium) with various $p_{t}$ cuts for $E_{cm}=m_Z$ and $m_b$ = 4.9 GeV. } \label{dyptcutbb2}
\end{figure*}

Finally, we present the charmonium and bottomonium $p_T$ distributions with various $y$ cuts and the $y$ distributions with various $p_T$ cuts for the channel $e^+e^- \to Z^0 \to |H_Q\rangle+Q\bar{Q}$ in Figs.\ref{dyptcutcc1}, \ref{dyptcutcc2}, \ref{dyptcutbb1} and \ref{dyptcutbb2}. In these figures, both color-singlet and color-octet contributions are included. Here, for the $\psi_Q$ production through $e^+e^- \to Z^0 \to \psi_Q +Q\bar{Q}$, the components $|(Q\bar{Q})[\left(1^1S_0^{\bf (8)}\right)]g\rangle, |(Q\bar{Q})[\left(1^3S_1^{\bf (8)}\right)]g\rangle$ and $|(Q\bar{Q})[\left(1^3S_1^{\bf (1)}\right)]\rangle$ have been summed up. For the $P$-wave quarkonium production through $e^+e^- \to Z^0 \to (h_Q, \chi_{QJ})+Q\bar{Q}$, the contributions to $h_Q$ and $\chi_{QJ}$ production, including both color-singlet and color-octet components, have been summed up.

\section {Summary}

In this paper, we have studied the charmonium and bottomonium production through the $e^+e^-$ annihilation in the leading $\alpha_s$-order. A detailed discussion on the heavy quarkonium production at the super $Z$ factory via the two types of semi-exclusive channels, $e^{+}e^{-}\rightarrow |H_{Q\bar{Q}}\rangle+X$ with $X=Q\bar{Q}$ or $gg$, has been presented.

As usual, total cross sections are dominated by the color-singlet $1S$-level quarkonium states. However the color-singlet quarkonium states at the $2S$-level and the $1P$-level, and the color-octet quarkonium state $|(Q\bar{Q})[1^3S_1^{({\bf 8})}]g\rangle$ can also provide sizable contributions. Total cross sections for the color-singlet channels versus the collision energy $E_{cm}$ have been presented in Figs. \ref{ccY}, \ref{ccZ}, \ref{bbY} and \ref{bbZ}. The curves for the color-octet $S$-wave states have the same shapes as for the corresponding color-singlet $S$-wave states. In the low collision energy region, the cross sections are dominated by the processes via the $\gamma^*$ propagator (in agreement with observations at the B factories): the main contributions are around $6-20$ GeV for charmonium production and $10-40$ GeV for bottomonium production. Around the collision energy $E_{cm}=m_{Z}$, due to the $Z^0$-boson resonance effect, the cross sections become much larger for the processes via the $Z^0$ propagator. Then, at even higher energies, the total cross section drops down logarithmically for all the processes. By varying $E_{cm}$ within the range of $(1 \pm 3\%) m_Z$, the total cross sections for $J/\psi$ and $\Upsilon$ production drop to about $17\%$ of their peak values.

At the super $Z$ factory, the heavy quarkonium production processes $e^+e^- \to |H_{Q\bar{Q}}\rangle +X$ are dominated by those via the $Z^0$ propagator. Compared to the quarkonium production at the $B$ factories (at an energy of about $10.6$ GeV), much higher cross sections for $e^+e^- \to |H_{Q\bar{Q}}\rangle+X$  are expected at the super $Z$ factory and at the GigaZ program of the ILC.

For the luminosity ${\cal L} = 10^{34} {\rm cm}^{-2} {\rm s}^{-1}$ ($\simeq 10^5 {\rm pb}^{-1}/{\rm year}$ integrated over one year~\cite{hantao}) we expect the following quarkonium yields per year:
\begin{eqnarray*}
N_{J/\psi}  &=& 2.74\times 10^5  \;, N_{\psi'} = 1.16\times 10^5  \;, N_{\eta_c} = 1.76\times 10^5 , \nonumber\\
N_{\eta_c'} &=& 1.11\times 10^5  \;, N_{h_c} = 2.41\times 10^4  \;, N_{\chi_{c0}} = 7.37\times 10^4 , \nonumber\\
N_{\chi_{c1}} &=& 1.55\times 10^5  \;,  N_{\chi_{c2}} = 2.17\times 10^5
\end{eqnarray*}
for the process $e^+e^- \to |H_{c\bar{c}}\rangle+c\bar{c}$ and
\begin{eqnarray*}
N_{J/\psi}  &=& 3.91\times 10^3  \;, N_{\psi'} = 2.47\times 10^3  \;,\; N_{\eta_c} = 1.72\times 10^4 , \nonumber\\
N_{\eta_c'} &=& 1.09\times 10^4
\end{eqnarray*}
for the process $e^+e^- \to |H_{c\bar{c}}\rangle+gg$. For the bottomonium production we expect
\begin{eqnarray*}
N_{\Upsilon}  &=& 2.22\times 10^4  \;, N_{\Upsilon'} = 1.08\times 10^4  \;, N_{\eta_b} = 1.94\times 10^4 , \nonumber\\
N_{\eta_b'} &=& 9.60\times 10^3  \;, N_{h_b} = 7.22\times 10^2  \;, N_{\chi_{b0}} = 1.58\times 10^3 , \nonumber\\
N_{\chi_{b1}} &=& 2.52\times 10^3  \;, N_{\chi_{b2}} = 2.90\times 10^2
\end{eqnarray*}
for the process $e^+e^- \to |H_{b\bar{b}}\rangle+b\bar{b}$ and
\begin{eqnarray*}
N_{\Upsilon}  &=& 6.55\times 10^3  \;, N_{\Upsilon'} = 3.24\times 10^3  \;, N_{\eta_b} = 8.18\times 10^3 , \nonumber\\
N_{\eta_b'} &=& 4.05\times 10^3
\end{eqnarray*}
for the process $e^+e^- \to |H_{b\bar{b}}\rangle+gg$.

These numbers show that the super Z factory will represent an excellent platform for studying the heavy quarkonium properties, complementing those performed at the $B$ factories BaBar and Belle and at the hadronic colliders Tevatron and LHC. To compare with the conventionally measured data on the prompt $J/\psi$ or $\eta_{c}$ ($\psi_b$ or $\eta_b$) production, we need to consider the feeddown contributions from all the higher charmonium (bottomonium) states. These merging contributions can be done by using our present results for each state with the help of the measured branching ratios~\cite{pdg} as ${\cal B}[\psi(2s)\rightarrow J/\psi]$, ${\cal B}(\chi_{cJ}\rightarrow J/\psi)$, and etc..

For the charmonium and bottomonium production channels via the $Z^0$ propagator, there is approximate ``spin degeneracy". Taking the charmonium production channels as an example, we obtain $$\frac{\sigma_{e^+e^- \to Z^0 \to \eta_c+c\bar{c}}}{\sigma_{e^+e^- \to Z^0 \to J/\psi+c\bar{c}}} \simeq 96\%$$ and $$\frac{\sigma_{e^+e^- \to Z^0 \to \chi_{c0}+c\bar{c}}}{\sigma_{e^+e^- \to Z^0 \to \chi_{c1}+c\bar{c}}} \simeq 91\%.$$ Such ``spin degeneracy" is confirmed by analyzing the quarkonium $p_T$- and $y$- distributions, and also by a cross check using the fragmentation approach.

We have also discussed the uncertainties related to the knowledge of the effective quark masses. For the charmonium production channel $e^{+}e^{-} \rightarrow Z^{0}\rightarrow |H_{c\bar{c}}\rangle+X$, when $X=c\bar{c}$, the uncertainties associated to the variation $m_c=1.50\pm0.15$ GeV are $\sim40\%$ for the $S$-wave case, and $\sim70\%$ for the $P$-wave case; when $X=gg$, the uncertainties for both the $S$-wave and $P$-wave cases are $\sim 15\%-19\%$. For the bottomonium production channel $e^{+}e^{-} \rightarrow Z^{0}\rightarrow |H_{b\bar{b}}\rangle+X$, when $X=b\bar{b}$, the uncertainties caused associated to the variation $m_b=4.90\pm0.15$ GeV are $\sim11\%$ for the $S$-wave case, and $\sim15\%-18\%$ for the $P$-wave case; when $X=gg$, the uncertainties for both the $S$-wave and $P$-wave cases are $\sim 6\%$. To reduce the theoretical uncertainties on the predictions for the super $Z$ factory, it will be important to perform a next-to-leading-order calculation for the channel $e^{+}e^{-} \rightarrow Z^{0}\rightarrow |H_{c\bar{c}}\rangle+X$, which is in progress.

\section{ACKNOWLEDGMENTS}

This work was supported in part by the Fundamental Research Funds for the Central Universities under Grant No. CDJXS12300004 and CQDXWL-2012-Z002, the Program for New Century Excellent Talents in University under Grant No. NCET-10-0882, and the Natural Science Foundation of China under Grant No. 11075225 and No. 11275280. The authors are grateful for the anonymous referee's comments and suggestions that substantially improve the paper.

\appendix

\section{Phase space splitting for the color-singlet case}

Generally, we can factorize the $2\to3$ phase into two parts
\begin{eqnarray}
&&\quad{\rm d}\Phi_{3}(p_{1}+p_{2};p_3,p_4,p_5) \nonumber\\
&=& (2\pi)^4 \delta^{(4)} \left(p_{1}+p_{2}-\sum_{i=3}^{5} p_i \right) \prod_{j=3}^{5} \frac{{\rm d}^3 p_j}{(2\pi)^32 E_j} \nonumber\\
&&= {\rm d} \Phi_2(q;p_1,p_2)\frac{{\rm d}q^2}{2\pi} {\rm d}\Phi_{3}(q;p_3,p_4,p_5) ,
\end{eqnarray}
where $q^2= (\sum_{i=3}^{5}E_i)^2 -|\sum_{i=3}^{5}\vec{p}_i|^2$ with $p_i=(E_i,\vec{p}_i)$. At the same time, the hard scattering amplitude can be rewritten as
\begin{eqnarray}
&&\quad\left|{\cal M}_{2\to3}\right|^{2} \nonumber\\
&=& \left|{\cal M}^\mu_{2\to1}\left(-g_{\mu\nu}+\frac{q_\mu q_\nu}{M_Z^2}\right) {\cal M}_{1\to3}^\nu\right|^{2} \nonumber\\
&=& \left|\sum\limits_{\lambda}{\cal M}^\mu_{2\to1}{\cal M}^\nu_{1\to3} \epsilon^{*}_{\mu}(q,\lambda) \epsilon_{\nu}(q,\lambda)\right|^2 \nonumber\\
&=& \frac{1}{3}\sum \limits_{\lambda1,\lambda_2} \left|{\cal M}^\mu_{2\to1} \epsilon^*_\mu(q,\lambda_1)\right|^2 \left|{\cal M}^\nu_{1\to3} \epsilon_\nu(q,\lambda_2)\right|^2 ,
\end{eqnarray}
where $\epsilon(q,\lambda)$ stands for the $Z^0$ polarization vector. Then the process $e^+e^- \to |H_{cc}\rangle+c\bar{c}$ can be divided into two parts: the $2\to1$ process ($e^+e^- \to Z^0$) and the $1\to3$ process ($Z^0 \to |H_{Q\bar{Q}}\rangle+Q\bar{Q}$). The phase space of the $2\to1$ process is easily calculable, while for the $1\to3$ process it is the same as for the $Z^0$ decay into three final particles.

\end{document}